\documentclass{aa}
\usepackage{graphicx}
\usepackage{txfonts}
\usepackage{todonotes}
\usepackage[normalem]{ulem}
\usepackage{amstext}
\usepackage[breaklinks=true]{hyperref}
\usepackage{xspace}
\usepackage{multirow}

\newcommand{\orcit}[1]{\protect\href{https://orcid.org/#1}{\protect\includegraphics[width=8pt]{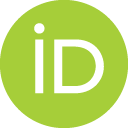}}}

\newcommand{\gdr}[1]{\gaia~DR#1\xspace}
\newcommand{\gedr}{\gaia~EDR3\xspace}

\newcommand{\gaia}{\textit{Gaia}\xspace}
\newcommand{\gband}{$G$--band\xspace}

\newcommand{\bprp}{\ensuremath{G_{\rm BP}-G_{\rm RP}}\xspace}

\newcommand{\xp}{BP and RP\xspace}
\newcommand{\xpor}{BP or RP\xspace}
\newcommand{\wrt}{with respect to\xspace}

\newcommand{\pupil}{\ensuremath{P_\tau}}
\newcommand{\gaiaxpy}{\texttt{GaiaXPy}\xspace}

\newcommand{\hip}{{ Hipparcos}\xspace}
\newcommand{\tyc}{{ Thyco}\xspace}
\newcommand{\tyctwo}{{ Thyco~2}\xspace}

\newcommand{\secname}{Sect.}
\newcommand{\equref}[1]{Eq.~\ref{eq:#1}}
\newcommand{\equrefs}[2]{Eqs.~\ref{eq:#1}-\ref{eq:#2}}
\newcommand{\secref}[1]{\secname~\ref{sec:#1}}
\newcommand{\asecref}[1]{Section~\ref{sec:#1}}
\newcommand{\appref}[1]{Appendix~\ref{sec:#1}}
\newcommand{\figref}[1]{Fig.~\ref{fig:#1}}
\newcommand{\afigref}[1]{Figure~\ref{fig:#1}}

\newcommand{\tabref}[1]{Table~\ref{tab:#1}}
\newcommand{\instref}[1]{\inst{\ref{inst:#1}}}
\begin{document} 
\authorrunning{P.~Montegriffo~et~al.}
\title{\gaia Data Release 3: External calibration of BP/RP low-resolution spectroscopic data}

\author{
P.~Montegriffo\orcit{0000-0001-5013-5948}\instref{oabo}\fnmsep\thanks{Corresponding author: P.~Montegriffo\newline
e-mail: \href{mailto:paolo.montegriffo@inaf.it}{\tt paolo.montegriffo@inaf.it}}
\and
F.~De~Angeli \orcit{0000-0003-1879-0488}\instref{ioa}
\and
R.~Andrae\orcit{0000-0001-8006-6365}\instref{mpia}
\and
M.~Riello\orcit{0000-0002-3134-0935}\instref{ioa}
\and
E.~Pancino\orcit{0000-0003-0788-5879}\inst{\ref{inst:oafi},\ref{inst:asi}}
\and
N.~Sanna\orcit{0000-0001-9275-9492}\instref{oafi}
\and
M.~Bellazzini\orcit{0000-0001-8200-810X}\instref{oabo}
\and
D.~W.~Evans\orcit{0000-0002-6685-5998}\instref{ioa}
\and
J.~M.~Carrasco\orcit{0000-0002-3029-5853}\instref{ub}
\and
R.~Sordo\orcit{0000-0003-4979-0659}\instref{oapd}
\and
G.~Busso\orcit{0000-0003-0937-9849}\instref{ioa}
\and
C.~Cacciari\orcit{0000-0001-5174-3179}\instref{oabo}
\and
C.~Jordi\orcit{0000-0001-5495-9602}\instref{ub}
\and
F.~van~Leeuwen\instref{ioa}
\and
A.~Vallenari\orcit{0000-0003-0014-519X}\instref{oapd}
\and
G.~Altavilla\orcit{0000-0002-9934-1352}\inst{\ref{inst:oaroma},\ref{inst:asi}}
\and
M.~A.~Barstow\orcit{0000-0002-7116-3259}\instref{lei}
\and
A.~G.~A.~Brown\orcit{0000-0002-7419-9679}\instref{leiden}
\and
P.~W.~Burgess\instref{ioa}
\and
M.~Castellani\orcit{0000-0002-7650-7428}\instref{oaroma}
\and
S.~Cowell\instref{ioa}
\and
M.~Davidson\instref{ifa}
\and
F.~De~Luise\orcit{0000-0002-6570-8208}\instref{oate}
\and
L.~Delchambre\orcit{0000-0003-2559-408X}\instref{liege}
\and
C.~Diener\instref{ioa}
\and
C.~Fabricius\orcit{0000-0003-2639-1372}\instref{ub}
\and
Y.~Fr\'{e}mat\orcit{0000-0002-4645-6017}\instref{obsbe}
\and
M.~Fouesneau\orcit{0000-0001-9256-5516}\instref{mpia}
\and
G.~Gilmore~\orcit{0000-0003-4632-0213}\instref{ioa}
\and
G.~Giuffrida\instref{oaroma}
\and
N.~C.~Hambly\orcit{0000-0002-9901-9064}\instref{ifa}
\and
D.~L.~Harrison\orcit{0000-0001-8687-6588}\inst{\ref{inst:ioa},\ref{inst:ifc}}\and
S.~Hidalgo\orcit{0000-0002-0002-9298}\instref{iac}
\and
S.~T.~Hodgkin\orcit{0000-0002-5470-3962}\instref{ioa}
\and
G.~Holland\instref{ioa}
\and
S.~Marinoni\orcit{0000-0001-7990-6849}\inst{\ref{inst:oaroma},\ref{inst:asi}}
\and
P.~J.~Osborne\instref{ioa}
\and
C.~Pagani\orcit{0000-0001-5477-4720}\instref{lei}
\and
L.~Palaversa\orcit{0000-0003-3710-0331}\inst{\ref{inst:zag},\ref{inst:ioa}}
\and
A.~M.~Piersimoni\orcit{0000-0002-8019-3708}\instref{oate}
\and
L.~Pulone\orcit{0000-0002-5285-998X}\instref{oaroma}
\and
S.~Ragaini\instref{oabo}
\and
M.~Rainer\orcit{0000-0002-8786-2572}\instref{oafi}
\and
P.~J.~Richards\instref{stfc}
\and
N.~Rowell\orcit{0000-0003-3809-1895}\instref{ifa}
\and
D.~Ruz-Mieres\orcit{0000-0002-9455-157X}\instref{ioa}
\and
L.M.~Sarro\orcit{0000-0002-5622-5191}\instref{daie}
\and
N.~A.~Walton\orcit{0000-0003-3983-8778}\instref{ioa}
\and
A.~Yoldas\instref{ioa}
}


\institute{
INAF -- Osservatorio di Astrofisica e Scienza dello Spazio di Bologna, Via Gobetti 93/3, 40129 Bologna, Italy
\label{inst:oabo}
\and
Institute of Astronomy, University of Cambridge, Madingley Road, Cambridge CB3 0HA, UK\label{inst:ioa}
\and
Max-Planck-Institute for Astronomy, K\"onigstuhl 17, 69117 Heidelberg, Germany\label{inst:mpia}
\and
INAF -- Osservatorio Astrofisico di Arcetri, Largo E. Fermi, 5, 50125 Firenze, Italy\label{inst:oafi}
\and
Space Science Data Center - ASI, Via del Politecnico SNC, 00133 Roma, Italy\label{inst:asi}
\and
Institut de Ci\`encies del Cosmos (ICC), Universitat de Barcelona (IEEC-UB), c/ Mart\'{\i} i Franqu\`es, 1, 08028 Barcelona, Spain
\label{inst:ub}
\and
INAF - Osservatorio Astronomico di Padova, Vicolo Osservatorio 5, 35122 Padova, Italy\label{inst:oapd}
\and
INAF -- Osservatorio Astronomico di Roma, via Frascati 33, 00078 Monte Porzio Catone (Roma), Italy\label{inst:oaroma}
\and
School of Physics \& Astronomy, University of Leicester, Leicester LE9 1UP, UK\label{inst:lei}
\and
Leiden Observatory, Leiden University, Niels Bohrweg 2, 2333 CA Leiden, The Netherlands\label{inst:leiden}
\and
Institute for Astronomy, School of Physics and Astronomy, University of Edinburgh, Royal Observatory, Blackford Hill, Edinburgh, EH9~3HJ, UK
\label{inst:ifa}
\and
INAF - Osservatorio Astronomico d'Abruzzo, Via Mentore Maggini, 64100 Teramo, Italy\label{inst:oate}
\and
Institut d'Astrophysique et de G\'eophysique, Universit\'e de Li\`ege, 19c, All\'ee du 6 Ao\^ut, 4000, Li\`ege, Belgium\label{inst:liege}
\and
Royal Observatory of Belgium, 3 avenue circulaire, 1180, Brussels, Belgium\label{inst:obsbe}
\and
Kavli Institute for Cosmology, Institute of Astronomy, Madingley Road, Cambridge, CB3 0HA, UK\label{inst:ifc}
\and
IAC - Instituto de Astrofisica de Canarias, Via L\'{a}ctea s/n, 38200 La Laguna S.C., Tenerife, Spain\label{inst:iac}
\and
Ruđer Bo\v{s}kovi\'c Institute, Bijeni\v{c}ka cesta 54, Zagreb, Croatia\label{inst:zag}
\and
STFC, Rutherford Appleton Laboratory, Harwell, Didcot, OX11 0QX, United Kingdom\label{inst:stfc}
\and
Department of Artificial Intelligence, Universidad Nacional de Educación a Distancia, c/ Juan del Rosal 16, E-28040 Madrid, Spain\label{inst:daie}
}

\date{Received April 27, 2022; accepted May 23, 2022}

\abstract
{\gaia Data Release 3 contains astrometry and photometry results for about $1.8$ billion sources based on observations collected by the European Space Agency (ESA) \gaia satellite during the first 34 months of its operational phase (the same period covered \gaia early Data Release 3; \gedr). Low-resolution spectra for 220 million sources are one of the important new data products included in this release.}
{In this paper, we focus on the external calibration of low-resolution spectroscopic content, describing the input data, algorithms, data processing, and the validation of the results. Particular attention is given to the quality of the data and to a number of features that users may need to take into account to make the best use of the catalogue.}
{We calibrated an instrument model to relate mean \gaia spectra to the corresponding spectral energy distributions (SEDs) using an extended set of  calibrators: this includes modelling of the instrument dispersion relation, transmission, and line spread functions. Optimisation of the model is achieved through total least-squares regression, accounting for errors in \gaia and external spectra.
}
{The resulting instrument model can be used for forward modelling of \gaia spectra or for inverse modelling of externally calibrated spectra in absolute flux units.}
{The absolute calibration derived in this paper provides an essential ingredient for users of BP/RP spectra. It allows users to connect BP/RP spectra to absolute fluxes and physical wavelengths.
} 
 

\maketitle


\section{Introduction}\label{sec:introduction}

The European Space Agency (ESA) mission \gaia \citep{Prusti2016} is designed to be self-calibrating for the large majority of its data products. For example, the core product of the mission, namely exquisitely accurate and precise astrometry for $\simeq 1.8$~billion celestial sources, is entirely based on observations obtained by the mission itself (relative positions at different epochs and the relative colours of the sources), while external data are only used for validation \citep{LL:LL-EDR3}. Analogously, the removal of any instrumental imprint and/or space- or time-dependent inhomogeneities from the mission all-sky photometry and spectrophotometry is achieved using repeated measurements of large sets of internal calibrators \citep{Riello2021, Carrasco2021, DeAngeli2022}.

However, within the production chain of photometry and spectrophotometry, there are two remarkable exceptions to this generally adopted approach:

    1) The physical flux scale, the main ingredient in the conversion of internally calibrated fluxes (expressed in $\rm e^-s^{-1}$) into physical units ($\rm W~m^{-2}nm^{-1}$), which is determined using an external set of spectrophotometric standard stars, the Gaia spectrophotometric standard stars \citep[SPSSs;][]{Pancino2021}. 
    
    2) The physical wavelength scale, required to convert internal pseudo-wavelength labels (pixels; see \secref{overview}) associated to fluxes in \xp spectra into wavelengths in physical units (nm), achieved (mainly) thanks to a set of external spectra of sources with strong emission lines at known wavelength.

As there is no way to infer these scales from \gaia data alone, it is necessary to make use of external calibration data.

The \xp Instrument Models (IMs), which include these two fundamental components, depend on a number of factors (the dispersion relation, the instrument response, and the line spread function (LSF); see Sects. \ref{sec:overview} and \ref{sec:instrumentModel}), which 
are derived using external data in the process known as \textit{absolute calibration}. The IM is the fundamental tool for forward modelling of \xp observations, starting from a theoretical model spectrum, with the main goal being to facilitate the inference of astrophysical parameters from their BP/RP spectra by comparison on the plane of observations \citep{Creevey2022}.
Once the parameters of the IM have been estimated (see \secref{processing}), the model can also be used in the opposite direction, that is, to transform an internally calibrated mean BP/RP spectrum \citep{DeAngeli2022} into a wavelength- and flux-calibrated spectrum that we call an externally calibrated spectrum (ECS). As the IM also includes the modelling of the LSF at any wavelength, its application significantly reduces the effect of photon mixing inherent to the slit-less spectra produced by the \xp spectrophotometers, enhancing the effective spectral resolution of ECS. It is important to realise that the IM solves for all the relevant factors (e.g. calibrations of flux and wavelength, and LSF) simultaneously, as they are deeply and inseparably entangled in BP/RP spectra. 

While \xp spectrophotometry has already been used for internal processing in previous releases \citep{Riello2021}, with \gaia Data Release 3 \citep[\gdr3,][]{DR3-top-level} the BP/RP spectra of about 220 million sources are released for the first time. 
These can be retrieved from the \gaia Archive\footnote{https://gea.esac.esa.int/archive/} as internally calibrated mean spectra in a continuous representation (see \citet{DeAngeli2022} for details) while for a subset of sources with $G<15$  they will also be provided as ECS sampled on a standard wavelength grid (see \secref{dataProducts} for more details on data formats). 
The Python package  \texttt{GaiaXPy} \citep{DeAngeli2022b} has been developed to help users to convert spectra from continuous to sampled representation in the internal or absolute flux scale (ECS). The tool also implements the IM  presented here to allow for the simulation of \gaia-like mean spectra from a
given spectral energy distribution (SED); for example a synthetic stellar spectrum or an absolute-flux-calibrated measured spectrum.

In this paper, we illustrate how the BP/RP IM is derived and how internally calibrated mean \xp spectra are converted into ECS, discussing the performances and the limitations of the final products. After giving an overview of the external calibration approach (\secref{overview}) and a description of the external calibrators (\secref{calibrators}), 
we describe the implementation of the IM (\secref{instrumentModel}) and the method implemented to reconstruct the ECS (\secref{basisInversion}). \asecref{processing} is dedicated to the description of the processing scheme and the main results are shown in \secref{results}, while \secref{validation} is dedicated to the validation of the calibrations. Finally, in \secref{caveats} we discuss a few known problems. 

\section{Overview of the problem}\label{sec:overview}

The instrument model (described in detail in \secref{instrumentModel}) allows us to relate a mean \xpor spectrum to the corresponding SED via an integral equation of the following kind:
\begin{equation}
\label{eq:xpCompact}
n_e(u) = \int\, I(u,\lambda)\cdot n_p(\lambda) \, {\rm d}\lambda 
,\end{equation}
where the observed spectrum $n_e(u)$ is the internally calibrated mean spectrum in units of $\textrm{e}^- \textrm{s}^{-1}$ ($u$ denotes a coordinate in data space, often referred to as a \emph{pseudo-wavelength}) and the kernel $I$ is a combination of few components:

    (i) the LSF, that is, the instantaneous one-dimensional intensity distribution in the spectrum of a monochromatic point source;

    (ii) the dispersion model, that is, the relation that links absolute wavelengths to pseudo-wavelength coordinates; and

    (iii)  the response model, which represents the ratio between the number of detected photons for wavelength interval and the number of photons per wavelength interval entering the telescope aperture.

As the detectors are photon-counting devices, all the relations are expressed in terms of $n_p(\lambda)$, the spectral photon flux distribution (SPD).
The SPD (hereafter expressed in units of $\rm photon~ m^{-2} s^{-1} nm^{-1}$) is related to the source SED via the equation
\begin{equation}
    n_p(\lambda) = \frac{10^{-8}\lambda}{hc} f(\lambda)
,\end{equation}
where $hc$ is the product of the Planck constant and the vacuum speed of light, and the SED $f(\lambda)$ is expressed in units of $\rm W ~m^{-2} nm^{-1}$ (the factor $10^{-8}$ compensates for per-nanometre flux units; all other measurements and constants are expressed in S.I. units).
\par~\par
The external calibration concept is based on two key assumptions: 
the first is that  all differential effects that impact raw observations (spatial variations of the instrument across the focal plane, different observing configurations, time evolution of instrument characteristics, etc.)  have been removed by the internal calibration chain \citep{DeAngeli2022} that precedes the external calibration; 
the second is that the common reference system defined by the internal calibration is similar to the physical instrument in some unspecified point of the focal plane.

These assumptions underlie the design of the external calibration strategy:
   
   a) The external calibration model is unique for all sources and in particular does not depend on source luminosity or colour. The first assumption above, although plausible, cannot be a priori guaranteed and any infringement of it will manifest as systematic differences within different classes of sources.
   
   b) Also, the IM can be parametrised in terms of corrections to the \textit{nominal} instrument, which is defined by our pre-launch knowledge and laboratory measurements made on the satellite components.

This second point helps to cope with degeneracies that are present in possible IM solutions: 
in principle the optimisation of the IM parameters could be derived from an arbitrarily large number and variety of standard stars (sources with known SED from independent ground- or space-based observations) by matching the predictions of the model with the corresponding observed BP/RP mean spectra. 
As discussed by \citet{Weiler2020}, the traditional approach of deriving a simple response of the instrument as a function of wavelength ---by computing the ratio between the observed spectrum and the SED for a limited set of (possibly featureless) calibrators--- will not work for a \gaia-like instrument because of the rather large width of the LSF compared to the wavelength scale of the response variations. As a consequence, the derived response changes with the spectral type of the calibrator: the LSF must be taken into account, creating the need for a much larger set of calibrators. 

\section{Calibrators}\label{sec:calibrators}

A reliable calibrator must satisfy several stringent requirements: it must be an isolated and point-like source with stable flux and high signal-to-noise ratio (S/N), and it should not be subject to strong interstellar extinction in order to avoid polarisation that could cause variations in the measured flux with the observing geometry and so on  \citep[see][and references therein]{Pancino2012,Pancino2021}. 

The data set of spectrophotometric standard stars (SPSSs)\footnote{\url{http://gaiaextra.ssdc.asi.it:8080/}}, expressly built over the years for the calibration of \gaia photometric and spectroscopic data, is composed of 111 stellar sources, calibrated to the CALSPEC\footnote{\url{https://www.stsci.edu/hst/instrumentation/reference-data-for-calibration-and-tools/astronomical-catalogs/calspec}} scale \citep{bohlin14,bohlin20} with flux accuracy of about $1\%$ \citep{Pancino2012,Altavilla2015,Pancino2021,Altavilla2021}. The SPSSs have been monitored for short-term constancy at the 0.005~mag level over a few hours \citep{Marinoni2016}. The current SPSS release is based on about 25\% of the spectra collected for the project; a more complete release, which will contain about 200 SPSSs, will be used to calibrate future {\em Gaia} releases. The SPSS data set was recently complemented \citep{Pancino2021} by a second set of 60 stars with looser requirements on the absolute accuracy (up to 5\%) and flux stability (up to variations of about 0.05~mag) but 
including stellar types not contained in the SPSS set (bright O, B and late M stars), the passband validation library (PVL). These were originally intended to be used for validation purposes only, but a subset of these were eventually included in some phases of the actual calibration of the IM. 
The consequence of the severe criteria applied to the selection of primary calibrators
is that the resulting stellar spectra are not independent from a mathematical point of view: their principal components span only a subspace of all possible spectral shapes and consequently not all the necessary instrument components would be constrained by these, allowing for degeneracies in the solutions. 
Implementing the IM as a perturbation of the nominal model confers the advantage that unconstrained components have a reliable a priori estimation.
However, to enforce more observational constraints to the IM, 
the set of primary calibrators has been extended by adding a set of secondary calibrators, including a wide variety of sources featuring strong emission lines over the entire wavelength range (mostly 
quasi-stellar objects (QSOs) and Wolf Rayet stars). Although these objects often show variability and require a special treatment in the processing (see \secref{processing}), they are essential to provide strong constraints to the wavelength and LSF calibrations.
For the  \gdr{3,} a total of 211 peculiar sources have been selected from the literature, including 188 QSOs from the Sloan Digital Sky Survey \citep[SDSS; see][and references therein]{Lyke2020}, 17 young stellar objects (YSOs) from the X-Shooter spectral library \citep[][]{Verro2021}, and six emission line sources (ELSs) from STELIB  \citep{LeBorgne2003}. 
Most of these objects have SEDs only partially covering the \gaia wavelength range: this limitation had consequences on the processing strategy, as described in \secref{processing}.
\begin{figure}
\begin{center}
\centerline{
\includegraphics[width=\columnwidth/2]{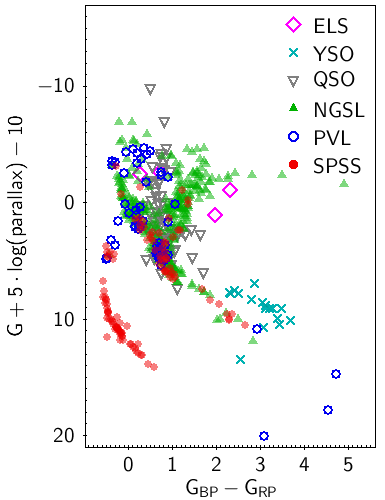}
\includegraphics[width=\columnwidth/2]{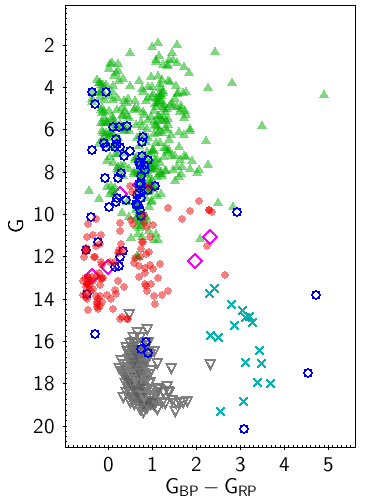}}
\caption{Colour--magnitude diagrams for the whole set of calibrators used in external calibration processing and validations. On the vertical axes are  absolute (\emph{left}) and apparent (\emph{right}) $G$ magnitudes.}
\label{fig:calibratorsCmd}
\end{center}
\end{figure}
Finally, we also 
used the catalogue of the Next Generation Spectral Library  \citep[NGSL,][]{ngsl16} for validation
purposes, which consists of 348 bright sources with magnitude ranging from $G=1.97$ and $G=12.0$.
\afigref{calibratorsCmd} shows the whole pool of calibrators and validation sources in a colour--magnitude diagram either in terms of absolute (\emph{left}) or apparent (\emph{right}) $G$ magnitudes as function of \bprp colour.


\section{Instrument model}\label{sec:instrumentModel}

The \gaia satellite observes the sky spinning around its axis \citep{Prusti2016}: the light collected by its telescopes is projected onto the focal plane where an array of charge-coupled device (CCD) detectors make measurements while operating in time-delay integration (TDI) mode. In the case of \gaia spectrophotometers, two slit-less prisms disperse the light onto two separate rows of seven CCDs each, both covering the focal plane in the \emph{across scan} (AC) direction. The dispersion direction, referred as \emph{along scan} (AL) being aligned with the transit direction of the projected light, is perpendicular to the AC direction. Each spectrophotometer covers part of the spectral wavelength interval, and the two ranges  partially overlap: the blue photometer (BP) covers the nominal range [330, 680] nm while the red photometer (RP) covers the range [640, 1050] nm. 
For each observed source, to limit the telemetry of the satellite, only a small window around the position of the source is actually read and transmitted to Earth during each transit.
Windows are $60\times12$ pixels wide in each of the AL and AC directions. The maximum exposure time of an observation ($\sim4.4$ s) is fixed by the velocity by which the image transits on a CCD, but, to avoid saturation, it can be reduced according to the magnitude of the source by limiting the activation of the reading window to only a portion of the CCD  with
\emph{gates} \citep{DeAngeli2022}.
Sources brighter than $G\simeq11.5$ are transmitted as 2D windows (called window class 0, WC0) while for fainter sources, data are binned in the AC direction producing 1D spectra of 60 samples (WC1). Each source is observed several times during the lifetime of the mission under many different observing configurations and conditions \citep[see][for an exhaustive description]{Carrasco2021}. The internal calibration \citep{DeAngeli2022} has the complex task of calibrating all these configurations to reduce spectra to a common reference system, the \emph{mean} instrument. 

If we assume that the dispersion of the prism is perfectly aligned with the AL direction, then we can model the dispersed image of a point-like source in the data space as
\begin{equation}
    \Im(u, w) =  \pupil\, \int\limits_0^\infty n_p(\lambda) \, P_\lambda\left(u - u_d(\lambda), w \right) \, R(\lambda) \, {\rm d}\lambda
    \label{eq:2Dmodel}
,\end{equation}
where 
$u$ is the continuous coordinate in data space in the AL direction, $w$ is the continuous coordinate in data space in the AC direction, $P_\tau$ is the telescope pupil area, $n_p(\lambda)$ is the SPD of the source,
$u_d(\lambda)$ is the dispersion function, $P_\lambda(u, w)$ is the effective monochromatic point spread function (PSF) at wavelength $\lambda$, and $R(\lambda)$ is the overall instrument response function.
This relation assumes indirectly that non-linear effects, such as those produced by charge transfer inefficiency (CTI) effects\footnote{CTI, by delaying the release of the charge by a physical pixel as CCD charges scroll in the AL direction, would result in a deformation of the source spectrum that depends not only on the source SED but also on the \emph{scene} of the observation, i.e. the temporal sequence of sources observed by that particular pixel immediately before the considered transit.}, are not important and can be neglected.
As all internally calibrated spectra are binned to 1D windows, we can integrate the previous equation in the AC direction, obtaining the following model, which is suitable for describing a mean spectrum:
\begin{equation}
   n_{e}(u)=  \pupil\, \int\limits_0^\infty n_p(\lambda) \, L_\lambda\left(u - u_d(\lambda)\right) \, R(\lambda) \, {\rm d}\lambda
   \label{eq:model}
,\end{equation}
where $n_{e}$ is given in units of $\rm e^- s^{-1}$ and $L_\lambda(u)$ is the effective monochromatic LSF at wavelength $\lambda$ obtained by integrating $P_\lambda$ in the AC direction. This is the explicit form of \equref{xpCompact}.
A detailed description of each factor of the model is given in the following subsections.

\subsection{Dispersion model}\label{sec:dispersion}

Airbus Defence and Space (DS), the company in charge of developing and building the \gaia satellite, provided nominal dispersion functions  based on chief-ray analysis for the \xp prisms in units of millimetres as a function of wavelength, by fitting a sixth-degree polynomial to the unperturbed \gaia optical design. For each field of view (FoV), dispersion functions are provided for the centre of each CCD (the dispersion varies in the AC direction) in the form of the coefficients $A_i$ of the expansion 
\begin{equation}
\label{eq:nominalDf}
AL(\omega) - AL(\omega_{ref}) = \sum_{i} A_i \omega^i,
\end{equation}
where 
\begin{itemize}
\item $AL(\omega)$ denotes the AL image position in mm, 
\item $\omega = 1/\lambda$ in $\rm nm^{-1}$ denotes the wavenumber, and 
\item $\omega_{ref} = 1/440~\rm nm^{-1}$ for BP and $\rm 1/800~nm^{-1}$ for RP, corresponding roughly to the central wavelength of each instrument. 
\end{itemize}
\begin{figure}[t]
    \centering
    \includegraphics[width=(\columnwidth)]{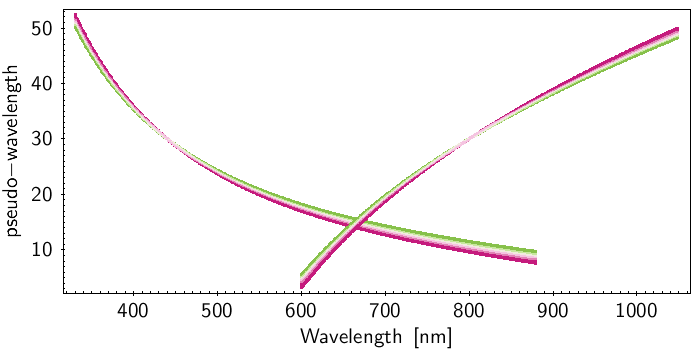}
    \caption{Pre-launch nominal dispersion relations for \xp instruments. Different curves refer to all FoV/CCD row combinations.}
    \label{fig:dispersion}
\end{figure}
The AL position in pixel units $u$ is obtained by dividing \equref{nominalDf} by the pixel size of a CCD in the AL direction $P_{AL}$.
Nominal dispersion curves for all FoV/CCD row combinations are shown in \figref{dispersion}.
The dispersion curve varies across the focal plane due to the tilt of the prisms \wrt the focal plane assembly: the comparison of the dispersion 
functions for different AC positions shows that the functions are related through a linear scaling  to a high
degree and are virtually independent from the FoV.
We can therefore arbitrarily assume the coefficients for any of the CCD rows or FoV and model the generic dispersion function as:
\begin{equation}
\label{eq:dispModel}
   u_d(\lambda) =\sum_{k=0}^{N_u-1}  d_k  \cdot  \left[ \frac{1}{P_{AL}}\sum_{i=0}^{N}  A_i \frac{1}{\lambda^i} \right]^k
,\end{equation}
where
$u_d(\lambda)$ denotes the AL image position in pixel units, coefficients $A_i$ are arbitrarily assumed as those of CCD row 4, and FoV 1 
and $d_k$ are the IM parameters to be optimised in the calibration process. 
For \gdr3, we assume a number of parameters  $N_u=3$.
Lower order parameters can be interpreted as follows: 
parameter $d_0$ represents the zero point of the dispersion relation and by construction is the reference AL position $u_{ref}$ corresponding to the reference wavenumber $\omega_{ref}$; 
and parameter $d_1$ is the scale of the dispersion relation.
The nominal values are $d_0 = 30$, $d_1 = 1.0,$ and $d_2 = 0.0$ for both \xp instruments.

\subsection{The line spread function model}\label{sec:lsf}

\begin{figure}
\begin{center}
\centerline{
\includegraphics[width=\columnwidth/2]{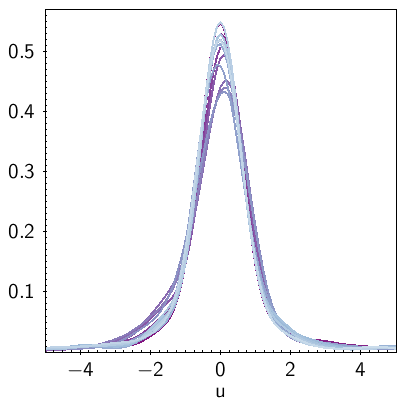}
\includegraphics[width=\columnwidth/2]{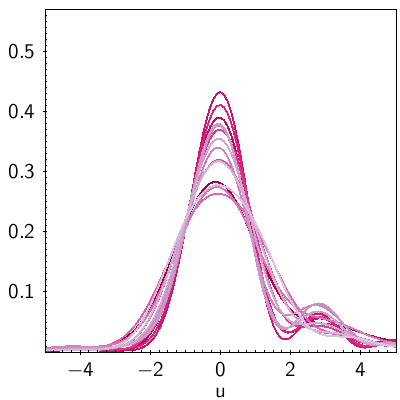}}
\caption{Example of pre-launch nominal monochromatic LSF computed at wavelength $\lambda = 440$ nm  (\emph{left}) and $\lambda = 800$ nm  (\emph{right}). Different curves represent the LSF for each FoV/CCD row combination.}
\label{fig:lsfNom}
\end{center}
\end{figure}

The LSF model is the only component of the IM that cannot be implemented as a simple perturbation of a nominal model,  as in the cases of the dispersion and response models.
Numerical monochromatic LSFs were provided by Airbus DS for testing purposes for each combination of FoV and CCD row; \figref{lsfNom} shows two sets of these LSFs computed at wavelength $\lambda = 440$ nm for the BP instrument  (\emph{left}) and $\lambda = 800$ nm for the RP instrument (\emph{right}).
These models were built upon the optical PSF model of the telescope, which included optical aberrations based on laboratory measurements of the wavefront error (WFE) maps made on the telescope mirrors. The great variations in the shapes of the LSF shown in the figure are essentially due to variations in the WFE map from one FoV/CCD pair to the other. The problem is that these 
WFE maps are not applicable to the flying instrument because several factors (changes in physical conditions, mechanical stress of the launch, defocusing of the instrument, etc.) lead to them changing in an unpredictable way.

The strategy adopted for the current model implementation, which is explained in more detail in \appref{lsfDetailed}, is to create a large sample of theoretical PSFs based on the optical design of \gaia, including randomly generated realistic WFE maps: these optical PSFs are then converted to effective PSFs that include a number of effects (charge diffusion, smearing introduced by TDI, pixel integration) to account for the discretised nature of the data. Effective PSFs are then marginalised in the AC direction to obtain a set of numerical LSFs
sampled on a two-dimensional grid in spatial and wavelength coordinates. These numerical LSFs are then used to build a set of two-dimensional basis functions to allow the LSF to be modelled with a minimum number of free parameters for a given accuracy. The reduction to the 2D basis functions is achieved by means of generalised principal component analysis \citep[GPCA,][]{ye}, a fast and efficient algorithm for 2D image compression used to concentrate relevant information of a given data set in a small number of dimensions. Unlike the usual principal component analysis, GPCA is able to preserve the spatial locality of pixels in an image by projecting the images to a vector space that is the tensor product of two lower dimensional vector spaces.
\begin{figure}
\begin{center}
\centerline{
\includegraphics[width=\columnwidth/2]{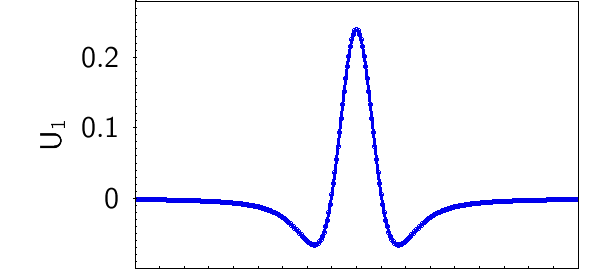}
\includegraphics[width=\columnwidth/2]{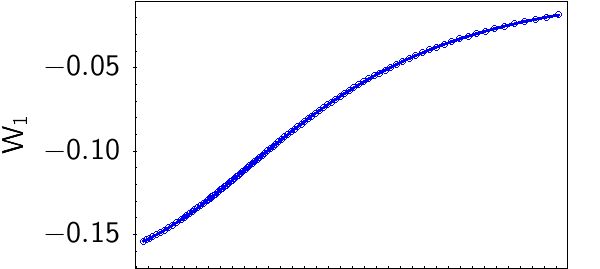}}
\centerline{
\includegraphics[width=\columnwidth/2]{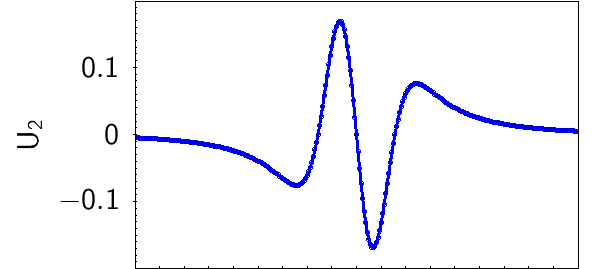}
\includegraphics[width=\columnwidth/2]{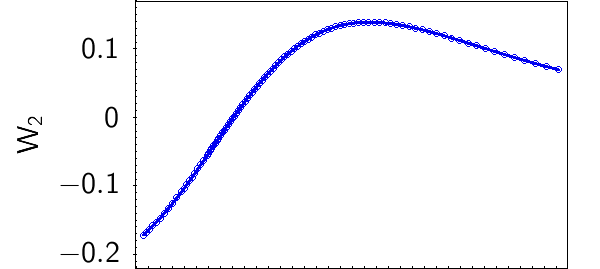}}
\centerline{
\includegraphics[width=\columnwidth/2]{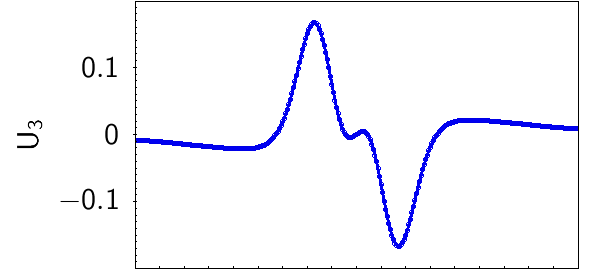}
\includegraphics[width=\columnwidth/2]{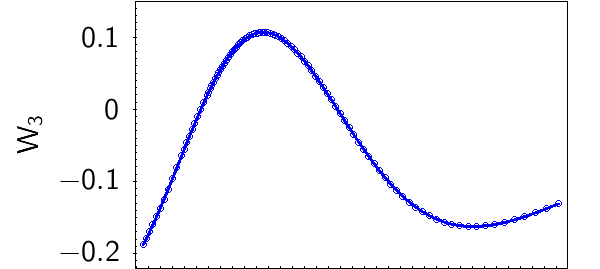}}
\centerline{
\includegraphics[width=\columnwidth/2]{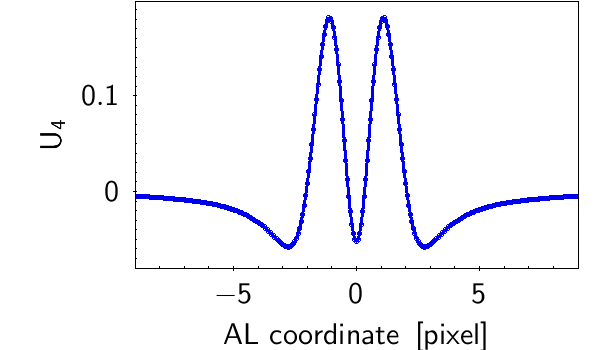}
\includegraphics[width=\columnwidth/2]{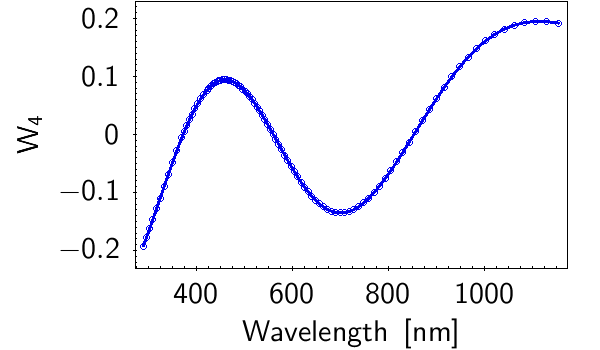}}
\caption{LSF basis functions.
\emph{Left}: First four basis functions of  matrix U as a function of the AL coordinate.
\emph{Right}: First four basis functions of matrix W  as a function of wavelength.}
\label{fig:lsfBases}
\end{center}
\end{figure}
These two vector spaces are designed by matrices $U$ and $W$ whose columns represent the basis functions with which the dependencies are modelled along the spatial coordinate $u$ and the wavelength coordinate $\lambda,$ respectively.
\afigref{lsfBases} shows the first four bases for vector space U (\emph{left}) and W (\emph{right}). 
 The U and W bases are interpolated to continuous variables ($u$, $\lambda$) by 1D interpolation.
To ensure that the interpolation for the U bases satisfies the `shift invariant sum' condition, which preserves the underlying function normalisation independently from the subpixel position of the sampling grid,  these bases were then fitted with an S-spline model \citep{LL:LL-084}. The interpolation for W bases is achieved by a cubic spline.
The model for the LSF is finally  given by
\begin{equation}
\label{eq:fullLsf}
L(u, \lambda) = \overline{L}(u, \lambda) + \sum_{m=1}^{\ell_1}  \sum_{n=1}^{\ell_2} d_{m,n} \cdot U_{m}(u) \cdot W_{n}(\lambda)
,\end{equation}
where $\overline{L}(u, \lambda)$ is the numerical mean LSF of the theoretical set and $d_{m,n}$ are the IM parameters that are fitted during the external calibration processing. As explained in \appref{lsfDetailed}, the LSF wavelength modelling requires roughly half the dimensions needed for AL dependency modelling, hence $\ell_1\simeq 2\ell_2$. Reiterating the caveat expressed at the beginning of this section regarding the absence of a proper \emph{nominal} LSF model, when all parameters $d_{m,n}$ are set to zero the LSF coincides with the mean numerical model  $\overline{L}(u, \lambda)$.


Equation \ref{eq:fullLsf} represents an undispersed LSF centred on the origin of the $U$ bases. 
The dispersed LSF in the data space 
can be obtained by shifting the origin by an amount given by the dispersion relation as seen in \equref{model}:
$L\left(u-u_d(\lambda), \lambda\right)$. However, it is worth noting that 
the LSF origin does not necessarily coincide with its centroid: the centroid is an intrinsic property of the LSF, and in the case of a symmetric LSF is naturally given by the point of symmetry $\xi$, that is $L(\xi-u) = L(\xi+u)~ \forall u$, but in general the LSF is not symmetric and its shape can change with wavelength, introducing some degeneracy between the chromaticity and the dispersion.
\begin{figure}[t]
    \centering
    \includegraphics[width=(\textwidth-4mm)/2]{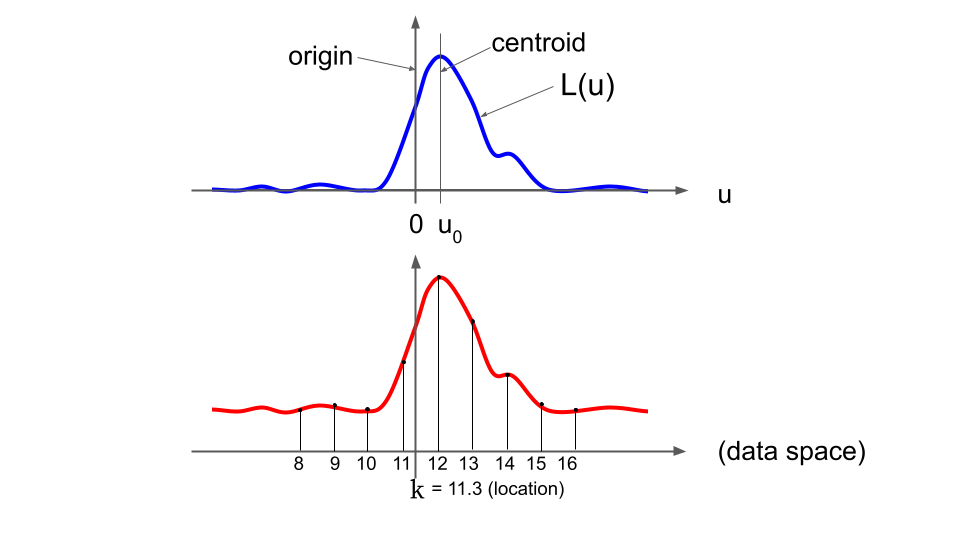}
    \caption{Definition of centroid, origin, and location. \emph{Top}: A schematic monochromatic LSF at a given wavelength $\lambda_0$ with origin $u = 0$ and centroid $u=u_0$. \emph{Bottom:} Location ${\kappa}$ of the LSF in the data stream of sample values: a narrow emission line at wavelength $\lambda_0$ is overlying a continuous signal at $k = \kappa$.}
    \label{fig:centroid}
\end{figure}
As sketched in \figref{centroid}, the dispersion model of \equref{dispModel} provides the location of the LSF origin: 
as far as the instrument model is concerned, the 
degeneracy between dispersion and chromaticity has little significance because the dispersion model and the LSF origin are consistently defined; however, physical interpretation of the data requires a dispersion relation that gives the centroid of the monochromatic LSF as a function of wavelength. This dispersion relation is provided as a lookup table where  the centroid $u_0$ is computed at each wavelength $\lambda$ by solving the non-linear equation:
\begin{equation}
\label{eq:centroid}
\int\limits_{-\infty}^\infty L(u_0 + u, \lambda) \, w(u/s) \, {\rm d}u = 0
,\end{equation}
where 
the weighting function $w$ is the Tukey's bi-weight: 
\begin{equation}
\label{eq:tukeysbiweight}
w(z) = \left\{
\begin{array}{l@{\quad}l}
z \, \left(1-z^2\right)^2 & {\rm if} ~|z| < 1\\~\\
0                & {\rm otherwise,} 
\end{array}\right.
\end{equation}
and the scale parameter $s = 2.7$ is a value
suitable for the \gaia case \citep{LL:LL-068}.
This dispersion function is provided for both BP and RP instruments as a single CSV file tabulated for wavelengths ranging from 320 nm to 1100 nm in steps of 0.5 nm\footnote{Available at \url{https://www.cosmos.esa.int/web/gaia/dr3-bprp-instrument-model}}.

\subsection{Response model}

The response $R(\lambda)$  defined in \secref{overview}, as the ratio between the number of detected photons and the number of photons entering the telescope aperture per wavelength interval, is modelled as the product of the individual responses of each physical element (e.g. primary mirror, secondary mirrors) hit along the optical path. It changes across the focal plane, and depends on the observing configuration of each source and transit (gates, window class) and on time (contamination and decontamination issues). Assuming that all these dependencies have been accounted for by the internal calibration, we are left with a function of the wavelength alone. 

\begin{figure}[t]
    \centering
    \includegraphics[width=(\textwidth-4mm)/2]{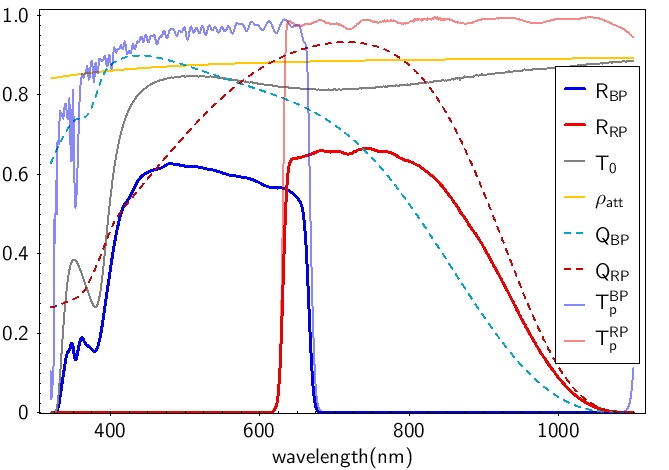}
    \caption{Pre-launch nominal responses for \xp instruments.}
    \label{fig:nomResp}
\end{figure}

The nominal pre-launch response curve for the mean \xp instruments can be described   \citep{Jordi2006} as the product of the following elements: 
\begin{equation}
\label{rnominal}
R_{N}(\lambda) = T_0(\lambda) \rho_{att}(\lambda) Q(\lambda) T_p(\lambda),
\end{equation}
where
\begin{itemize}
  \item $T_0(\lambda)$ is  the telescope (mirrors) reflectivity; 
  \item  $\rho_{att}(\lambda)$ is the attenuation due to rugosity (small-scale variations in smoothness of the surface) and molecular contamination of the mirrors; 
  \item $Q(\lambda)$ is the typical CCD quantum efficiency curve; 
  \item $T_p(\lambda)$ is the prism (fused silica) transmittance curve including the filter coating.
\end{itemize}
These quantities were initially measured by Airbus DS during on-ground laboratory test campaigns and are plotted in \figref{nomResp}. 
As can be seen, the steepest features of these curves are the BP and RP cut-offs produced by the prism transmittance curves, and the steep BP drop around $\simeq 400$~nm which is mainly due to mirror reflectivity. Laboratory measurements 
showed that the precise location, that is, in wavelength, of the cut-off 
varies across the focal plane due to the uneven thickness of the prism coating (which is a few nm in both instruments).
Moreover, combining measurements taken all over 
the focal plane  results in a further smearing of the nominal curve.
Therefore, a suitable modelling of the actual response  cut-offs 
has been achieved by assuming the nominal curves, degrading the wavelength resolution by convolution with a rectangular window of width varying with the spectral dispersion per pixel, and  
re-shaping the cut-off mathematically with a two-parameter Gauss error function for RP and a two-parameter complementary error function for BP to control the wavelength position $\lambda_{C}$ and the slope $\sigma_{C}$ of 
these features (the tabulated transmittance curve is in practice truncated just before the cut-off and multiplied by the error function to mimic the cut-off shape).
Nominal values for $(\lambda_{C},\,\sigma_{C})$ are $(667.9,\,4.71)$ and $(631.0,\,4.0)$ for BP and RP, respectively. 
The corresponding curves are represented in \figref{nomResp} as blue (BP) and red (RP) thick lines.
It is well known that the actual on-board overall response was heavily affected by rapid and discontinuous variations 
due to water vapour contamination of the satellite instrument components  \citep{2016A&A...595A...1G} and to the various decontamination campaigns. The internal calibration initialises the internal reference system using only high-quality data collected in periods of low and slowly varying contamination \citep{DeAngeli2022}, which ensures that the mean instrument response is not too different from the nominal $R^*_N(\lambda; \lambda_{C},\sigma_{C})$. Nevertheless, to be able to model any deviation from the nominal curve, the response is multiplied by a parametric function: 
\begin{equation}
    R(\lambda) = R^*_N(\lambda; \lambda_{C},\sigma_{C}) \cdot R_d(\lambda; r_i).
    \label{eq:shapeParams}
\end{equation}
Parameters $r_i$ are referred to as response \emph{shape} parameters to distinguish them from the two response \emph{cut-off} parameters $\lambda_C$ and $\sigma_C$; their nominal values are all zeros. 
The distortion model $R_d$ is implemented as the exponential of a linear combination of a set of basis functions $\mathcal{R}$ in the AL sampling space $u$:
\begin{equation}
    R_d(u) = e^{\sum r_i \mathcal{R}_i(u)}
    \label{eq:respShape}
,\end{equation}
which is transformed to wavelength space $\lambda$ through the dispersion relation:
\begin{equation}
    R_d(\lambda) = R_d\left(u_d(\lambda)\right)\frac{{\rm d} u}{{\rm d}\lambda}
.\end{equation}
The exponential form guarantees the non-negativity of the overall model, while modelling in the sample space ensures the natural instrument spectral resolution, avoiding over-fitting where the spectral resolution is lower. 
The basis functions $\mathcal{R}_i$ used for the \gdr3 models are spline functions of second order with
an initial uniform knot spacing in $u$ that becomes non-uniform in later processing stages 
(see \secref{processing} for details).
The full set of IM response parameters to be optimised is 
therefore $(r_i, \lambda_C, \sigma_C)$: when they are set to their nominal values, the response model closely resembles the nominal model.


\section{Basis inversion and SED reconstruction}\label{sec:basisInversion}

Once the instrument model is defined, \equref{model} can be used to estimate the spectral photon flux distribution $n_p(\lambda)$ corresponding to an observed spectrum $n_e(u)$. 
Obtaining the SPD/SED allows the user to inspect \gaia spectra in a format that is more intuitive and of common usage.
It may be convenient to define an 
\emph{effective} spectral photon distribution as:
\begin{equation}
    n_p^*(\lambda) = n_p(\lambda) \cdot R(\lambda)
    \label{eq:effective}
,\end{equation}
so that
\begin{equation}
\label{eq:modelEffective}
n_{e}(u)=  \pupil\, \int\limits_0^\infty n_p^*(\lambda) \, L_\lambda\left(u - u_d(\lambda)\right) \, {\rm d}\lambda
.\end{equation}
The effective spectrum is the observed spectrum deconvolved by the LSF function and transformed to the wavelength space through the dispersion relation (and scaled by some factor). Therefore, its shape will preserve the basic features of the observed spectrum.
However, \equref{modelEffective} is a Fredholm integral equation of the first kind, which is difficult to solve for the unknown $n_p^*(\lambda)$ because such integral equations are often ill-posed problems: large variations in the solution $n_p^*(\lambda)$ can occur for a slightly perturbed observable $n_{e}(u)$ (as is the case here, as   $n_e$ is affected by noise). 
 However, as BP/RP mean spectra are modelled as a linear combination of basis functions \citep{Carrasco2021}:
\begin{equation}
    n_e(u) = \sum_n b_n  \varphi_n(u)
,\end{equation}
an interesting solution can be found by modelling the effective spectral photon distribution as a linear combination of the {same} spectral coefficients $b_n$ with a particular set of bases:
\begin{equation}
    n_p^*(\lambda) = \sum_{n=1}^N b_n \phi_n(\lambda)
,\end{equation}
where the $\phi_n$ bases satisfy the following condition:
\begin{equation}
\label{eq:invBases}
\varphi_n(u)=  \pupil\, \int\limits_0^\infty \phi_n(\lambda) \, L_\lambda\left(u - u_d(\lambda)\right) \, {\rm d}\lambda
.\end{equation}
In practice, the externally calibrated spectral photon distribution (and related SED) corresponding to each pair of observed \xp spectra can be reconstructed 
by finding a set of proper functions whose images through the dispersed LSF model are the bases of the internal representation: the great advantage of this approach with respect to solving \equref{modelEffective} directly is that it requires inverting the integral equation for a set of analytic functions that are by definition noise-free. We refer to these functions $\phi_n$ as the \emph{inverse bases}\xspace hereafter. 

To choose the most suitable representation for the inverse bases, it is useful to review the representation used for internally calibrated mean spectra described in \citet{Carrasco2021} and \citet{DeAngeli2022} and summarised here for convenience. The basis functions implemented for \gdr3 are orthonormal Gauss-Hermite functions $\varphi_n(\theta)$ where a linear transformation is set between the pseudo-wavelengths axis $u$ and the argument of the Hermite functions $\theta$ as:
\begin{equation}
    \theta = \frac{u - \Delta \theta}{\Theta} 
.\end{equation}
The number of bases for \gdr3 has been set to $N=55$ for both \xp spectra.
An optimisation post-process is applied to mean spectra basis functions to concentrate most of the information in the lower order spectral coefficients: this optimisation takes the form of a rotation of the bases specified by a square matrix $V_C$. This rotation has no consequence for the basis inversion algorithm described here because, whenever \equref{invBases} is satisfied, the inverse bases for the optimised bases are simply obtained by applying the same rotation matrix to the inverse basis set $\phi_n$.
The model chosen to represent each $\phi_n$ function is a linear combination of the same bases that model their image through the instrument model, that is, a linear combination of Hermite functions:
\begin{equation}
    \label{eq:inverseBasisModel}
    \phi_n(\theta) = \sum_{k=1}^K h_{k,n} \cdot \varphi_k(\theta)
,\end{equation}
with the same mapping between axes $u$ and $\theta$ as adopted for the bases used to represent internally calibrated spectra:
\begin{equation}
   \label{eq:invBasesModel}
   \varphi_n\left(\frac{u - \Delta \theta}{\Theta}\right)=  \pupil\, \int\limits_0^\infty \sum_{k=1}^K h_{k,n}\varphi_k\left(\frac{u_d(\lambda) - \Delta \theta}{\Theta}\right) \, L_\lambda\left(u - u_d(\lambda)\right) \, {\rm d}\lambda
.\end{equation}
The reason for choosing such a model is that, as in the case of the effective photon distribution, left and right bases should share the same basic features,
as the function on the left hand side of the equation is a smeared version of that on the right hand side
once mapped to the same axis $u$.

We could solve \equref{invBasesModel} for coefficients $h_{k,n}$ in a least-squares sense by sampling $\varphi_n$ on a sufficiently dense and extended grid on the $\theta$ axis, but in this case 
the optimal number of bases $K$ of the model would be undefined and it would not be clear whether or not a limit to this number were set by some hidden condition.
A more appealing possibility is to project \equref{invBasesModel} into the coefficient space $b_n$ of mean spectra, where the $n^{th}$ function $\varphi_n$ is represented by definition by a vector of  coefficients that are all null except the $n^{th}$ one equal to unity. 
 In matrix notation, let $\textbf{b}$ denote the array of coefficients,
 $\textbf{s}$ a mean spectrum sampled on a given grid $\textbf{u}$ of $U$ points,
 and $\mathrm{D}\! \in\mathbb{R}^{U\times N}$  the design matrix whose element $D_{nu}$ is the value of the $n^{th}$ Hermite function evaluated at the $u^{th}$ pixel grid point. Consequently, 
\begin{equation}
    \label{eq:sampledSpec}
    \textbf{s} = \mathrm{D} \cdot \textbf{b}
,\end{equation}
and
\begin{equation}
   \label{eq:coeffsProjection}
   \textbf{b} = \mathrm{D}^\dagger \cdot \textbf{s}
,\end{equation}
where 
$\mathrm{D}^\dagger\! \in\mathbb{R}^{N\times U}$ is the 
pseudo-inverse of the design matrix (see  \appref{projCoeffs} for details).
The integrals of \equref{invBasesModel} can be computed numerically by trapezoidal integration over a fine regular wavelength grid  
with $\Lambda$ points and step $\delta\lambda$ extending over the wavelength interval where the response is not null: let $\mathcal{L}\! \in\mathbb{R}^{U \times \Lambda}$ represent the instrument dispersed LSF model kernel sampled over that discrete grid, with the $(i, j)^{th}$ element being 
\begin{equation}
   \mathcal{L}_{i, j} = \pupil \cdot L_{\lambda_j}\left(u_i - u_d(\lambda_j)\right) \, \delta\lambda
.\end{equation}
We can write an equation like \equref{invBasesModel} for each of the left $N$ Hermite basis functions: if the left term is interpreted as a column of the design matrix $\mathrm{D}$, then all the $N$ relations can be condensed into one single equation:
\begin{equation}
    \label{eq:invBasesModelMatrix}
    \mathrm{D} =  \mathcal{L} \cdot \mathrm{D}_\varphi \cdot \mathrm{H}
,\end{equation}
where $\mathrm{D}_\varphi \in\mathbb{R}^{\Lambda \times K}$ is the design matrix for the right $\varphi_k$ bases sampled on the wavelength integration grid, while matrix $\mathrm{H} \in\mathbb{R}^{K \times N}$ contains in its columns the set of coefficients $h_{k,n}$ that define the shape of each inverse basis.
By multiplying \equref{invBasesModelMatrix} by $\mathrm{D}^\dagger$ from the left we finally obtain:
\begin{equation}
    \mathrm{I}_N =  \mathrm{D}^\dagger \cdot \mathcal{L} \cdot \mathrm{D}_\varphi \cdot \mathrm{H}
,\end{equation}
where the left hand term is the $(N\times N)$ identity matrix.
By setting
\begin{equation}
    \mathrm{B} =  \mathrm{D}^\dagger \cdot \mathcal{L} \cdot \mathrm{D}_\phi
,\end{equation}
it is evident that the problem can be solved if $\mathrm{B}$ is a square matrix with $K=N$, that is, the number of inverse bases is equal to the number of bases for mean spectra representation.
In this case, the matrix $\mathrm{H}$ that defines the basis functions for the externally calibrated spectra 
is simply given by:
\begin{equation}
    \mathrm{H} =  \mathrm{B}^{-1}
.\end{equation}


The model for inverse bases consists of a matrix of coefficients H for each of the \xp instruments. Let $\boldsymbol{\lambda}$ be the wavelength grid over which we sample the externally calibrated spectrum corresponding to a pair of \xp observed spectra and $\Lambda$ be the dimension of vector $\boldsymbol{\lambda}$.
We can build a design matrix $\mathrm{D}_\varphi \in\mathbb{R}^{\Lambda \times N}$  by sampling the $N$ Hermite functions on the grid:
\begin{equation}
    \boldsymbol{\theta} = \frac{u_d(\boldsymbol{\lambda}) - \Delta \theta}{\Theta}
.\end{equation}
If $V_C$ is the orthogonal rotation matrix that defines the optimisation of the basis functions for the internally calibrated mean spectra, we obtain that the sampled effective spectral photon distribution is defined as
\begin{equation}
    \mathbf{n_p^*} = \left(\mathrm{D}_\varphi \cdot \mathrm{H} \cdot V_C^T\right) \cdot \mathbf{b}
.\end{equation}
If we build two more design matrices 
$\mathrm{D^P}_\varphi$ and $\mathrm{D^E}_\varphi$
whose elements are respectively defined as
\begin{equation}
    {\mathrm{D}_\varphi^P}_{\,i, j} = {\mathrm{D}_\varphi}_{\,i, j} \,\frac{1}{ R(\lambda_i)}
    \label{eq:designSpd}
,\end{equation}
and 
\begin{equation}
    {\mathrm{D}_\varphi^E}_{\,i, j} = {\mathrm{D}_\varphi}_{\,i, j} \,\frac{1}{ R(\lambda_i)} \,\frac{10^8 \mathrm{h} \mathrm{c}}{\lambda_i}
    \label{eq:designSed}
,\end{equation}
where $R(\lambda)$ is the instrument response and $hc$ the product of the Planck constant and the vacuum speed of light, then we get
\begin{equation}
    \mathbf{n_p} = \left(\mathrm{D}_\varphi^P \cdot \mathrm{H} \cdot V_C^T\right) \cdot \mathbf{b}
    \label{eq:reconstructedSpd}
,\end{equation}
which is the SPD in units of $\rm photons\, s^{-1} m^{-2} nm^{-1}$, 
and
\begin{equation}
    \mathbf{f_\lambda} = \left(\mathrm{D}_\varphi^E \cdot \mathrm{H} \cdot V_C^T\right) \cdot \mathbf{b}
    \label{eq:reconstructedSed}
,\end{equation}
which is the SED in units of $\rm W m^{-2} nm^{-1}$.

As the design matrices of \equrefs{designSpd}{designSed} depend on the inverse of the instrument response function, the sampling wavelength grid must be limited to the range $[330, \, 650]$ nm for BP and $[635, \, 1050]$ nm for RP in order to avoid large errors in the reconstructed
spectra.
\par
The  \xp instruments produce two partially overlapping SPDs and SEDs: these are combined into a single distribution by
computing a weighted mean in the overlapping region $[\lambda_{lo},\, \lambda_{hi}]$, 
where the weight varies linearly with wavelength: 
\begin{equation}
    w_{BP}(\lambda) = 1 - \frac{\lambda - \lambda_{lo}}{\lambda_{hi} - \lambda_{lo}}
,\end{equation}
and
\begin{equation}
    w_{RP}(\lambda) = 1 - w_{BP}(\lambda)
,\end{equation}
for $\lambda_{lo} < \lambda < \lambda_{hi}$.
\par
\xp spectra coefficients are accompanied by the covariance matrix $\mathrm{K_{bb}}$: this is used to calculate the covariance matrix for the sampled spectrum, which in the case of the SED is computed as
\begin{equation}
   \mathrm{K_{ff}} = \left(\mathrm{D}_\varphi^E \cdot \mathrm{H} \cdot V_C^T\right) \cdot \mathrm{K_{bb}}  \cdot \left(\mathrm{D}_\varphi^E \cdot \mathrm{H} \cdot V_C^T\right)^T
   \label{eq:ecsCovMatrix}
.\end{equation}
The square roots of the diagonal elements of $\mathrm{K_{ff}}$ are the errors associated with the sampled SED.


\section{Processing}\label{sec:processing}

\begin{figure*}[!t]
    \centering
    \includegraphics[width=(\textwidth-4mm)]{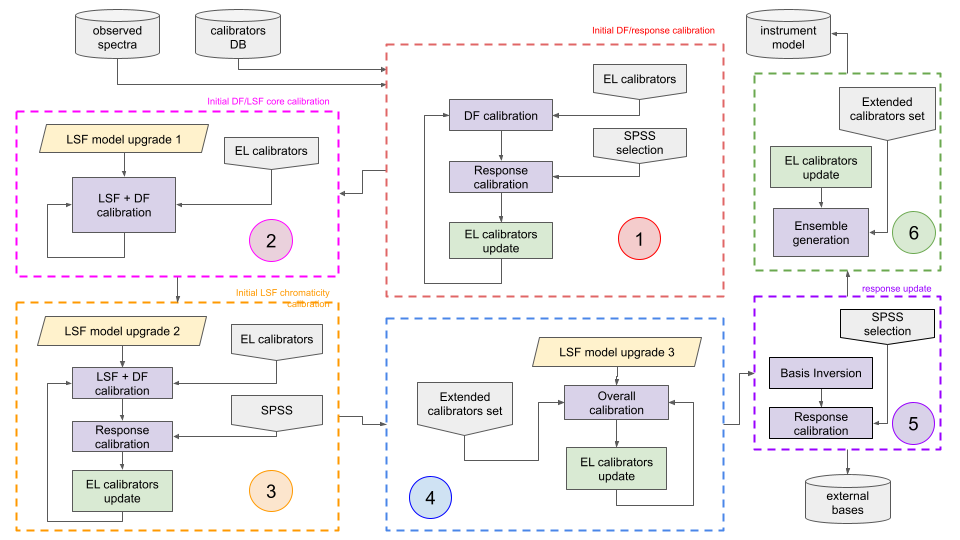}
    \caption{Processing scheme.}
    \label{fig:processing}
\end{figure*}
The instrument model has been designed to reproduce the \emph{nominal} instrument model when all parameters are set to their nominal values: these parameters are $d_i$ for the dispersion  (\equref{dispModel}), $d_{m,n}$ for the LSF (\equref{fullLsf}), and $r_i, \lambda_C, \sigma_C$ for the response model (\equrefs{shapeParams}{respShape}).
The concept of \emph{nominal} refers to the dispersion function and to the overall response curve, for which pre-launch laboratory measurements are available. The LSF model instead is initialised as the mean of a large number of theoretical LSF models: 
for this reason it corresponds to a flat WFE map and is symmetric.
Model parameter optimisation can be fulfilled with a sufficient number of calibrators by 
minimising in a least squares sense 
a $\chi^2$ -based cost function
\begin{equation}
    \chi^2 = \textbf{r}^T \cdot W \cdot \textbf{r}
,\end{equation}
where the array of residuals $\textbf{r}$ and the weight matrix $W$ can be evaluated
in two different spaces:
\begin{enumerate}
    \item Sample space: the observed BP/RP mean spectrum is sampled according to \equref{sampledSpec}:
    \begin{equation}
        \textbf{r} =  \mathrm{D} \cdot \textbf{b} - \mathcal{I}_{u,\lambda} \cdot \bf{n_p}
        \label{eq:sampleSpace}
    ,\end{equation}
    where $\mathcal{I}_{u,\lambda}$ represents the instrument matrix sampled on the wavelength array of the source SPD and on the same AL grid as the mean spectrum; $(\mathcal{I}_{u,\lambda}\bf{n_p})$ represents the discretised version of \equref{xpCompact}, and the weight matrix is computed as
    \begin{equation}
        W = \left( \mathrm{D} \cdot \textrm{K}_\textrm{bb} \cdot \mathrm{D}^T 
        + \mathcal{I}_{u,\lambda} \cdot \textrm{K}_\textrm{pp} \cdot \mathcal{I}_{u,\lambda}^T\right)^{-1}
    ,\end{equation}
    where $K_\textrm{bb}$ is the covariance matrix of the spectra coefficients, the product $\mathrm{D} K_\textrm{bb} \mathrm{D}^T$ is the pixel covariance matrix of the sampled BP/RP spectrum, and $K_\textrm{pp}$ represents the covariance matrix of the source SPD.
    
    \item Coefficient space: the model prediction is projected in the coefficient space according to \equref{coeffsProjection}
    \begin{equation}
        \textbf{r} =  \textbf{b} - \mathrm{D}^\dagger \cdot \mathcal{I}_{u,\lambda} \cdot \bf{n_p}
        \label{eq:coeffResiduals}
    ,\end{equation}
    and the weight matrix is the inverse of the sum between coefficients and projected SPD covariances:
    \begin{equation}
        W = \left( \textrm{K}_\textrm{bb} + 
        \left(\mathrm{D}^\dagger \cdot \mathcal{I}_{u,\lambda}\right)
        \cdot \textrm{K}_\textrm{pp} \cdot
        \left(\mathrm{D}^\dagger \cdot \mathcal{I}_{u,\lambda}\right)^T
        \right)^{-1}
        \label{eq:coeffFullCovariance}
    .\end{equation}
\end{enumerate}

In all scenarios, we include the usage of the covariance matrices for both the observed BP/RP spectra and the calibrator SPDs to allow for a total least-squares regression analysis of the data \citep{huffell}. Although the SPD covariance matrix would be extremely important to properly account for systematic effects of wavelength calibration errors, especially in regions where the SPD changes steeply with the wavelength for certain spectral types, only errors on fluxes are usually available in the literature. Therefore, only BP/RP covariances are full matrices
while SPD covariances are simply diagonal matrices populated with the corresponding variances on the sampled flux. Moreover, total regression is generally highly demanding in terms of computational cost because it requires an evaluation and inversion of the covariance matrix at each step of the solver, and has therefore only effectively been taken into account in the final stages of the calibration, as is clarified below.

When optimisation is carried out in sample space,
the AL grid used to compare sampled spectra is usually oversampled by some factor \wrt the \gaia pixel, and therefore the coefficients covariance matrix transformed to sample space $(\mathrm{D} \cdot \textrm{K}_\textrm{bb} \cdot \mathrm{D}^T)$ (hereafter \emph{samples covariance}) does not have full rank and cannot generally be inverted. A first obvious solution is to take into account only the diagonal elements of the samples covariance, even if the off-diagonal elements are not negligible because mean spectra are continuous functions and hence random noise will manifest as random wiggles
in the form of long-range correlations between pixels.
A formal solution for a correct weighting scheme in sample space is to compute the weight matrix as:
\begin{equation}
W = {D^\dagger}^T \cdot \mathrm{K_{bb}}^{-1}\cdot D^\dagger
\label{eq:fullWeight}
,\end{equation}
where the pseudo-inverse of the samples covariance matrix $D^\dagger$ has been defined in \equref{pseudoInverse}. 
This approach was tested but some occasional numerical instabilities discouraged us from using it for the present calibrations. 
Cost evaluation in coefficient space would solve the problem of the invertibility of the matrix, allowing for a full exploitation of spectra covariances. However, this approach could not be followed for the current release because of the incomplete wavelength coverage of many emission line calibrators used in the optimisation process (\equref{coeffsProjection} implicitly requires the sampled spectrum 
to extend over the entire AL range). For  calibrators with incomplete wavelength coverage, $\chi^2$ computation was limited to the available section of the model spectrum, excluding a safety margin of a couple of pixels where truncation occurs because the redistribution of light produced by LSF will cause some systematic difference between the partial spectrum and the one that would have been obtained by a complete SPD.
For this reason, we decided to carry out the cost evaluation in sample space.

Provided that the cost function is not linear \wrt the model parameters, the  optimisation process can be carried out using an implementation of the differential evolution algorithm (DEA) as described by \citet{storn}: although this class of algorithms has been shown to achieve global optimisation with a natural ability to escape  local minima traps in the $\chi^2$ space, 
we find it convenient to proceed with the bootstrapping of the model by limiting the number of free parameters at the first stages of the processing,  and 
gradually enhancing the model complexity (i.e. increasing the number of parameters) only when convergence is progressively achieved. This led to a complex processing scheme that is sketched in \figref{processing}. 

\begin{figure}[t]
    \centering
    \includegraphics[width=(\textwidth-4mm)/2]{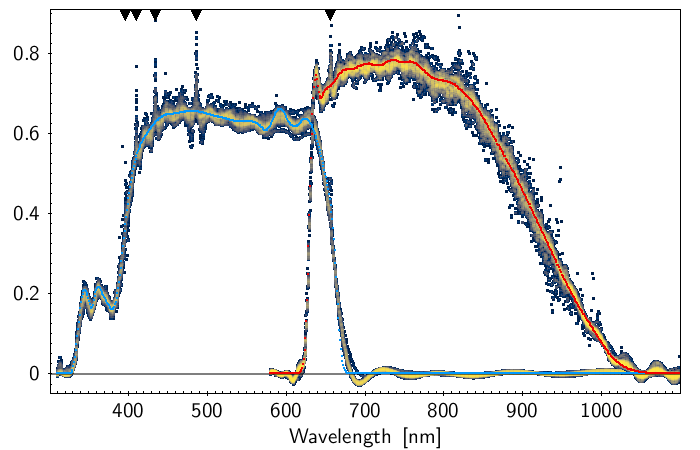}
    \centering
    \includegraphics[width=(\textwidth-5mm)/2]{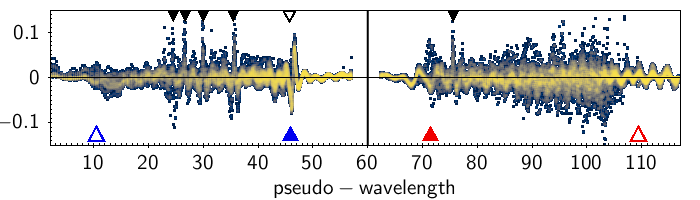}
    \caption{\xp response curves traced by the ratio between effective SPD computed from \xp spectra and source SPDs from ground-based observations. \emph{Top}: Data are plotted against wavelength; black triangles mark the position of Balmer lines.  \emph{Bottom}: Residuals between data and the response models are plotted against pseudo-wavelengths. BP  pseudo-wavelengths have been swapped left-to-right, while RP data are shifted by 60 samples. Blue- and red-filled triangles mark the position of BP and RP cut-of,f respectively, while blue and red open triangles show where the response drops to zero.}
    \label{fig:response}
\end{figure}

Calibrators are divided into two groups, absolute flux or primary calibrators (labelled \emph{SPSS}) and secondary calibrators, sources with emission line features (labelled {EL calibrators}). As explained in \secref{calibrators}, the latter group contains potentially variable sources. To overcome this problem, 
an update process equivalent to a grey flux calibration is implemented for these calibrators: the input SPD of each EL source is scaled by a parameter that is evaluated at each calibration cycle to  
minimise the squared residuals between the current model prediction and the corresponding observed mean spectrum.
In the first stage of the processing, the response model is initialised with a low number of shape parameters (see \equref{shapeParams}, 8 parameters for BP, 5 for RP), while cut-off parameters are set to their nominal values, the dispersion model degree is set to 1, and the LSF model is symmetric. This stage is designed to provide a reliable initialisation of the dispersion relation and the response shape: the cost evaluation is limited to the central region of the spectra, avoiding the wings and the cut-off regions; the optimisation of the dispersion parameters, based on the EL calibrators, is alternated with the optimisation of the response shape, which is achieved by using a selection of featureless SPSSs. 
Each optimisation cycle typically consists of approximately 3000-4000 DEA iterations  involving about $50$  walkers (different realisations of model parameters, initially distributed randomly around the starting set of parameters). The iterations stop when the individual costs from all the walkers converge to a common value. The set of parameters with the lowest cost is used to initialise the subsequent optimisation cycle.
Once convergence is reached, the second stage begins, which entails modelling the shape of the central part of the LSF: the LSF model is initialised with $(\ell_1, \ell_2) = (4, 1)$ bases,  and LSF and dispersion parameters are optimised together to model any possible asymmetry of the LSF core. A second upgrade to the LSF model is made in stage 3 where the number of bases is increased to $(\ell_1, \ell_2) = (4, 2)$ to allow modelling of any chromaticity effect: an LSF and dispersion parameter fit is alternated with response model adjustment and EL source update. In stage 4, the LSF and dispersion models are set to their final configurations, the number of response shape parameters are doubled while the cut-off parameters are left free to change; all model parameters are optimised together using an extended set of calibrators (EL + SPSS). The number of bases is set to $(\ell_1, \ell_2) = (7, 3)$ for BP and is left unchanged for RP; in both instruments the dispersion model degree is set to 2.
In stage 5, the basis inversion process is performed to allow the reconstruction of the effective SPD for the SPSS: these are divided by the corresponding source SPDs to obtain the data shown in \figref{response}. 
This plot was obtained using a subset of 41 SPSSs selected to be as featureless as possible and, given the definition of effective spectra in  \equref{effective}, it traces the overall instrument response curves. 
The top plot shows data plotted as a function of wavelength. The bottom plot shows the residuals between data and model represented as function of the AL coordinate (or pseudo-wavelength); given that the dispersion direction of the BP instrument is inverted \wrt that of the RP instrument, BP data have been mirrored horizontally in order to sort the displayed data according to wavelength. RP data are offset by 60 samples in pseudo-wavelength to avoid superposition with BP. 
It is possible to recognise the signs left in the data by the first Balmer absorption lines ($H_\alpha$ through $H_\epsilon$) as peaks highlighted in the plots. Interestingly, a wavy regular pattern is visible at all wavelengths with a nearly constant frequency in the pseudo-wavelength space. The origin of this pattern is not fully understood; it could be related to wiggles in the mean spectra \citep{DeAngeli2022}.
Using this data, we upgraded the instrument response distortion model by increasing the number of spline knots to model these wiggles, especially in the range $[500,800]$ nm, carefully excluding the signature left by the $H_\alpha$ line (the response model must not be a function of any source astrophysical parameter). The total number of response parameters is 26 for BP and 23 for RP. The model curves represented in the top plot were computed after this upgrade.
The last stage of the processing is dedicated to the creation of the \emph{ensemble} of instrument models: the DEA solver is run on the last instrument models until parameter relaxation is achieved (typically after about $400$ iterations), and then all the 50 walkers are saved into the database providing the 50 instrument models.
\begin{figure*}[t!]
\centerline{
    \includegraphics[width=(\textwidth)/2]{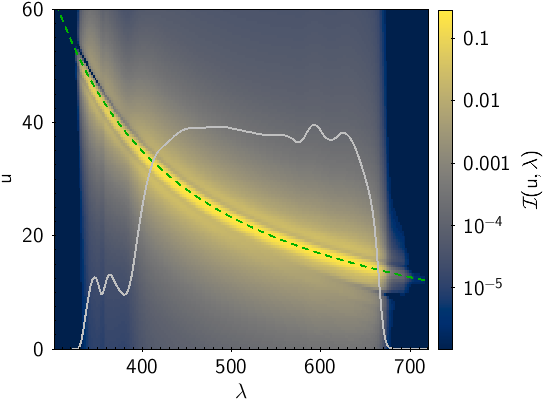}
    \includegraphics[width=(\textwidth)/2]{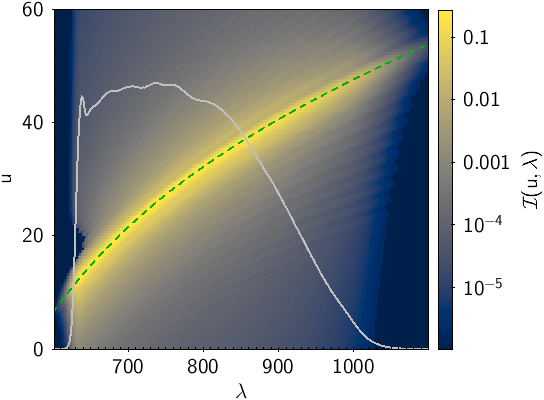}}
    \caption{Visual representation of instrument matrix for BP (\emph{left panel}) and RP (\emph{right panel}) instruments. Green dashed curves are the dispersion relations, and white curves sum up the matrix columns and represent the response curves.}
    \label{fig:instrMatrix}
\end{figure*}
The walker with the lowest chi-square is chosen to represent the instrument model, which enables forward modelling to simulate mean \xp spectra or inverse modelling to provide SEDs through the inverse basis representation. The ensemble instead is used to derive the uncertainties in the simulated spectra. The instrument model ensemble and the inverse bases are used in the \gaiaxpy tool \citep{DeAngeli2022b} to simulate mean spectra and to generate sampled \gaia \xp calibrated spectra in the absolute system.


\section{Results}\label{sec:results}
\begin{figure}[]
\centerline{
    \includegraphics[width=(\columnwidth)]{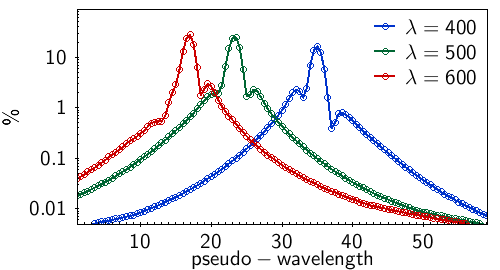}}
\centerline{
    \includegraphics[width=(\columnwidth)]{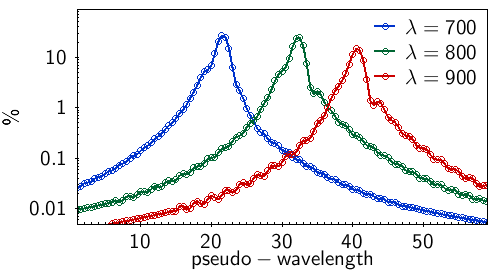}}
    \caption{Three selected columns of the instrument matrix for BP (\emph{top panel}) and RP (\emph{bottom panel}) in logarithmic scale. The curves represent the
    dispersed monochromatic LSFs at three different wavelengths rescaled by 100 to represent a percentage distribution.}
    \label{fig:imSlicesLsf}
\end{figure}
\begin{figure}[]
\centerline{
    \includegraphics[width=(\columnwidth)]{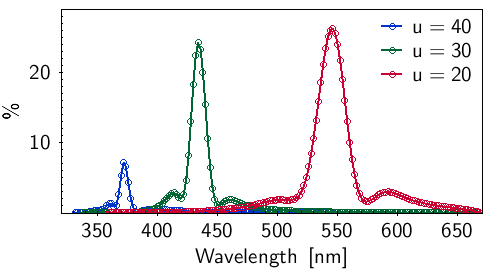}}
\centerline{
    \includegraphics[width=(\columnwidth)]{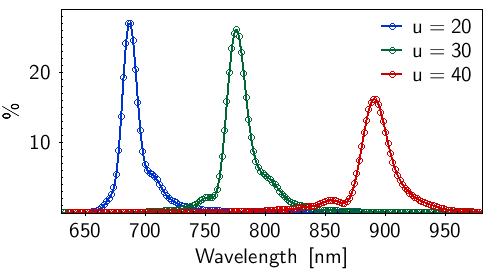}}
    \caption{Three selected rows of the instrument matrix for BP (\emph{top panel}) and RP (\emph{bottom panel}). The curves represent the percentage as function of wavelength of the incoming photons that are accumulated in the corresponding data sample.}
    \label{fig:imSlices}
\end{figure}

The final \xp instrument models were used to create \figref{instrMatrix} where the corresponding instrument matrices $\mathcal{I}_{u,\lambda}$ are represented: this plot shows 
at a glance how \xp mean spectra  are simulated for a given SPD.
The instrument matrix can be used to express \equref{xpCompact} in a discretised form as already done in \equref{sampleSpace}: a mean spectrum is the row-by-column product of the instrument matrix with the SPD (sampled on the same wavelength grid of $\mathcal{I}_{u,\lambda}$).  
\begin{figure*}[!h]
\centerline{
    \includegraphics[width=(\columnwidth)]{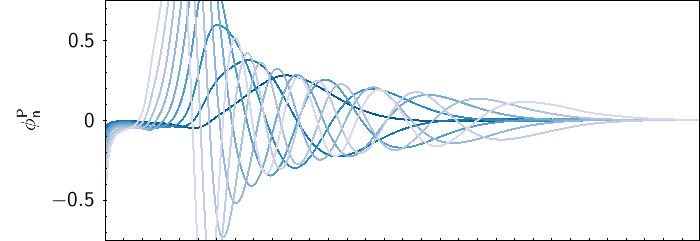}
    \includegraphics[width=(\columnwidth)]{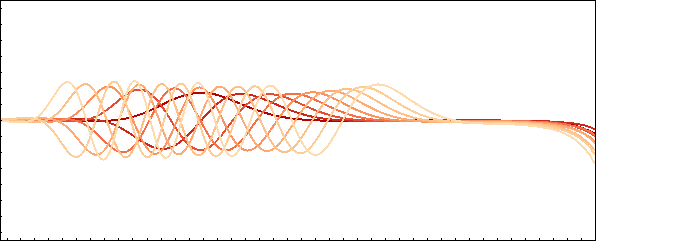}}
\centerline{
    \includegraphics[width=(\columnwidth)]{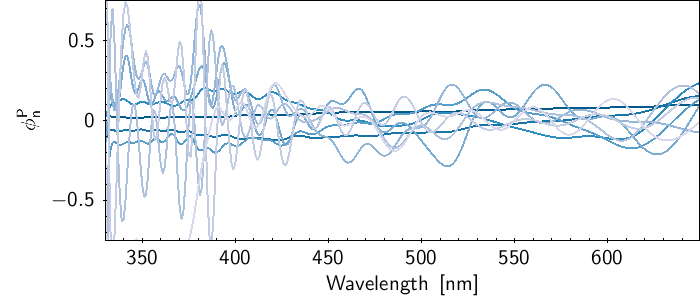}
    \includegraphics[width=(\columnwidth)]{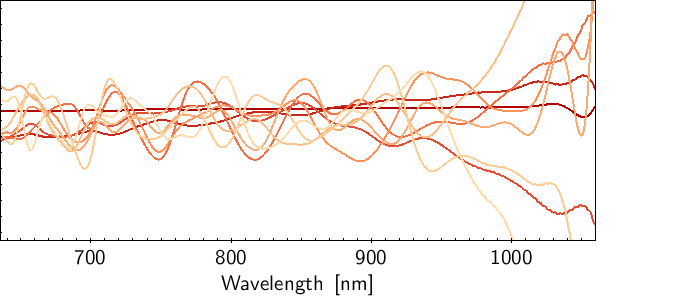}}
    \caption{First nine inverse basis functions for BP (\emph{left}) and RP (\emph{right}) instruments. The top panels show the inverse bases corresponding to the canonical Hermite functions, while the bottom panels show the inverse bases corresponding to the optimised bases.}
    \label{fig:extPpsBases}
\end{figure*}
\begin{figure*}
\centerline{
    \includegraphics[width=(\columnwidth)]{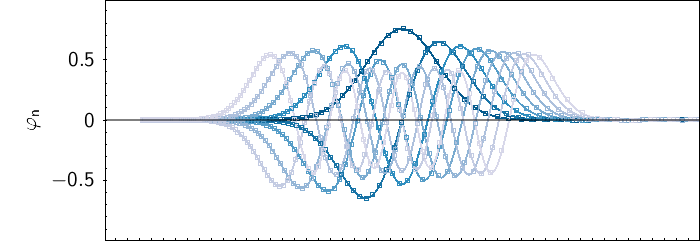}
    \includegraphics[width=(\columnwidth)]{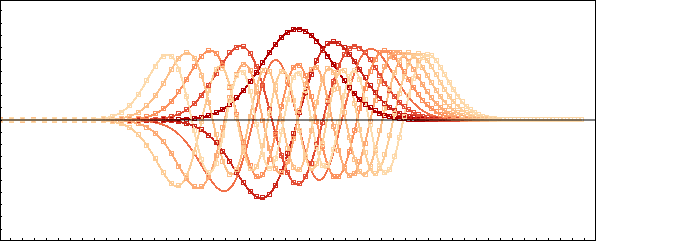}}
\centerline{
    \includegraphics[width=(\columnwidth)]{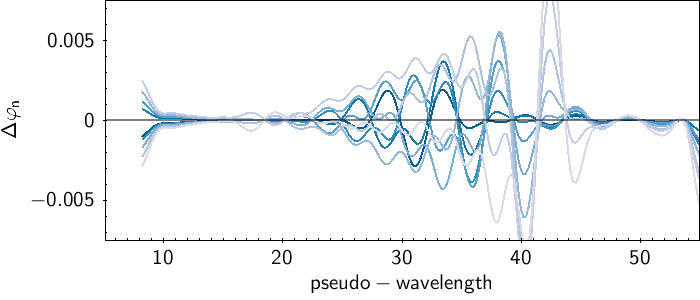}
    \includegraphics[width=(\columnwidth)]{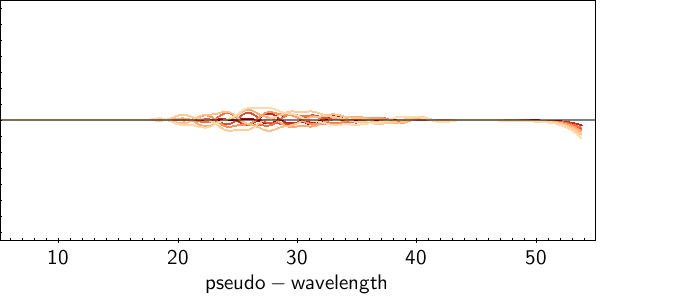}}
    \caption{Reconstruction error of the Hermite functions.
    \emph{Top panels}: Comparison between the first nine canonical Hermite functions (continuous lines) and the image of the inverse, non-optimised basis functions (open squares) for  BP (\emph{left}) and RP (\emph{right}) instruments as a function of the pseudo-wavelength u.
    \emph{Bottom panels}: Residuals between the two families of curves shown in the top panels.}
    \label{fig:hermBasesNoOpt}
\end{figure*}
\begin{figure*}
\centerline{
    \includegraphics[width=(\columnwidth)]{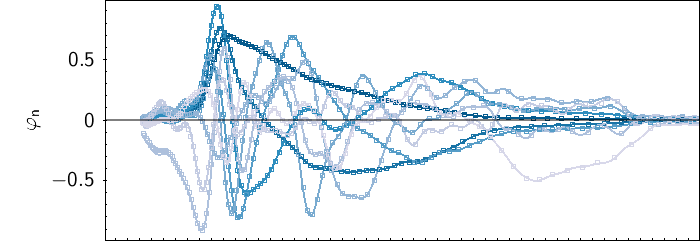}
    \includegraphics[width=(\columnwidth)]{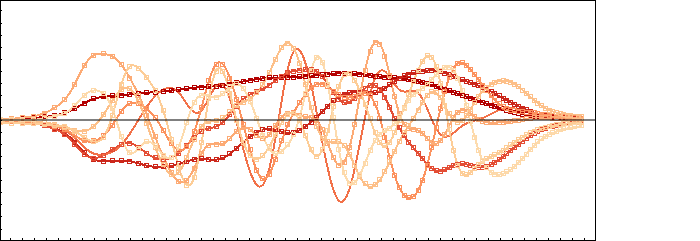}}
\centerline{
    \includegraphics[width=(\columnwidth)]{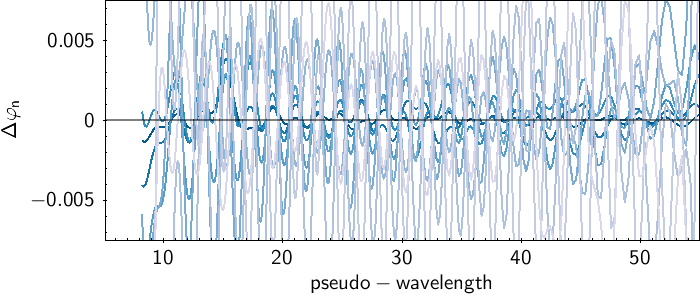}
    \includegraphics[width=(\columnwidth)]{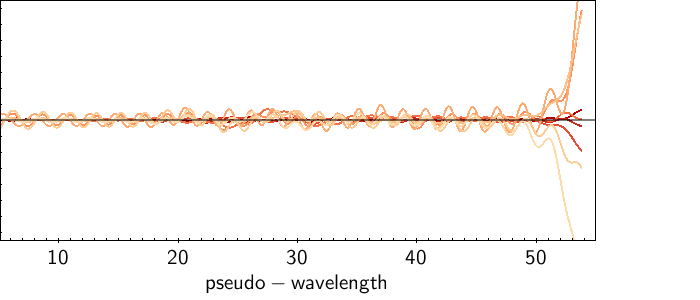}}
    \caption{Same as \figref{hermBasesNoOpt} but for the optimised basis functions.}
    \label{fig:hermBasesOpt}
\end{figure*}
Provided that: \textit{(i)} columns and rows represent small intervals of wavelengths and AL sample coordinates respectively, \textit{(ii)}
each column represents a monochromatic dispersed LSF scaled by the response (the response curve is over-plotted in white) at that wavelength and the telescope pupil area, and (\textit{iii)} the loci of LSF maxima, highlighted by the dashed green lines, correspond to the dispersion functions, then 
we can visualise the mean spectrum formation by splitting the incoming SPD into packets of photons according to their wavelength. These are then distributed following the corresponding matrix column profile, and the final mean spectrum is given by the accumulation on the $u$ axis of each dispersed packet. One of the elements that catches the eye is that the dispersion direction is opposite for the two instruments, and hence the wavelength decreases from left to right in the BP spectra plots.
Furthermore, in \figref{instrMatrix}, the RP instrument exhibits a pattern of ripples that run roughly parallel to the dispersion relation. Intensity and distance from the LSF crest grow with wavelength: these ripples are the signature
left by diffraction patterns of theoretical LSF in the  $U$ bases of the LSF model \citep{LL:PMN-012}; the amplitude of the diffraction pattern scales linearly with $\lambda$ and this is the reason why it is virtually invisible below 700 nm.
Figure \ref{fig:imSlicesLsf} shows the dispersed monochromatic LSF in logarithmic scale for wavelengths $\lambda = (400, 500, 600)$ nm for BP (\emph{top panel}) and $\lambda = (700, 800, 900)$ nm for RP (\emph{bottom panel}).
The structure of the RP ripples is clearly visible in the bottom plot.
On the other hand, we can figure out which wavelength ranges of the SPD contribute more to the photon budget of a particular sample of the mean spectrum by looking at the corresponding matrix row: this contribution for samples at $u = (20, 30, 40)$ is visualised in \figref{imSlices} for both instruments as the percentage of the incoming photons that are detected in each data sample. The width of the distribution sets the level of smearing of the spectrum at that wavelength. As can be seen, there are sections of mean spectra (as the case shown for BP at $u=20$) that receive photons from a very large wavelength range: this explains why it is so important to also resolve for the LSF while calibrating mean spectra.

\subsection{Inverse basis functions}\label{sec:invbases}

Figure \ref{fig:extPpsBases} shows the first few inverse bases $\phi^P_n$ for the BP and RP instruments (in units of $\rm photons\, s^{-1}m^{-2}nm^{-1}$) as a function of wavelength. The top panels show the inverse bases of the canonical Hermite functions while the bottom panels represent the inverse bases for the optimised basis functions. The amplification of the bases below $\lambda\lesssim400$ nm and above $\lambda\gtrsim900$ nm is due to the normalisation of the bases by the response function $R(\lambda)$ (see \equref{designSpd}). For the same reason, it is pointless to look at the bases outside the represented wavelength range because they diverge as the response goes to zero.

It is interesting to compare the image through the instrument model of the inverse bases $\varphi_n^\dagger$ (hereafter referred to as the \emph{reconstructed} bases), 
\begin{equation}
\label{eq:varphiImage}
   \varphi_n^\dagger(u)=  P_\tau\, \int\limits_0^\infty \phi^P_n(\lambda) \, L_\lambda\left(u - u_d(\lambda)\right) \, R(\lambda) \, {\rm d}\lambda
,\end{equation}
against the original bases $\varphi_n$. This provides a means to evaluate the accuracy of the inverse representation: in principle the reconstructed bases should be equal to the original ones.
This comparison is made in the top panels of \figref{hermBasesNoOpt} for the canonical Hermite functions and in \figref{hermBasesOpt} for the optimised versions. In both cases, the lines represent the original basis functions $\varphi_n$, while open squares represent the reconstructed bases  $\varphi_n^\dagger$. The bottom panels of the figures show the difference 
\begin{equation}
    \Delta \varphi = \varphi_n(u)-\varphi_n^\dagger(u)
.\end{equation}
In the case of non-optimised basis functions, the residuals for BP gradually increase at shorter wavelengths (from left to right) for higher order terms, becoming as high as $1\%$; in the RP case the residuals are much smaller, about $0.1\%$, over the entire wavelength range. In the case of optimised BP functions, the situation seems to worsen with residuals of  $\sim2\%$  over the entire pseudo-wavelength range, while for RP the quality of the comparison remains good. To understand this behaviour, it is convenient to quantify the quality of a reconstructed basis as a function of the order of the basis itself.
\begin{figure}
\centerline{
    \includegraphics[width=(\columnwidth-4mm)]{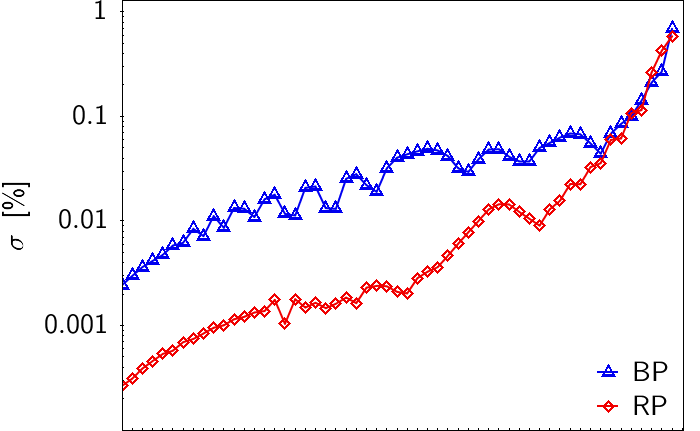}}
\centerline{
    \includegraphics[width=(\columnwidth-4mm)]{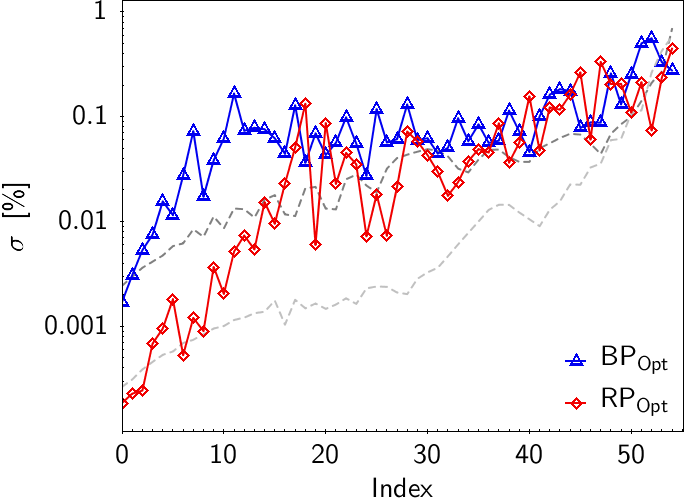}}
    \caption{Numerical reconstruction error of the inverse bases as a function of basis index for canonical Hermite functions (\emph{top panel}) and for optimised basis functions (\emph{bottom panel});  in the bottom panel, the grey dashed lines represent the curves of the top panel and are plotted to help in the comparison.}
    \label{fig:ExtBasesError}
\end{figure}
This is done in \figref{ExtBasesError} where, for each basis, we compute the quantity
\begin{equation}
    \sigma_n = \frac{\int_{u_0}^{u_1} \left|\varphi_n^\dagger(u)-\varphi_n(u)\right| \,{\rm d} u}{\int_{u_0}^{u_1} \left|\varphi_n(u)\right| \,{\rm d}u}
,\end{equation}
which is used as a proxy for the percentage error on the pseudo-wavelength integral of the basis function.
The reader must be aware that this is only an indirect way of assessing the accuracy of each inverse basis: we are evaluating the accuracy of external bases by forward modelling in the sampled spectra space where they are never used, and use them only in the inverse modelling of ECS. 
The plot in the top panel shows the reconstruction error for the non-optimised case, and one can see that the quality of the reconstruction is very high for both instruments (with numerical precision for RP outperforming the BP case by a factor of ten) for approximately the first  $47$ lower order bases,  and worsens very quickly for the higher order terms. We understand this behaviour as an intrinsic limit of the current inverse basis model  implementation (set in \equref{inverseBasisModel} with $K=55$): an inverse basis represents by definition a given Hermite function deconvolved by the LSF function, hence its model should have the ability to reproduce features with higher spatial frequencies \wrt its smeared representation. If we consider the $\phi_{54}$ case, despite the fact that it should have higher spatial frequencies than the  $\varphi_{54}$ Hermite function, we are effectively modelling it as a linear combination of the 
 $\varphi_0\cdots\varphi_{54}$ Hermite functions and therefore we are possibly missing the higher frequencies. 
This problem will be addressed in future releases; for now we note that this inadequacy of the model will mainly affect the accuracy in reproducing  
features with high spatial frequencies, such as narrow emission lines (introducing some systematic errors into the shape of the ECS), but it will not have consequences for the vast majority of the sources. As the optimisation process  consists in a rotation of the bases, this determines a redistribution of the error budget of the  higher
terms to lower order components (and this explains the behaviour of BP residuals in \figref{hermBasesOpt}): however, thanks to the efficiency of the optimised bases in concentrating most of the information in the lower order coefficients, this will have no effect on the overall quality of the ECS reconstruction.
In \secref{invForwExperiment}, we perform a test to evaluate the impact of this reconstruction error on real data.

\subsection{Spectral resolution}\label{sec:specresol}

The spectral resolution of a \gaia spectrum changes considerably across the wavelength range. This is even more evident in ECS where the resolution changes abruptly near the boundary between the \xp wavelength ranges. 
The full width at half maximum (FWHM) of the instrument LSF model provides a direct measurement of the spectral resolution of mean \xp spectra. However, the resolution of ECS should be increased by the deconvolution from the LSF smearing effect obtained by the bases inversion process. To evaluate the spectral resolution of ECS, we simulate the response of the instrument model to a monochromatic signal at a well-defined wavelength as follows:

    1) we produce a synthetic spectrum with a flat null continuum and a single emission line of zero intrinsic width (the Dirac delta function) centred at a given wavelength;

    2) we forward model the corresponding mean spectrum into the coefficients space;

    3) we reconstruct the corresponding ECS;

    4) we measure the FWHM of the line: this is the measurement of the `instrumental width' at the considered wavelength, that is the size of the spectral resolution element;

    5) and then we repeat with a synthetic spectrum whose emission line is displaced by $10$ nm until the whole \xp spectral range is covered.

\begin{figure}
    \centering
    \includegraphics[width=(\textwidth-4mm)/2]{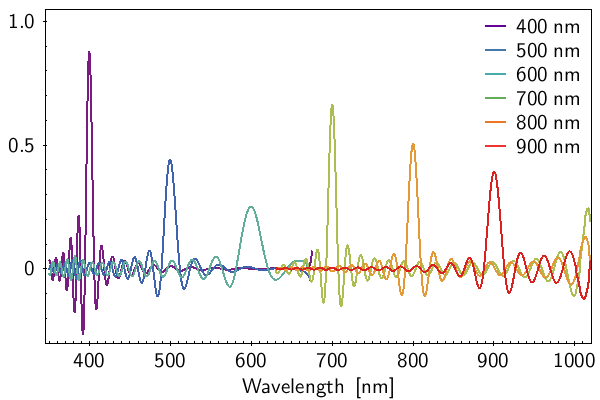}
    \caption{Response of the instrument model to a simulated monochromatic stimulus at different wavelengths.}
    \label{fig:spikeResponse}
\end{figure}
As we use the same instrument model in the forward and reverse directions, we are testing the theoretical resolution that would be obtained given `perfect' knowledge of the instrument model. The presence of undetected systematic differences between the instrument model and the actual instrument will possibly worsen the resolution of the current ECS \wrt this ideal case. Furthermore, the total regression results in Gaussian noise on the elements of the instrument matrix. In the inversion of the noisy instrument matrix, this well-behaved Gaussian random noise can become non-trivial noise and even create systematic errors due to the non-linearity of inversion. 
Figure \ref{fig:spikeResponse} shows the simulated ECS for a subset of the Dirac delta spectra. The reconstructed spectra show wiggles 
due to the inversion process. These will create artefacts in real ECS around small-scale features such as emission lines. As can be noticed, these wiggles tend to become very large at wavelengths greater than $1000$ nm due to the amplification due to the response normalisation (see \equrefs{designSpd}{designSed}). We show that the presence of such large fluctuations is common in ECS at wavelengths $\lambda \gtrsim 1030$ nm.

\begin{figure}
    \centering
    \includegraphics[width=(\textwidth-4mm)/2]{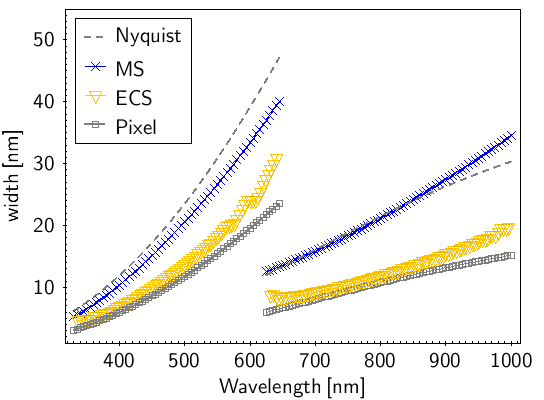}
    \caption{Full width at half maximum measured in nm as a function of wavelength for a monochromatic signal in \xp mean spectra (blue crosses) and in the externally calibrated spectra (yellow triangles) compared to the width of one CCD pixel (grey squares) and the width of two pixels to represent the Nyquist sampling limit (dashed grey line).}
    \label{fig:fwhm}
\end{figure}

The FWHM for internally calibrated BP/RP mean spectra (hereafter ICS) and for ECS are represented in \figref{fwhm} as a function of wavelength together with the CCD pixel scale that should be regarded as an upper limit to the resolution achievable by the basis inversion process (it is not possible to reconstruct a signal smaller than the pixel integration scale with the current approach). The plot also reproduces the scale length corresponding to $2$ pixels, which is the scale on which Nyquist sampling can take place: interestingly, the BP instrument suffers from a small under-sampling at all wavelengths (the FWHM being equal to $\sim 1.7$ pixel) and therefore does not satisfy the Nyquist criterion. This fact could play a relevant role in causing the 
rather high number of wiggles seen in BP mean spectra compared to the RP ones (see \citet{DeAngeli2022}).
Values for the FWHM of a monochromatic signal measured on ICS, ECS, and the corresponding spectral resolution are provided in \tabref{resolution} as a function of wavelength.
\begin{table*}
\begin{center}
\caption{Full width at half maximum of a monochromatic signal measured on ICS, ECS, and spectral resolution for ECS.}
\label{tab:resolution}
\begin{tabular}{c|c|c|c||c|c|c|c||c|c|c|c}
\hline
$\lambda$ & ICS & ECS & R$_\lambda$ & $\lambda$ & ICS & ECS & R$_\lambda$ & $\lambda$ & ICS &  ECS & R$_\lambda$\\
\hline
350.0 & 6.483 & 4.893 & 71.5 & 570.0 & 29.25 & 20.11 & 28.3 & 780.0 & 19.99 & 11.33 & 68.8\\
370.0 & 7.900 & 5.508 & 67.2 & 590.0 & 31.98 & 23.12 & 25.5 & 800.0 & 21.14 & 11.95 & 66.9\\
390.0 & 9.490 & 6.355 & 61.4 & 610.0 & 34.84 & 24.21 & 25.2 & 820.0 & 22.33 & 12.60 & 65.1\\
410.0 & 11.24 & 7.417 & 55.3 & 630.0 & 37.81 & 28.63 & 22.0 & 840.0 & 23.55 & 13.24 & 63.4\\
430.0 & 13.12 & 8.611 & 49.9 & 640.0 & 13.13 & 8.576 & 74.6 & 860.0 & 24.80 & 13.85 & 62.1\\
450.0 & 15.13 & 9.994 & 45.0 & 660.0 & 13.98 & 8.478 & 77.8 & 880.0 & 26.09 & 14.50 & 60.7\\
470.0 & 17.25 & 11.36 & 41.4 & 680.0 & 14.87 & 8.741 & 77.8 & 900.0 & 27.40 & 15.47 & 58.2\\
490.0 & 19.46 & 12.88 & 38.0 & 700.0 & 15.81 & 9.116 & 76.8 & 920.0 & 28.76 & 16.07 & 57.2\\
510.0 & 21.75 & 14.56 & 35.0 & 720.0 & 16.79 & 9.662 & 74.5 & 940.0 & 30.15 & 17.13 & 54.9\\
530.0 & 24.15 & 16.31 & 32.5 & 740.0 & 17.81 & 10.11 & 73.2 & 960.0 & 31.57 & 17.86 & 53.8\\
550.0 & 26.63 & 18.31 & 30.0 & 760.0 & 18.88 & 10.69 & 71.1 & 980.0 & 33.01 & 18.83 & 52.0\\
\hline
\end{tabular} 
\end{center}
\end{table*}

\begin{figure}[]
    \centering
    \includegraphics[width=(\textwidth-4mm)/2]{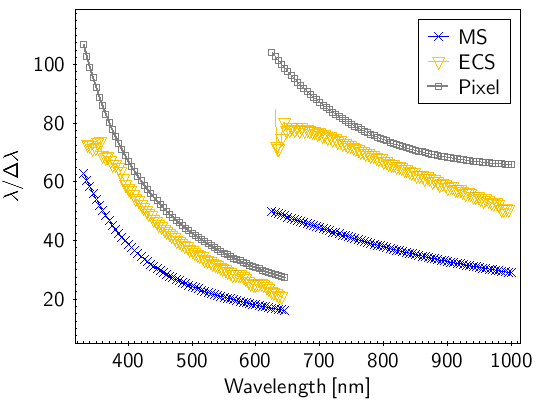}
    \caption{Spectral resolution as a function of wavelength for \xp mean spectra (blue crosses) and externally calibrated spectra (yellow triangles) compared to the resolution corresponding to the width of one CCD pixel (grey squares).}
    \label{fig:resolution}
\end{figure}
The spectral resolution is finally shown in \figref{resolution}: the RP instrument exhibits a more uniform variation \wrt wavelength compared to the BP case. The anomalous drop in RP resolution just below $650$ nm in the ECS curve as well as small fluctuations visible in the BP curve around $600$ nm are possibly due to ripples in the response profiles affecting the basis inversion process.
The final resolution of a typical \gaia ECS drops from $R_\lambda = 70$ to $R_\lambda \sim 22$ in the wavelength range $\lambda\in[330, 640]$ nm, and then suddenly rises to $R_\lambda \sim 78$ before decreasing smoothly to $R_\lambda \sim 55$ at longer wavelengths. 

\subsection{Wavelength calibration accuracy}\label{sec:waveAccuracy}

\begin{figure}[t]
    \centering
    \includegraphics[width=(\textwidth-4mm)/2]{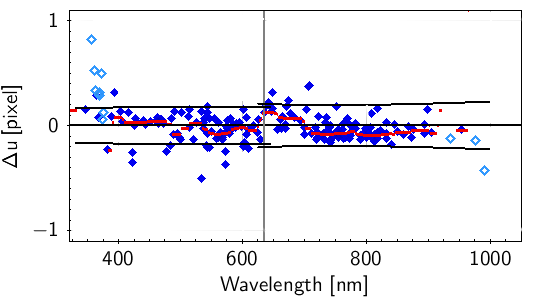}
    \centering
    \includegraphics[width=(\textwidth-4mm)/2]{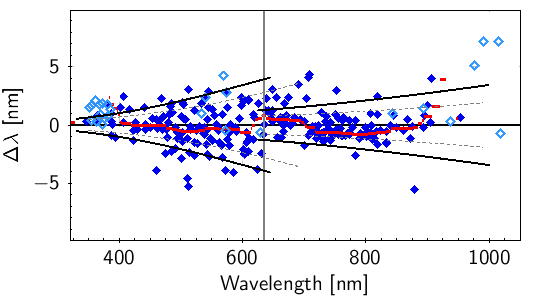}
    \caption{Wavelength calibration accuracy evaluated on a sample of QSOs with emission lines in their spectra. \emph{Top panel}: Difference in AL sample position of emission line peaks measured on \xp mean spectra and the expected position measured on simulated mean spectra (filled symbols). Open symbols refer to SEDs with incomplete wavelength coverage for which the expected position has been estimated based on the line wavelength at rest and the QSO redshift value. The vertical grey line separates BP from RP data. The black lines delimit the region within $\pm 10\%\,\mathrm{{FWHM}}$ of the LSF for mean spectra. \emph{Bottom panel}: Observed and expected emission line wavelengths as measured on the ECS. Dashed grey lines represents $\pm 10\%$ of the expected LSF FWHM for calibrated spectra, and black lines delimit the region within $\pm 10\%\,\mathrm{{FWHM}}$ of the LSF for mean spectra.}
    \label{fig:waveaccuracy}
\end{figure}
\begin{figure}[t]
    \centerline{
    \includegraphics[width=(\columnwidth)/3-4mm]{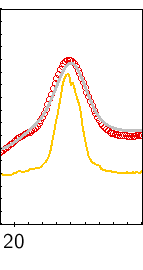}
    \includegraphics[width=(\columnwidth)/3-4mm]{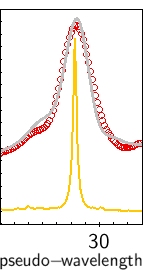}
    \includegraphics[width=(\columnwidth)/3-4mm]{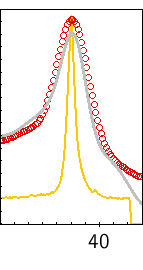}}
    \caption{Examples of emission lines in mean RP spectra. Red symbols are observed  spectra, the grey line is the model prediction, and the yellow curve is the source high-resolution SPD from SDSS (in arbitrary units). The asymmetries in the line profile are not accurately reproduced by the model predictions.}
    \label{fig:lsfRpExapmple}
\end{figure}

To assess the accuracy of the wavelength calibration, we compared the wavelength position of emission lines measured on the \xp mean spectra of a selected sample of QSO extracted from our calibrator set with the expected positions measured on simulated spectra computed with the instrument model and the corresponding source SEDs known from ground-based observations. 
We created a starting list of sources by selecting a sample of 102 SDSS QSOs showing strong emission lines in their \xp spectra and compiling a list of estimated wavelength positions for 263 lines based on the QSO SDSS redshift values and the wavelength at rest of each line. The differences between the AL position corresponding to the peak of each line in the mean spectra and in the simulated spectra are shown in the upper panel of \figref{waveaccuracy}.
Open symbols refer to spectra where the SDSS SED does not cover the full BP/RP wavelength range, and therefore we do not have a reliable simulation of the expected ICS: in these cases, the expected AL position is based on the initial estimate of the line wavelength. The red curve represents a smoothed median of the data while the black lines delimit the region within $\pm10\%$ of the LSF FWHM. As can be seen, at wavelengths below $400$ nm there is an indication of a lower accuracy of the dispersion relation due to the lack of a reliable number of QSOs with emission lines in this range employed in the instrument calibration process. We hope to resolve this issue in future releases. The other interesting element in this plot is the presence of a small systematic offset in the RP spectra (roughly equal to one-tenth of a pixel), possibly due to systematic error in the LSF model, which is unable to correctly reproduce the instrument chromaticity as can be seen in \figref{lsfRpExapmple} where three examples are shown (the LSF core is more asymmetric in RP than in BP). 
As the resolution of ECS is higher than that of ICS, we also checked the accuracy of wavelength calibration for the ECS case: starting from the estimated initial wavelength positions, we searched for local maxima in the ECS and in the corresponding simulated ECS (which is computed as described at beginning of \secref{validation}).
The wavelength positions of the peaks are then refined by computing the centroid of the lines accordingly to \equref{centroid}. 
The residuals between observed and expected wavelength positions are plotted in the lower panel of \figref{waveaccuracy}. Most of the points are enclosed between $\pm10\%$ of the expected FWHM for ECS, represented by the dashed lines (the dark lines represent  $\pm10\%$ FWHM for ICS). A systematic difference of about $1$ nm is confirmed in the RP wavelength range. Finally, we notice a lower precision in BP calibration that may be a consequence of LSF under-sampling seen in \secref{specresol}. We note that the number of measured lines in ECS is greater than that in ICS because several lines in ICS were lost because of the intrinsic shape of ICS (typically faint lines falling in regions with large variations of the instrument response).


\section{Validation}\label{sec:validation}

\begin{figure}[]
    \centerline{
    \includegraphics[width=(\textwidth)/4]{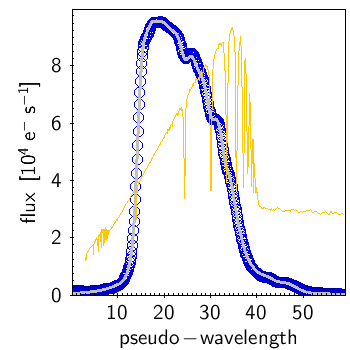}
    \includegraphics[width=(\textwidth)/4]{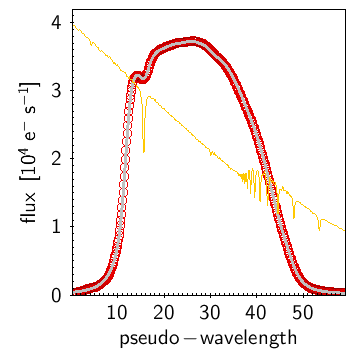}}
    \centerline{
    \includegraphics[width=(\textwidth)/4]{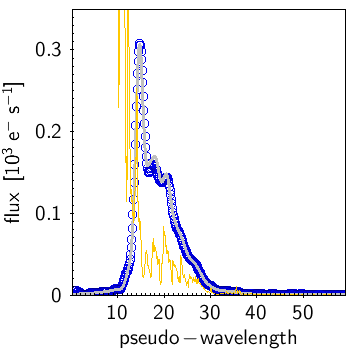}
    \includegraphics[width=(\textwidth)/4]{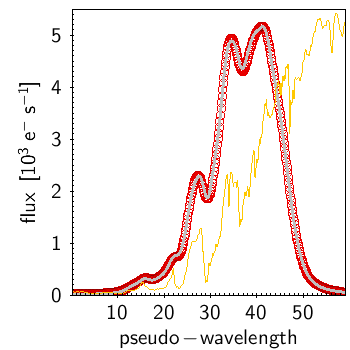}}
\caption{Comparison between mean and simulated spectra for source \texttt{Gaia\,DR3\,1435896975388228224} (\emph{top}) and source \texttt{Gaia\,DR3\,4339417394313320192} (\emph{bottom}), both belonging to the PVL dataset. Blue and red points represent the \gaia \xp mean spectra, while the grey curves represent the model predictions for the corresponding instrument. The superimposed yellow curves represent the source SPD arbitrarily rescaled in flux. We note that the wavelength in BP decreases from left to right.}
    \label{fig:samplSpecComp1}
\end{figure}
Validation of the instrument models and externally calibrated spectra is mainly based on comparisons between observations (i.e. \gaia mean BP/RP spectra) and model predictions for sources with known SED.
A test of this kind was also carried out by \citet{Andrae2022} where observed spectra of known solar twins were compared both to model spectra of the Sun and to IM predictions based on 
X-Shooter SEDs \citep{Verro2021} of the same sources, showing remarkably good agreement.
Hereafter, we use the term \emph{observation} to refer to \gaia spectra and \emph{simulation} to refer to external data simulated via the instrument model. The comparison data always rely on  external data sets.
In the following sections, we  use sources from the SPSS, PVL, and NGSL sets  for the comparisons. Sources with magnitude $G<4$ were filtered out from the NGSL sample to avoid saturation in BP/RP spectra (see \secref{saturation}).
Unless we specify otherwise, plots combine data from all three data sets.
The comparison can be made in terms of sampled internally calibrated mean BP/RP spectra or in terms of externally calibrated spectra. In the first case, the observations and  predictions are sampled on a common AL grid. However, results are often shown as a function of the more familiar absolute wavelength, with 640 nm being used as the boundary between BP and RP. 
Here, ECS are represented as SEDs: in this case, 
the expected distribution of a source is not the SED at its original high resolution, but rather its spectral resolution is degraded at the resolution equivalent to one \gaia pixel with the following procedure:
first we compute a degraded SED $f^\dagger_{X}(\lambda)$ for each \xp instrument: 
\begin{equation}
   f^\dagger_X(\lambda) = \int\limits_{\lambda_{lo}(\lambda)}^{\lambda_{hi}(\lambda)}  f(\lambda) \, {\rm d}\lambda
   \label{eq:sedDegradation}
,\end{equation}
where $f(\lambda)$ is the original SED and the integration limits are computed as:
\begin{eqnarray}
    \lambda_{lo}(\lambda) = \lambda - u^{-1}\left(u(\lambda)-0.5\right),\\
    \lambda_{hi}(\lambda) = \lambda + u^{-1}\left(u(\lambda)+0.5\right),
\end{eqnarray}
where $u(\lambda)$ is the BP/RP dispersion function and $u^{-1}$ the inverse dispersion function.  The two degraded SEDs $f^\dagger_{BP}$ and $f^\dagger_{RP}$ are then combined into a unique SED $f^\dagger(\lambda)$ following the same rule described in \secref{sedReconstruction} for the combination of BP/RP ECS.
Finally, a further validation comes from computing synthetic photometry on ECS and comparing the resulting magnitudes to photometric standards, as described in \secref{synthphot}.

\subsection{Sampled mean spectra comparisons}\label{sec:meanspec}

Figure \ref{fig:samplSpecComp1} shows the comparison between observed and simulated spectra for 
two \gaia sources chosen from the PVL set as an example: source \texttt{Gaia\,DR3\,1435896975388228224} (\emph{top}) and source \texttt{Gaia\,DR3\,4339417394313320192} (\emph{bottom}). BP spectra have wavelengths decreasing from left to right due to the dispersion direction of the prism. 
These two sources have magnitudes of $G=9.68$ and $G=13.81$ and colour indices of $\bprp =0.03$ and $\bprp=4.73,$ respectively. Because of the low resolution, the appearance of \gaia mean spectra is generally quite smooth and only a few conspicuous spectral features are visible as in the shown examples. The agreement between model (grey lines) and observed (blue and red open circles) is very good in this well-behaved example.
The complete set of plots for SPSS, PVL, and NGSL samples can be found in \citet{LL:PMN-016}. 
\begin{figure}[]
    \centering
    \includegraphics[width=(\textwidth-4mm)/2]{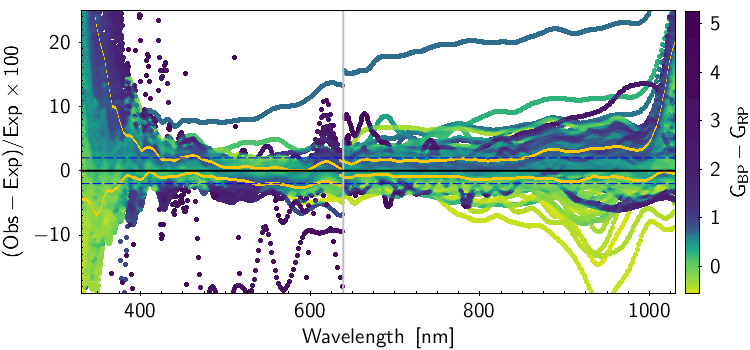}
    \caption{Difference between observed and expected mean flux as a function of wavelength for \xp sampled mean spectra for the whole SPSS, PVL, and NGSL sets. Yellow curves represent the $P_{16}$ and $P_{84}$ percentiles. The colour map encodes the \bprp colour index.}
    \label{fig:samplSpecComp}
\end{figure}

To make a more 
comprehensive  comparison, 
for each source we compute  the percentage difference between the observed spectrum and the model prediction  for the SPSS, PVL, and NGSL sets: these curves are plotted in \figref{samplSpecComp} as a function of wavelength, colour coded by \bprp.
The vertical grey line highlights the boundary between BP and RP data. 
The two yellow curves represent the $P_{16}$ and $P_{84}$ percentile distributions that are used as a proxy for the $\pm 1 \sigma$ distributions: for wavelengths higher than $\lambda\simeq 400$ nm the accuracy of the calibration is mostly enclosed in the  $\pm 2\%$ level marked by the two horizontal dashed blue lines, with some sources showing systematic offsets. Below $400$ nm there is a clear systematic error correlated to source colour: this is the most prominent systematic error known in current \gaia \xp spectra and is analysed in detail in \secref{systematics}. 
To eliminate sources with systematic errors in the absolute flux scale (which can be present at the $1\%$ level), we apply the same grey calibration described in \secref{processing}. 
\begin{figure}[]
    \centering
    \includegraphics[width=(\textwidth-4mm)/2]{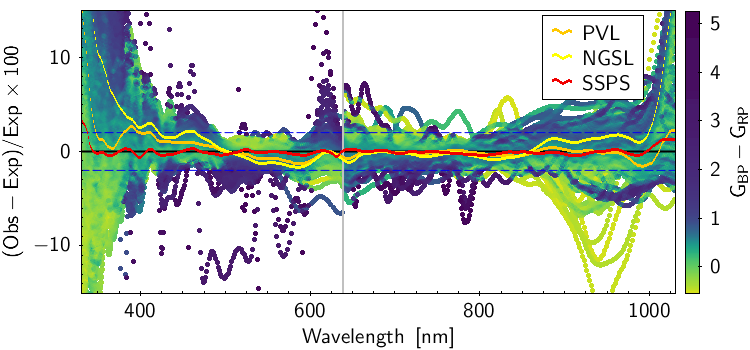}
    \caption{Difference between observed and expected mean flux as a function of wavelength for \xp sampled mean spectra. A grey calibration has been applied to spectra. The three curves represent the smoothed medians for SPSS (red), PVL (orange), and NGSL (yellow). The colourmap encodes the \bprp colour index.}
    \label{fig:samplSpecNormComp}
\end{figure}
Figure \ref{fig:samplSpecNormComp} shows residuals after such an 
additional step: precision is at the $\pm1\%$ level except for $\lambda\lesssim 400$ nm. The three curves in the plot represent the median distribution for SPSS (red), PVL (orange), and NGSL (yellow) sources. As can be seen, while the SPSS curve is almost completely flat, the median distributions for the other two control groups show significant deviations in the BP data as well as in the redder part of the RP spectra. As PVL and NGSL sources are generally brighter than SPSS, with the quality of PVL and NGSL being at least similar to SPSS, this suggests that the problem is unlikely due to a problem in PVL and NGSL data, but is rather related to magnitude-related systematic errors in \gaia mean spectra.
\begin{figure}[]
    \centering
    \includegraphics[width=\columnwidth]{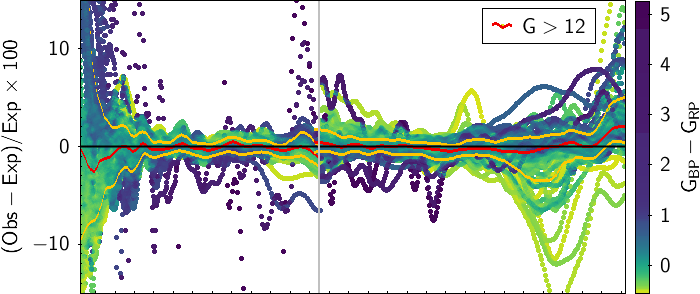}
    \includegraphics[width=\columnwidth]{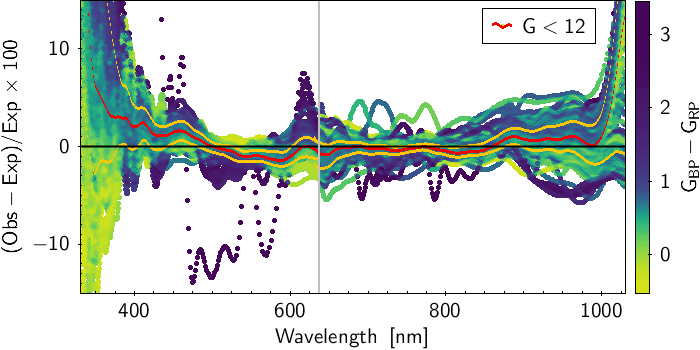}
    \caption{Quantities are the same as \figref{samplSpecNormComp} but data have been divided into two luminosity classes: sources with $G>12$ (\emph{top}) and $G<12$ (\emph{bottom}). Red and orange curves represent $P_{50}$, $P_{16}$ , and $P_{84}$ percentiles distributions,  respectively.}
    \label{fig:samplSpecCompByMag}
\end{figure}
This is confirmed by the two plots in \figref{samplSpecCompByMag} where sources with $G>12$ are shown in the top panel while the bottom panel includes only sources with $G<12$. Residuals from the first group are very flat and the precision is well below the $\pm1\%$ level, while for brighter sources the situation is more complex.

Figure 1 from \citet{DeAngeli2022} clearly illustrates the extreme complexity faced by the internal calibration when dealing with \xp data: sources brighter than $G=11.5$ are observed as 2D windows and, depending on their magnitude, under a high number of different observing configurations (so-called gates, which effectively limit the exposure time and the area of the CCD over which integration occurs, are activated to prevent chip  saturation of bright sources). 
These results may suggest that the internal calibration of bright sources is not fully converged on the common internal reference system, which is mostly dominated by faint sources observed mainly as 1D windows and under more uniform conditions.
Bearing in mind that one of the assumptions of the external calibration is that the instrument model does not depend on the source magnitude, by definition it cannot account for internal inhomogeneities in the data related to different observing conditions. It is expected that the external calibration model shows a   better fit to the faint end because of the magnitude distribution of the SPSSs, most of which fall in the well-behaved faint magnitude range. 

\subsubsection{Systematic effects}\label{sec:systematics}

\begin{figure}[]
    \centering
    \includegraphics[width=\columnwidth]{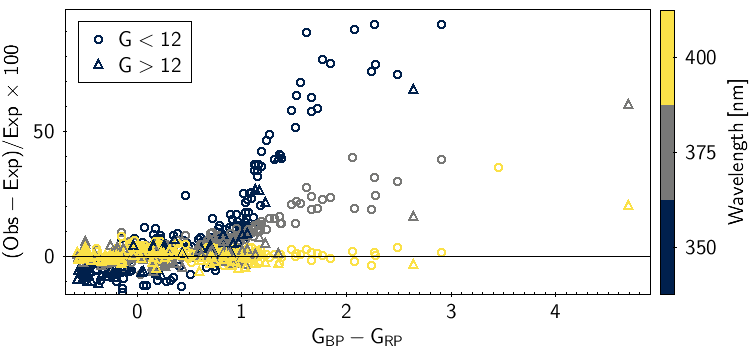}
    \caption{Percentage residuals measured at wavelength $\lambda=\{350,375,400\}$ nm plotted as a function of \bprp colour index. Open circles represent sources with $G<12$ while open triangles represent sources with $G>12$.}
    \label{fig:samplBlueColourTerm}
\end{figure}
The most prominent systematic effect visible in previous plots is a colour term affecting residuals at wavelengths $\lambda \lesssim 400$ nm. Figure \ref{fig:samplBlueColourTerm} shows the normalised residuals as a function of \bprp colour index; residuals are calculated at wavelengths 350, 375, and 400 nm. 
As can be seen, the colour term, which is virtually absent at $\lambda = 400$ nm, rapidly grows when wavelengths decrease, with observed fluxes for redder sources being higher than expected and fluxes of blue sources being slightly lower than expected. There is some evidence that the colour term is magnitude dependent, because it is smaller for
faint sources at the same wavelength. Although there may be issues in this wavelength range related to the wavelength calibration accuracy (see \secref{waveAccuracy}), as well as possible systematic effects in the LSF model, we believe that the presence of non-linear effects left by the internal calibration chain might not be excluded.

\begin{figure}[]
    \centering
    \includegraphics[width=\columnwidth]{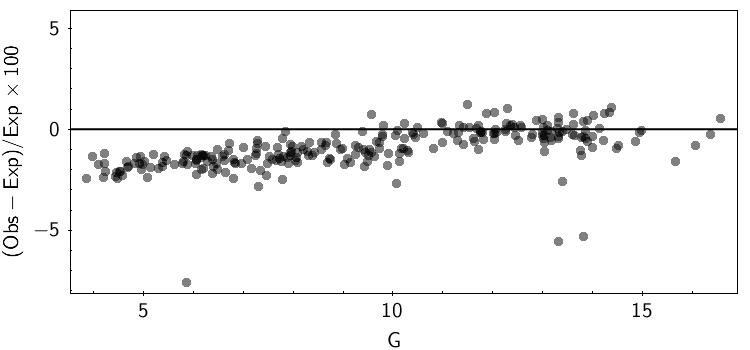}
    \caption{Percentage residuals as a function of G magnitude measured on BP spectra at wavelength $\lambda=580$ nm.}
    \label{fig:samplBpDip}
\end{figure}
A second magnitude-dependent effect is visible in BP spectra for wavelengths between 560 and 600 nm and is represented in \figref{samplBpDip}: the plot shows residuals measured at $\lambda=580$ nm as a function of $G$ magnitude of the source. While residual are flat for sources fainter than $G\simeq11$, a linear trend is clearly visible with fluxes becoming progressively lower than expected, up to $-2\%$ at the bright end. This systematic effect is located immediately before the BP/RP interface and could be related to internal LSF calibration issues that induce small deformations in the shape of the mean spectrum (as already seen in the residuals in the bottom panel of \figref{samplSpecCompByMag}).
\begin{figure}[]
    \centering
    \includegraphics[width=\columnwidth]{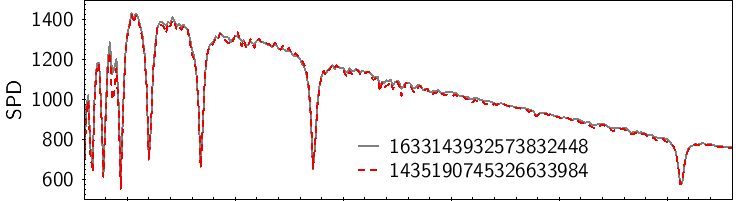}
    \includegraphics[width=\columnwidth]{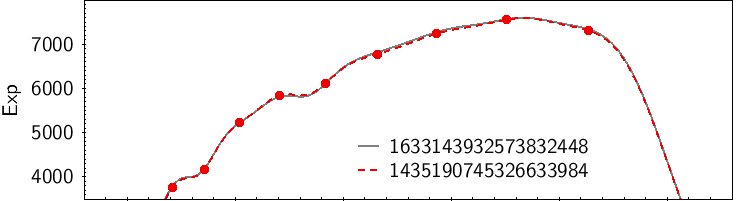}
    \includegraphics[width=\columnwidth]{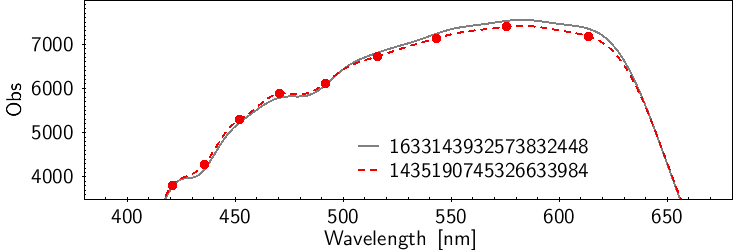}
    \caption{Comparison between flux distributions of two sources with the same SPD shape but different magnitude (source ID number shown in the legend). Fluxes for the second source have been normalised to match the same level of the first. The comparison is shown between high-resolution SPDs (\emph{top} panel), expected BP mean spectra (\emph{middle} panel), and observed mean BP spectra (\emph{bottom} panel).}
    \label{fig:samplBpDipExample}
\end{figure}
A further example is shown in \figref{samplBpDipExample} with the comparison between spectra of two PVL sources whose SPD shape is extremely similar but with different luminosity levels:  source \texttt{Gaia\,DR3\,1633143932573832448} at magnitude $G_A=12.46$ and source \texttt{Gaia\,DR3\,1435190745326633984} with magnitude $G_B=6.82$. In all the panels, the flux of the second source has been rescaled by a factor of $10^{-0.4 (G_A-G_B)}$ to put both fluxes on the same scale. 
As can be seen, while the expected BP mean spectra are almost indistinguishable, in the bottom panel the observed spectrum for the bright source is below the expected level.

\begin{figure}[]
    \centering
    \includegraphics[width=\columnwidth]{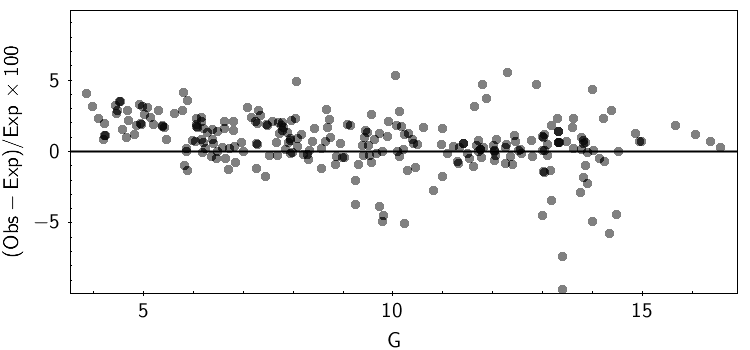}
    \caption{Percentage residuals as a function of $G$ magnitude measured at wavelength $\lambda=950$ nm.}
    \label{fig:samplRedDip}
\end{figure}
A similar magnitude-dependent systematic effect is also visible in the red part of RP spectra, even if data are more noisy in this second case: the plot in \figref{samplRedDip} shows residuals measured at $\lambda=950$ nm as a function of $G$ magnitude. 
In this case, residuals are almost flat for $G\gtrsim9,$ but at the brighter magnitude end, a linear trend can be seen as a rise with fluxes becoming higher than expected by $\sim 3\%$ at $G\simeq4$.

\subsubsection{Errors}\label{sec:errors}

\citet{DeAngeli2022} observed that mean spectra errors are generally underestimated for spectral coefficients with low index and are slightly overestimated for higher order coefficients. For most of the sources, this translates to a general underestimation of errors because most of the information is contained in the lower order coefficients. Moreover, these errors are only reliable for fainter sources.
To assess the validity of error on fluxes, we normalise the residuals between observed and expected mean fluxes by the error on fluxes obtained by summing in quadrature the error from the \xp mean spectra,
with the error on the model prediction coming from the SPD error budget and a calibration error obtained by the instrument models ensemble (see \secref{processing}).
\begin{figure}[]
    \centering
    \includegraphics[width=(\textwidth-4mm)/2]{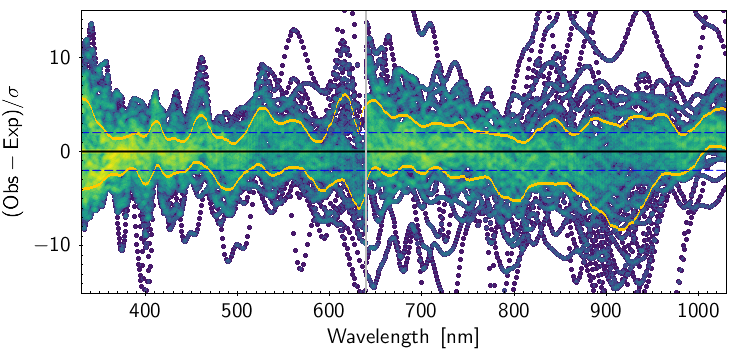}
    \caption{Difference between observed and expected mean flux normalised by the squared sum of errors as a function of wavelength for \xp sampled mean spectra. Yellow curves are the $P_{16}$ and $P_{84}$ quantile distributions while horizontal dashed blue lines set the $\pm2$ level.}
    \label{fig:samplSpecWithError}
\end{figure}
These normalised residuals are shown in \figref{samplSpecWithError} for sources with $G>12$. The yellow curves represent $P_{16}$ and $P_{84}$ percentiles used as a proxy for the $\pm1\sigma$ level. As can be seen, $\sigma\gtrsim2$ for most of the wavelength range (dashed blue horizontal lines are set at $\pm2$ to guide the eye): this means that differences are larger than the estimated uncertainties. It is worth pointing out that the considered error does not contain the contribution of covariances between samples: however, these covariances are not negligible because the mean spectra are represented as a continuous model, and therefore random noise takes the form of random wiggles that give rise to long-range correlations across several samples.
The solution is to evaluate a $\chi^2$ by projecting sampled spectra into coefficient space as described by \equref{coeffsProjection} and to use the full coefficient covariance matrix defined by \equref{coeffFullCovariance}.
\begin{figure}[]
    \centerline{
    \includegraphics[width=(\textwidth)/2]{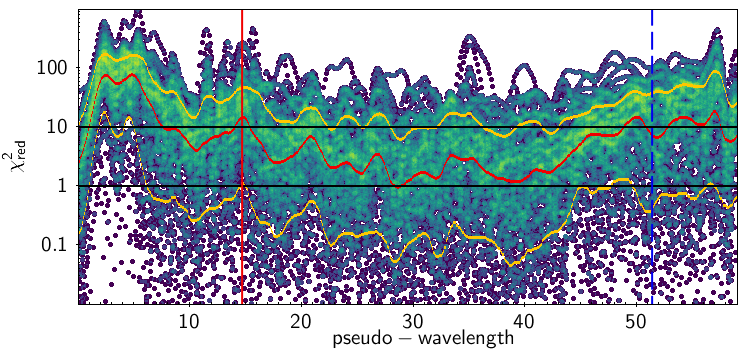}}
    \centerline{
    \includegraphics[width=(\textwidth)/2]{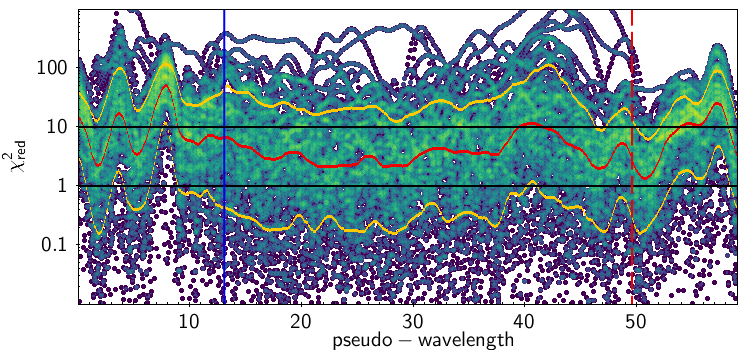}}
\caption{Contribution per sample to $\chi^2$ values for BP (\emph{top}) and RP (\emph{bottom}) computed for sources with $G>12$.}
    \label{fig:chiAllFaint}
\end{figure}
However, before doing so, it is useful to look at the full distribution of normalised residuals as a function of the AL sampling coordinate. In \figref{chiAllFaint}, we plot these residuals in the form
\begin{equation}
    \frac{\left(Obs - Exp\right)^2}{\sigma^2},
\end{equation}
interpreting this quantity as the contribution of each sample to the $\chi^2_{red}$ computation. The top panel shows residuals for the BP case: the vertical red line at $u\simeq14.7$ corresponds to the wavelength $\lambda = 640$ nm and roughly identifies the position of the filter cut-off while the vertical blue dashed line at $u\simeq51.4$ corresponds to wavelength $\lambda=330$ nm and roughly corresponds to the filter cut-off: outside this range the mean spectrum contains only the contribution coming from other parts of the spectrum due to the effect of LSF smearing. Red and yellow curves represent the $P_{50}$ and the $P_{16}$ and $P_{84}$ percentile distributions. 
A relevant piece of information coming from this plot is that while the $\chi^2$ contribution coming from the spectrum 
is reliable, with the median confined within the $1-10$ range,  we observe unrealistic values in the spectra
wings that would make the global $\chi^2$ useless. The situation is similar for the RP case shown in the bottom plot (the dashed blue line identifies the filter cut-off at $\lambda = 640$ nm, the red line the cut-off at $\lambda=1030$).
The discrepancies in the wings of spectra are driven by small systematic differences between observations and model predictions that are mostly due to the LSF model: the transmissivity of the instrument in that sample range is almost null and all the detected photons are due indeed to the smearing action of the LSF (they originate from different wavelengths); a small systematic effect in the wings of the LSF model can justify the observed discrepancies.
To evaluate the $\chi^2$ in the coefficient space, we then safely excluded a few pixels in the wings of sampled spectra
from the computation, projecting only the sample range $u=[9,53]$, and obtaining the distributions visualised in \figref{chisqHisto} for both instruments.
\begin{figure}[]
    \centerline{
    \includegraphics[width=(\textwidth)/4]{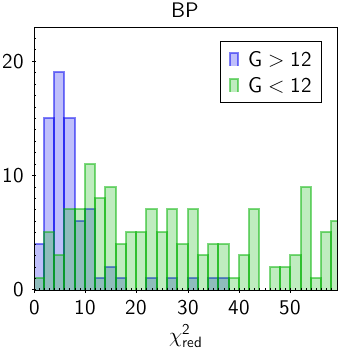}
    \includegraphics[width=(\textwidth)/4]{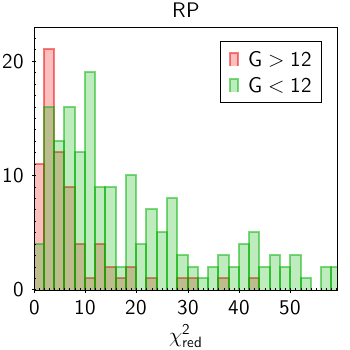}}
\caption{Reduced $\chi^2$ distributions for BP (\emph{left}) and RP (\emph{right}) computed on the set of SPSS, PVL, and NGSL spectra.}
    \label{fig:chisqHisto}
\end{figure}
We plot 
the distributions for sources with $G>12$ as blue and red histograms for BP (\emph{left}) and RP (\emph{right}), respectively, while brighter sources are represented as green distributions in both plots. The peak for the faint sources is located roughly at $\chi^2\simeq6$ for BP and $\chi^2\simeq3$ for RP: these rather large values may be in part due to an underestimation of spectra covariances, but there is evidence that the calibration error budget coming from the instrument models ensemble may also be underestimated because of the suboptimal procedure followed for its construction (differential evolution algorithm instead of a more appropriate Markov Chain Monte Carlo method).
Values for bright sources are meaningless because of the combined effect of systematic differences in the shapes of  spectra  seen in \secref{meanspec} and unreliable coefficient covariances.

\subsection{Comparing mean spectra in wavelength space}\label{sec:ecs}

\begin{figure}[]
    \centerline{
    \includegraphics[width=(\columnwidth)-5mm]{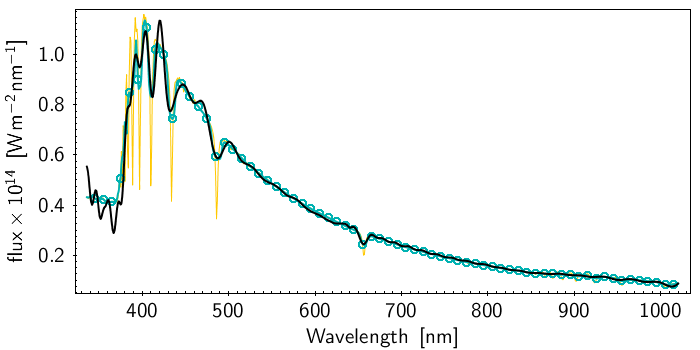}}
    \centerline{
    \includegraphics[width=(\columnwidth)-5mm]{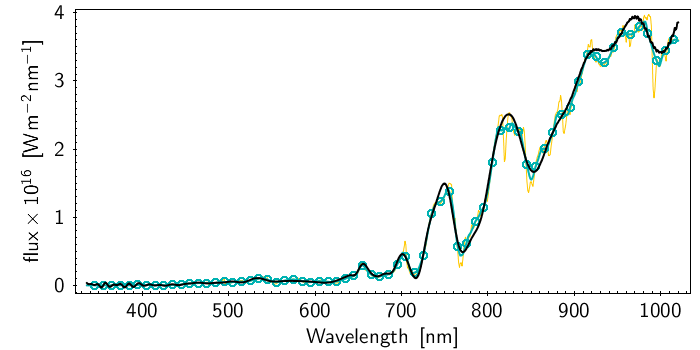}}
\caption{Comparison between ECS and model SED for source \texttt{Gaia\,DR3\,1435896975388228224} (\emph{top}) and source \texttt{Gaia\,DR3\,4339417394313320192} (\emph{bottom}). The black curve represents the \gaia ECS, the yellow curve represents the corresponding high-resolution SED from external data, and aquamarine open circles are the SED degraded at a spectral resolution corresponding to one \gaia pixel.}
    \label{fig:ecsComp}
\end{figure}
\afigref{ecsComp} shows the comparison between \gaia ECS and the corresponding reference SED from external data for the same sources as in \figref{samplSpecComp1}. The ECS is compared to the degraded version of the SED computed as described at the beginning of \secref{validation}. Although the agreement between the distribution is excellent, some wiggles are present, especially in the top plot in the blue part of the spectrum and around the Balmer absorption lines. The wiggles originate in the basis inversion process and their effect can vary a lot in ECS as can be seen by looking at the other examples in this section. 
\begin{figure}[]
    \centerline{
    \includegraphics[width=(\columnwidth)-5mm]{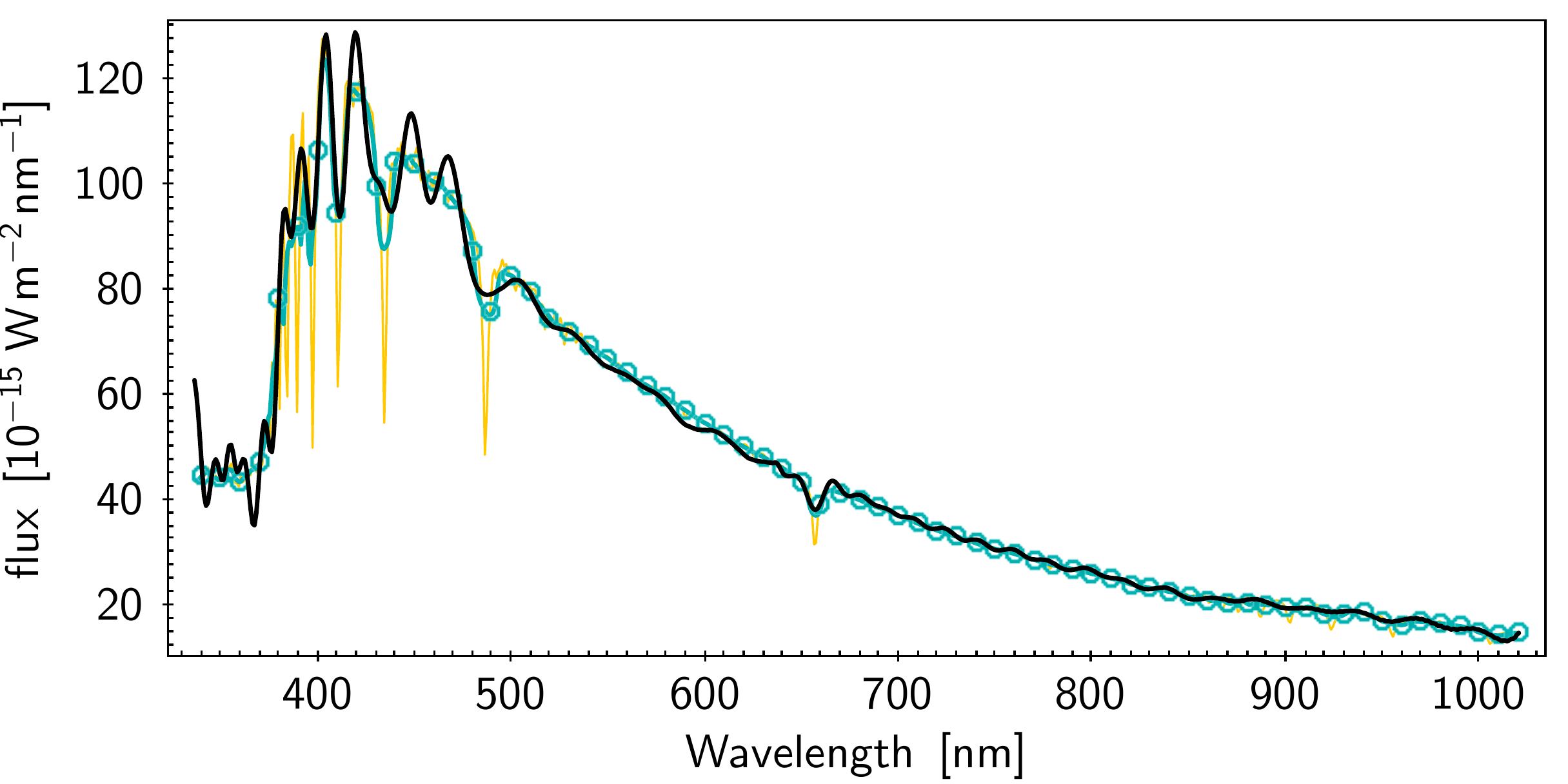}}
    \centerline{
    \includegraphics[width=(\columnwidth)-5mm]{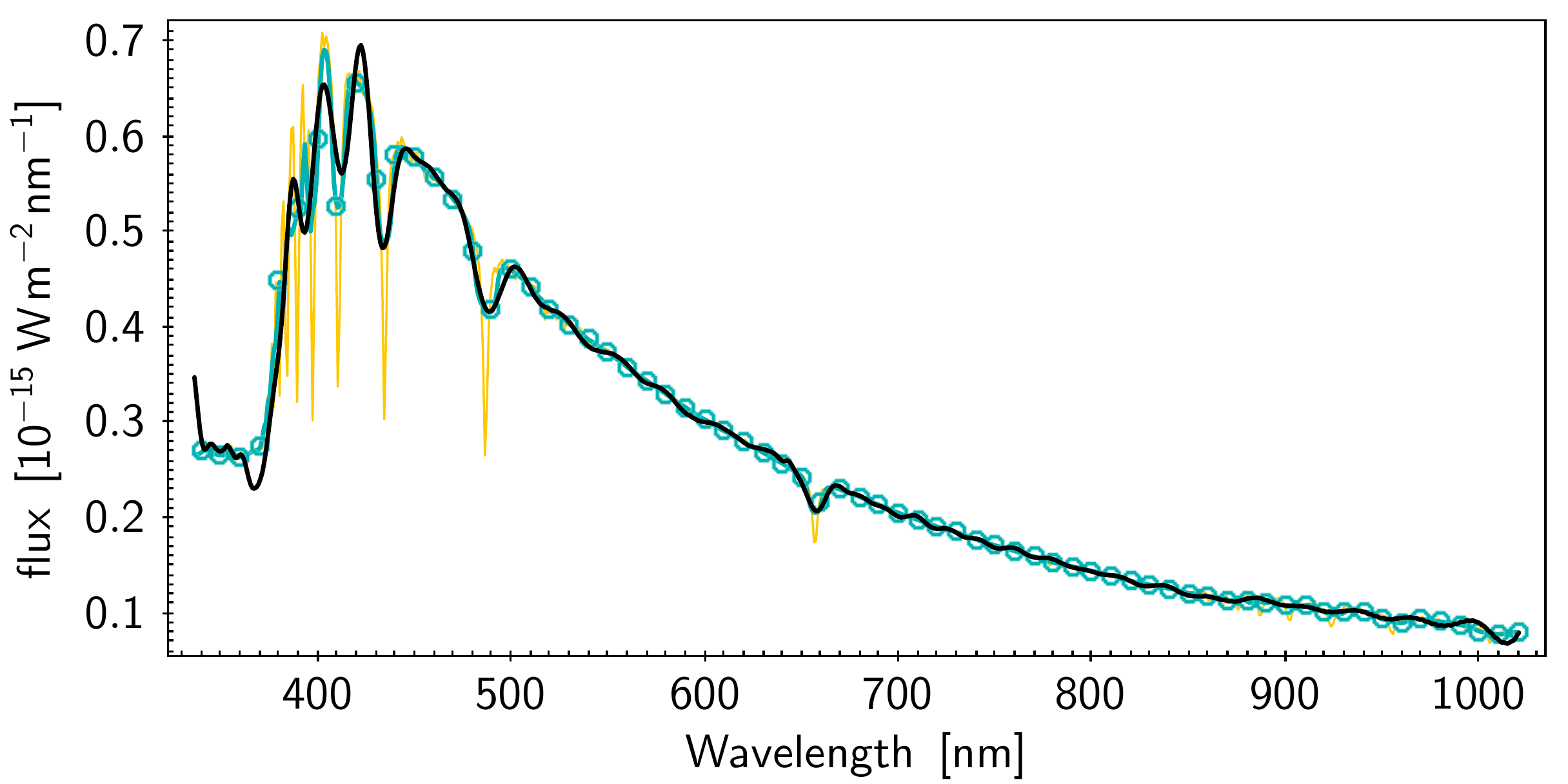}}
\caption{Comparison between ECS and model SED for source \texttt{Gaia\,DR3\,1435190745326633984} (\emph{top}) and source \texttt{Gaia\,DR3\,1633143932573832448} (\emph{bottom}), the same sources represented in \figref{samplBpDipExample}. The colour coding and symbols are the same as in \figref{ecsComp}.}
    \label{fig:ecsTwins}
\end{figure}

\afigref{ecsTwins} shows the ECS SED comparison for the same sources represented in \figref{samplBpDipExample}. These sources have reference SEDs with almost the same shape, but different magnitudes ($G=6.82$ in the top panel vs. $G=12.46$ in the bottom panel). The intensity of wiggles is much higher in the bright source \wrt the fainter one, especially for $\lambda\lesssim500$ nm, indicating that neither the shape of the spectrum nor the noise level necessarily determines the amount of wiggling.
\begin{figure}[]
    \centerline{
    \includegraphics[width=(\columnwidth)-5mm]{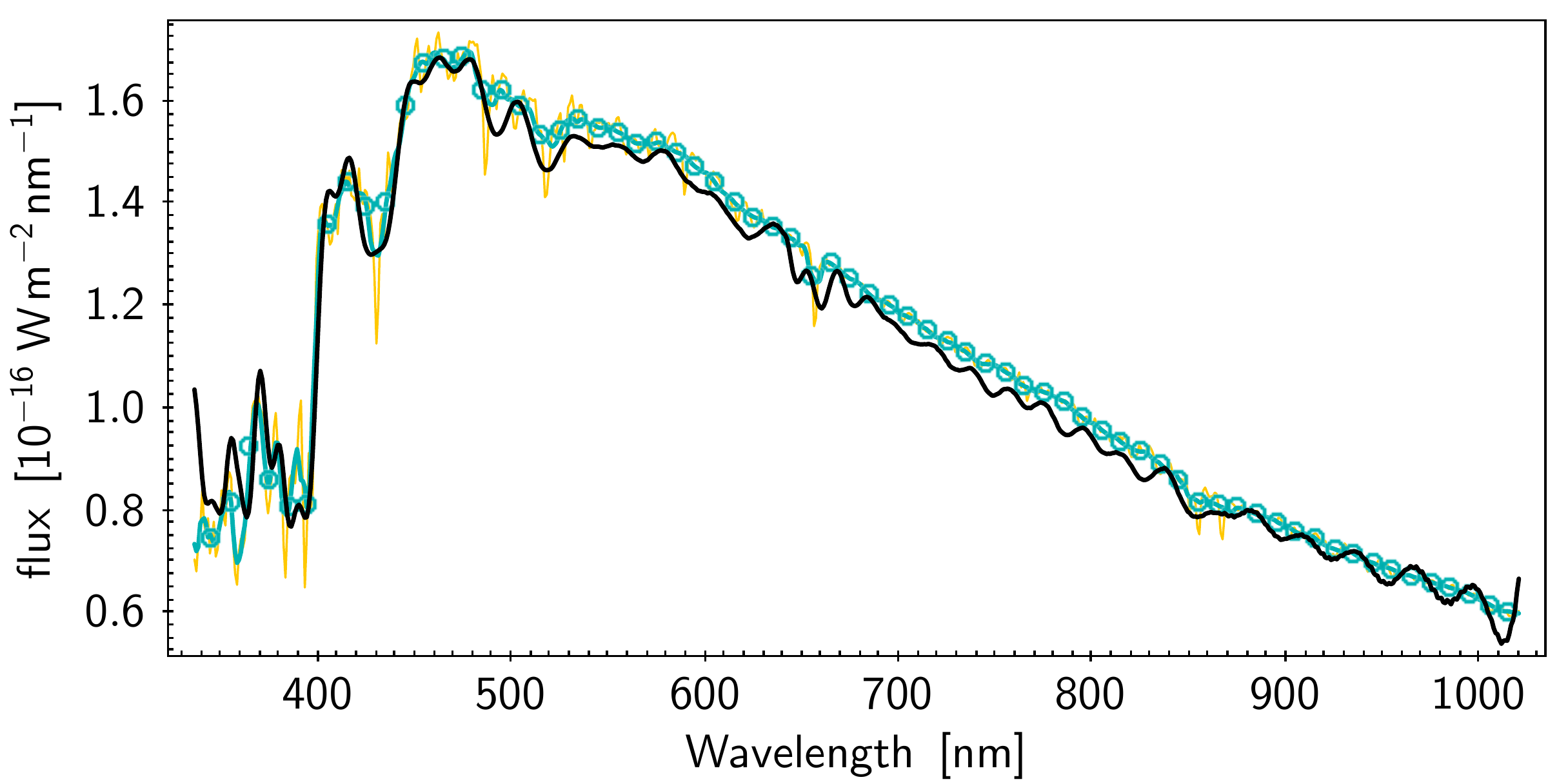}}
    \centerline{
    \includegraphics[width=(\columnwidth)-5mm]{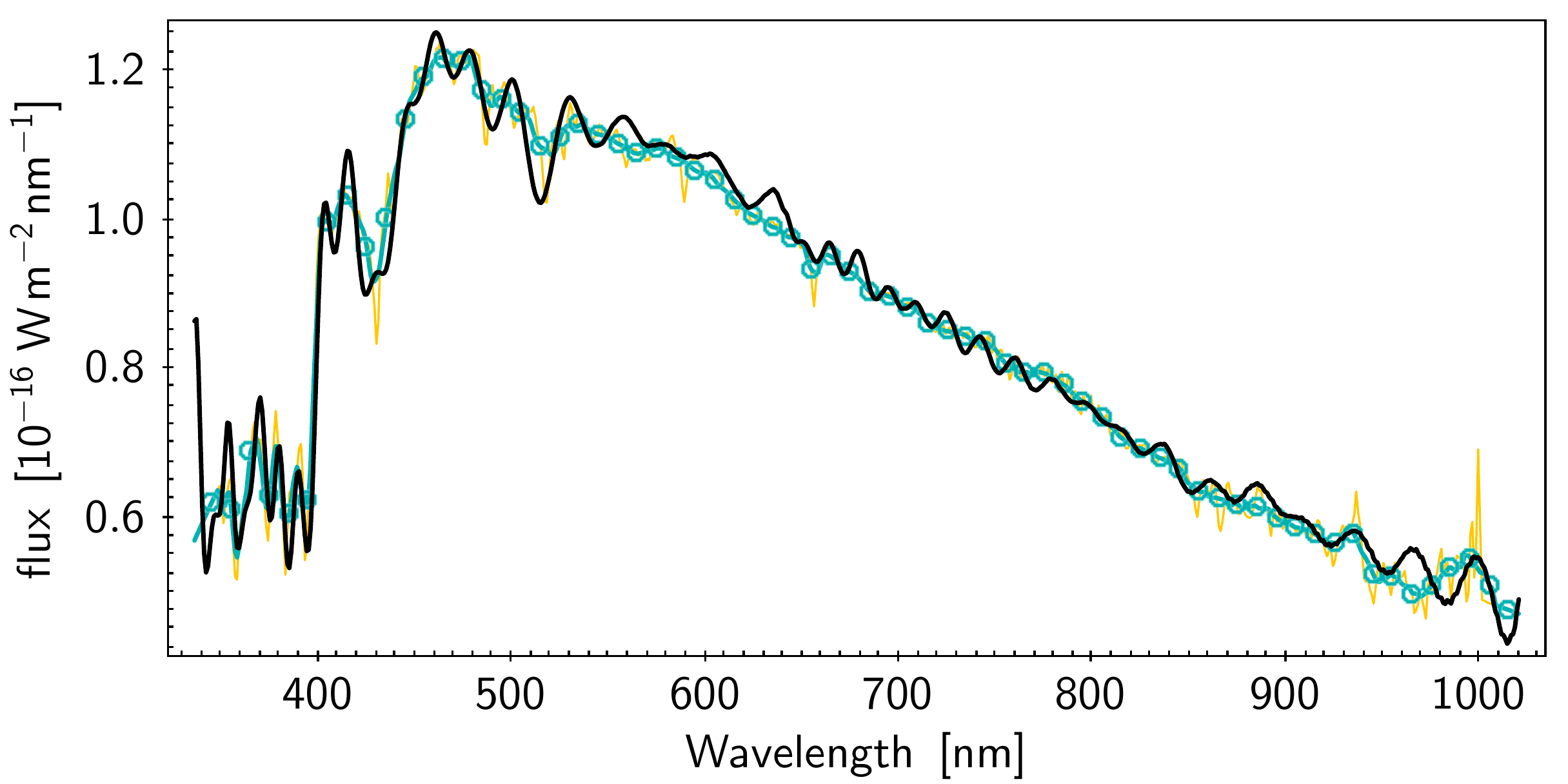}}
\caption{Comparison between ECS and model SED for source \texttt{Gaia\,DR3\,1399559249961569792} (\emph{top}) and source \texttt{Gaia\,DR3\,2323394345824851584} (\emph{bottom}), two SPSSs of similar spectral type and magnitude. The colour coding and symbols are the same as in \figref{ecsComp}.}
    \label{fig:ecsTwins2}
\end{figure}
\begin{figure}[]
    \centerline{
    \includegraphics[width=(\columnwidth)-5mm]{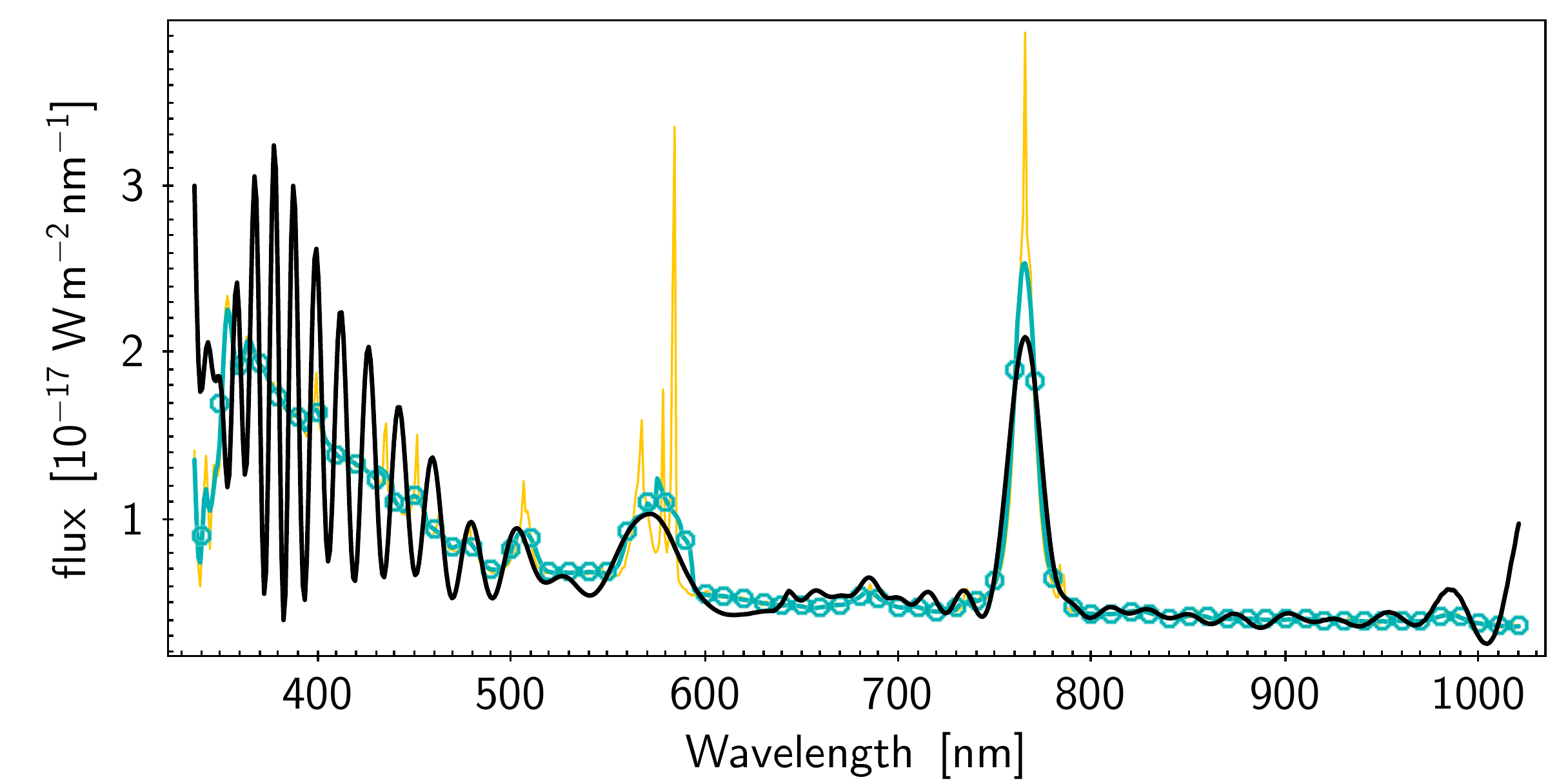}}
    \centerline{
    \includegraphics[width=(\columnwidth)-5mm]{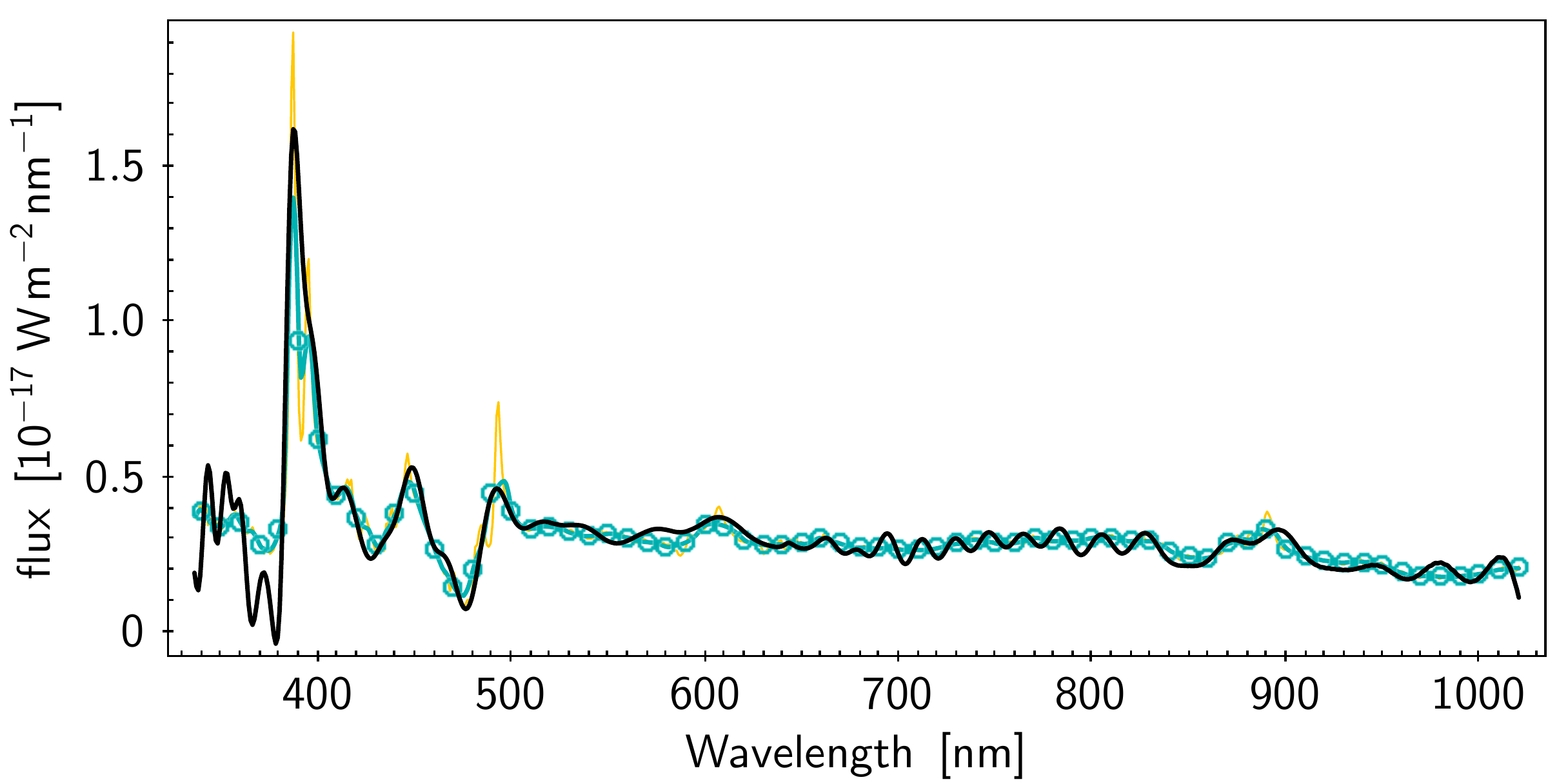}}
\caption{Comparison between ECS and model SED for source \texttt{Gaia\,DR3\,4007020769942990080} (\emph{top}) and source \texttt{Gaia\,DR3\,4467076569812517248} (\emph{bottom}), two QSOs from the SDSS catalogue. The colour coding and symbols are the same as in \figref{ecsComp}.}
    \label{fig:ecsQso}
\end{figure}
Another example of two sources with similar spectral type and comparable apparent magnitude but different wiggle intensity is given in \figref{ecsTwins2}: sources \texttt{Gaia\,DR3\,1399559249961569792} ($G=13.32$) and \texttt{Gaia\,DR3\,2323394345824851584} ($G=13.62$), both belonging to the SPSS catalogue.
Finally, an example of ECS comparison for two sources with emission lines is shown in \figref{ecsQso}: source \texttt{Gaia\,DR3\,4007020769942990080} (\emph{top}) and source \texttt{Gaia\,DR3\,4467076569812517248} (\emph{bottom}) with magnitudes of $G=16.45$ and $G=17.31$, respectively, both contained in the SDSS catalogue, exhibit an extremely different wiggle intensity in the blue part of their spectra, with the fainter one showing a better-behaved ECS (therefore wiggle intensity is not directly related to S/N).

\begin{figure}[]
    \centerline{
    \includegraphics[width=(\columnwidth)-5mm]{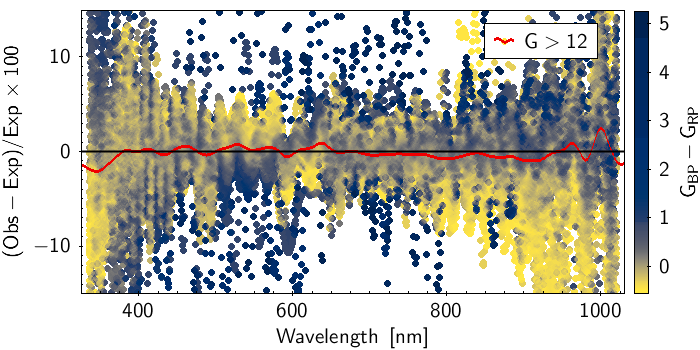}}
    \centerline{
    \includegraphics[width=(\columnwidth)-5mm]{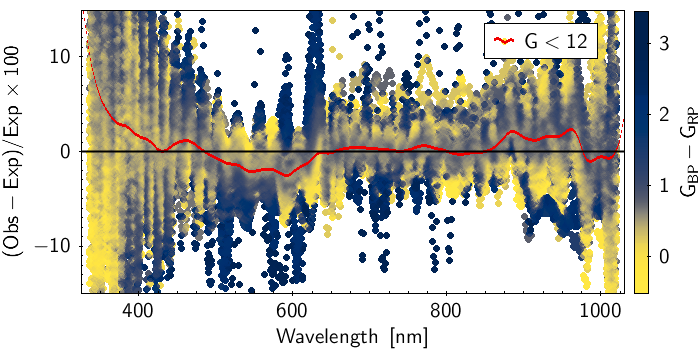}}
   \caption{Percentage residuals between ECS and reference fluxes for sources with $G>12$ (\emph{top}) and $G<12$ (\emph{bottom}), for the full set of SPSS, PVL, and NGSL samples. The colour map encodes the \bprp colour index. The red curve represents the median distribution smoothed by a Gaussian with $\sigma=10$ nm.}
    \label{fig:ecsPercRes}
\end{figure}
A more global analysis is presented in  \figref{ecsPercRes}, which shows the percentage residuals computed on the whole set of SPSS, PVL, and NGSL samples divided into two luminosity classes, sources with $G>12$ in the top panel and sources with $G<12$ in the bottom panel. 
If we compare these residuals with the equivalent quantities measured on ICS (\figref{samplSpecCompByMag}), the effect of wiggles is evident. Nevertheless, the red lines representing the median of the distributions, once smoothed by a Gaussian with $\sigma=10$ nm, show a substantially equivalent behaviour to that seen in the ICS residuals.  
\begin{figure}[]
    \centerline{
    \includegraphics[width=(\columnwidth)-5mm]{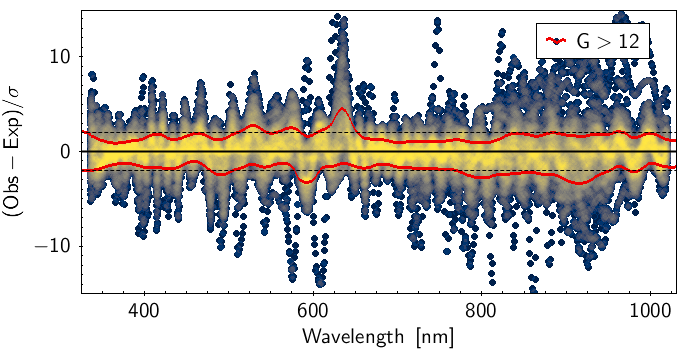}}
   \caption{Normalised residuals between ECS and reference fluxes for sources with $G>12$. The two red curves represent the $P_{16}$ and $P_{84}$ percentile distributions smoothed by a Gaussian with $\sigma=10$ nm. The $\pm2$ levels are also plotted as horizontal dashed black lines.}
    \label{fig:ecsNormRes}
\end{figure}
Finally, \figref{ecsNormRes} shows the normalised residuals for the faint sample. The error is computed here as the sum in quadrature of the contribution of the ECS error and the error on the model SED computed by propagating the errors on the full-resolution SED following \equref{sedDegradation} under the hypothesis that errors are independent.
Percentiles $P_{16}$ and $P_{84}$ (red curves), smoothed by a Gaussian with $\sigma=10$ nm, are used as a proxy to the actual standard deviation of the residuals which proves to be wider than expected. This result may in part be due to the lack of covariance terms for the high-resolution SED samples, which results in underestimation of the model error; however, the result is consistent with what is seen in \figref{samplSpecWithError} for the ICS comparison.
\begin{figure}[]
    \centerline{
    \includegraphics[width=(\columnwidth)/3]{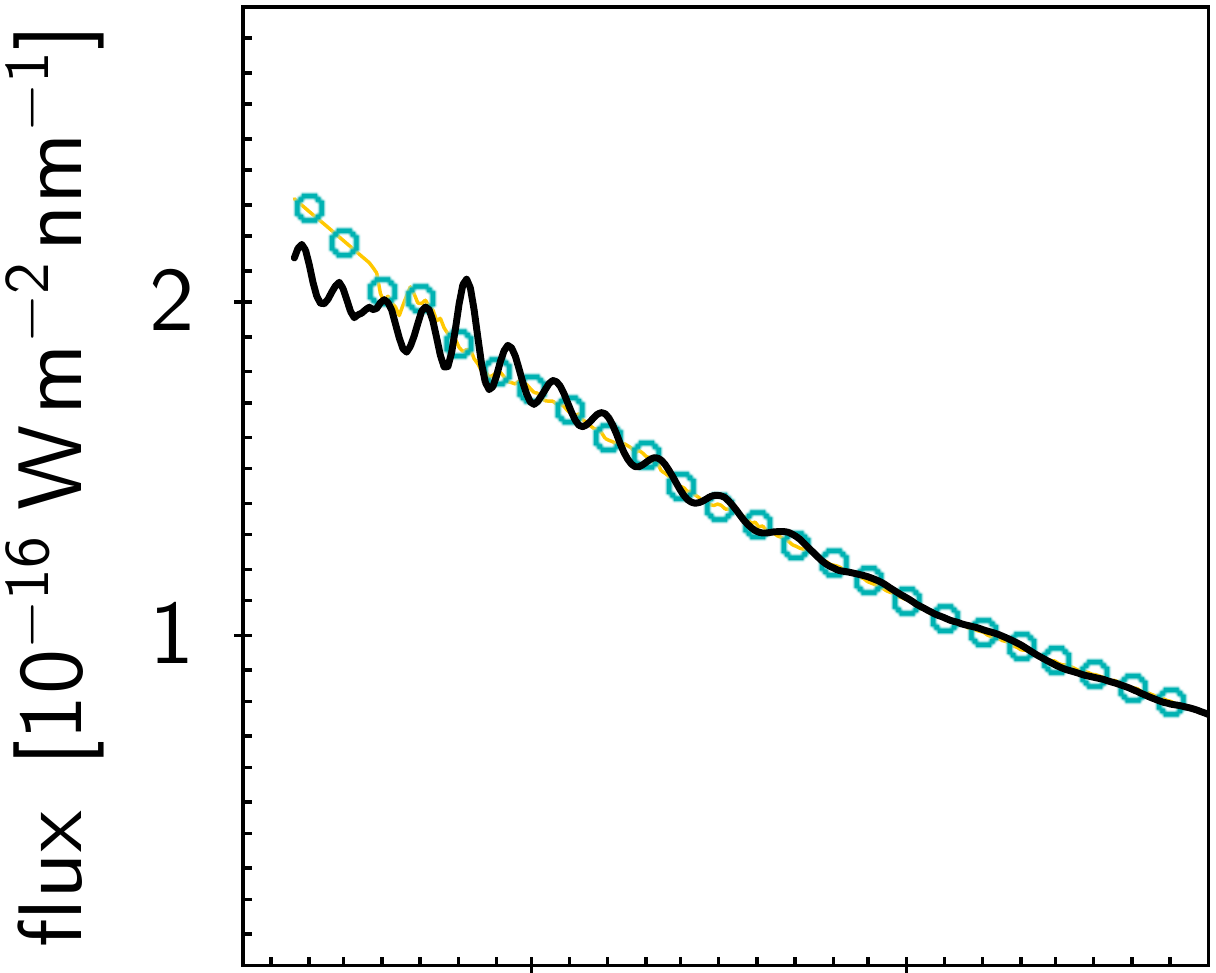}
    \includegraphics[width=(\columnwidth)/3]{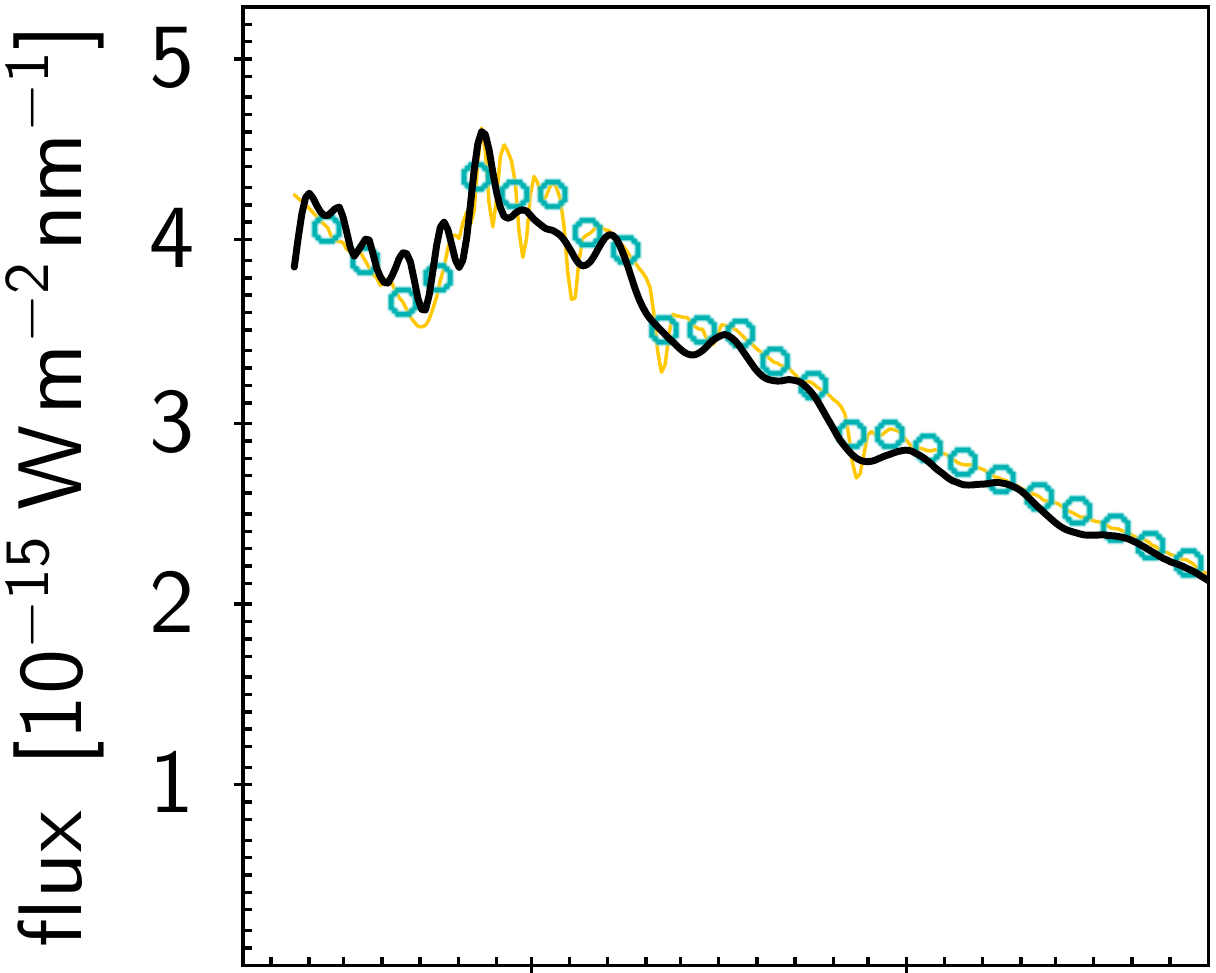}
    \includegraphics[width=(\columnwidth)/3]{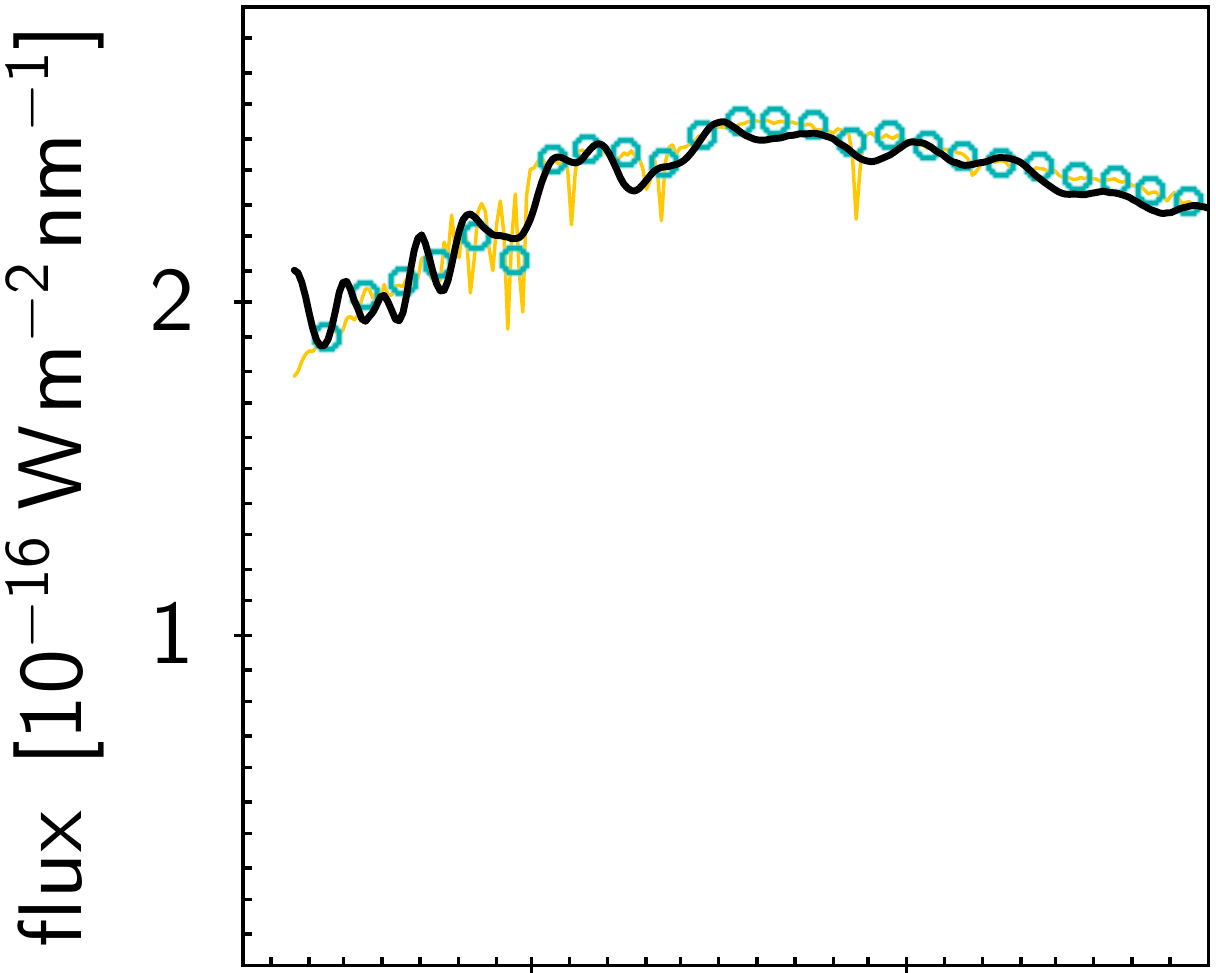}}
    \centerline{
    \includegraphics[width=(\columnwidth)/3]{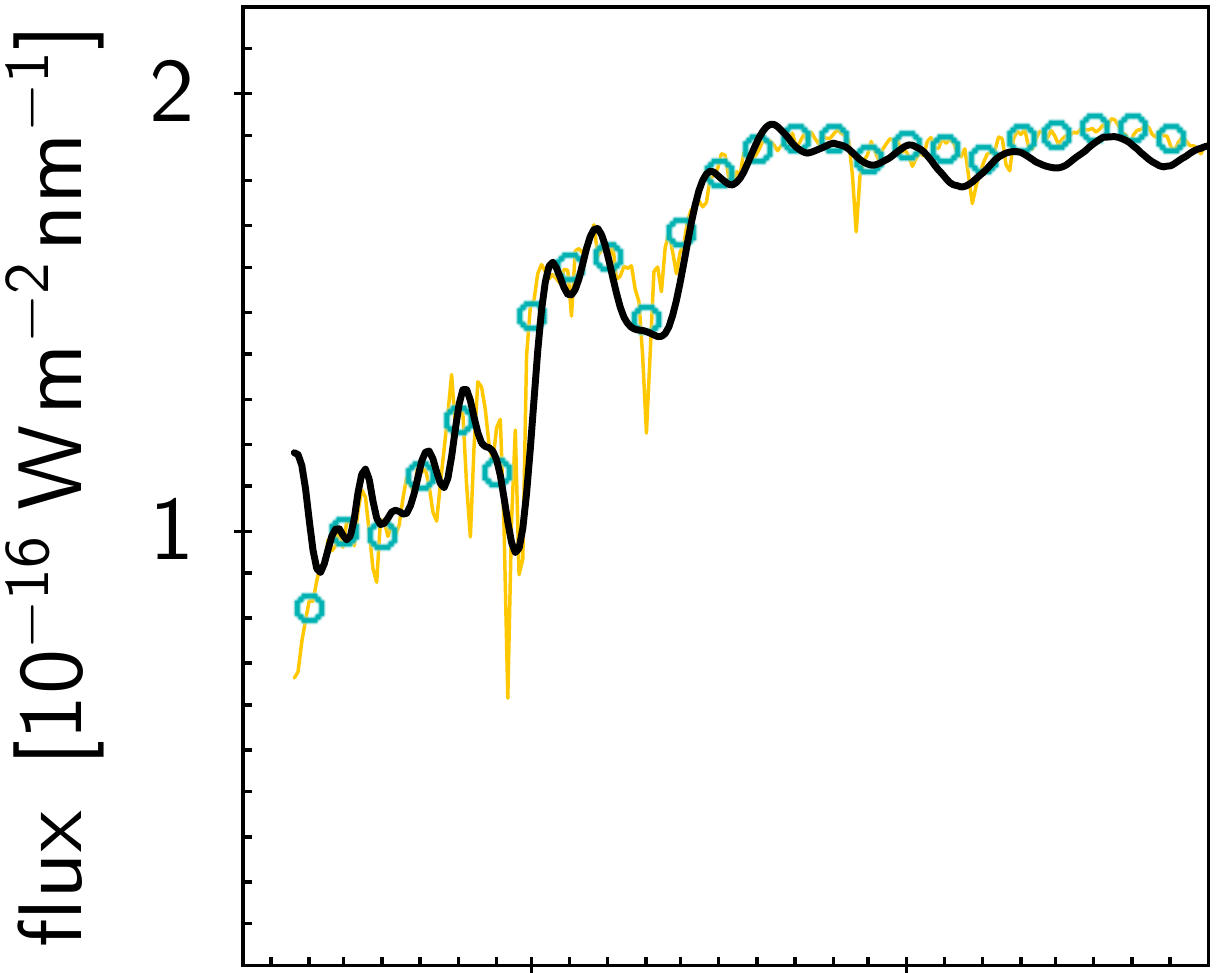}
    \includegraphics[width=(\columnwidth)/3]{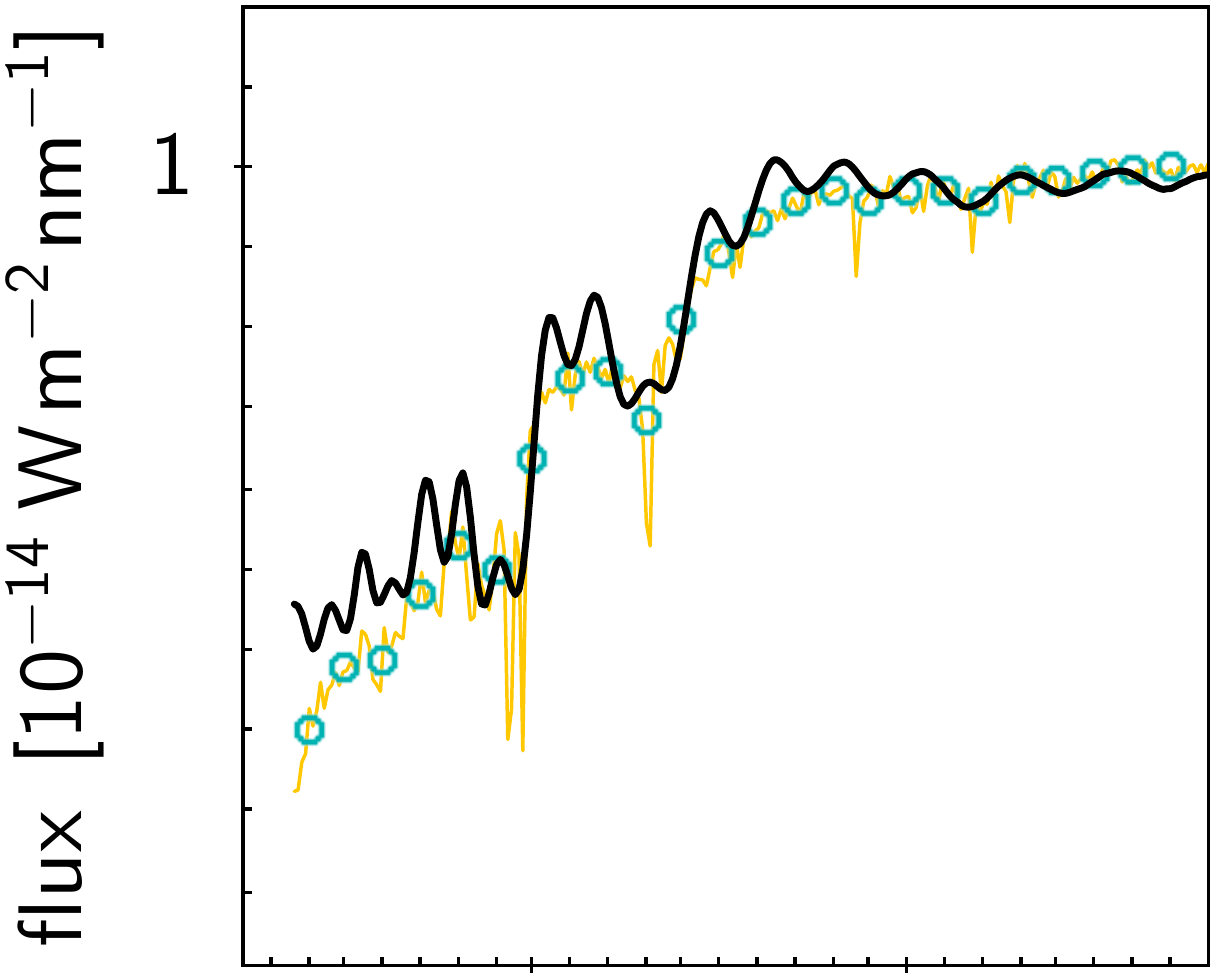}
    \includegraphics[width=(\columnwidth)/3]{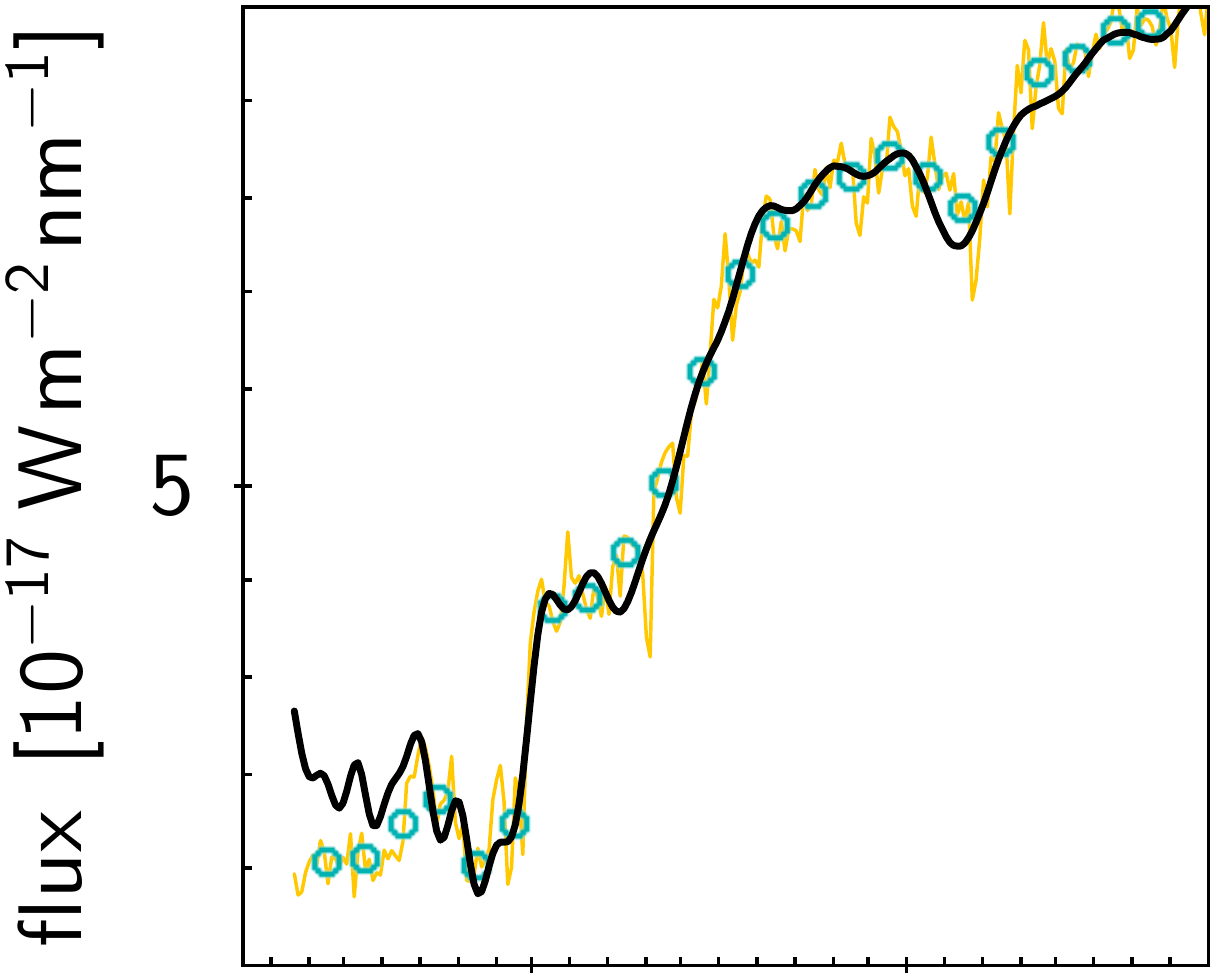}}
    \centerline{
    \includegraphics[width=(\columnwidth)/3]{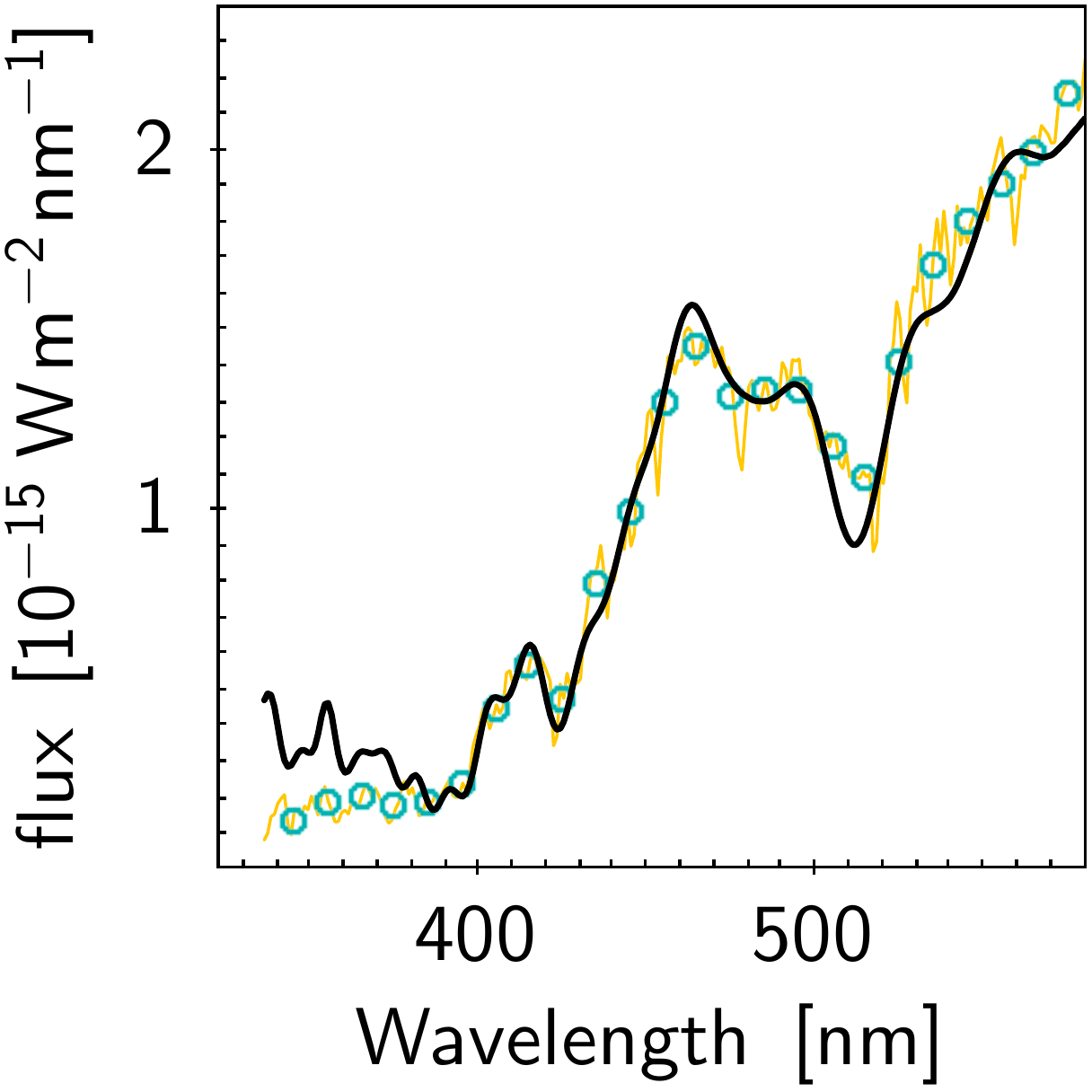}
    \includegraphics[width=(\columnwidth)/3]{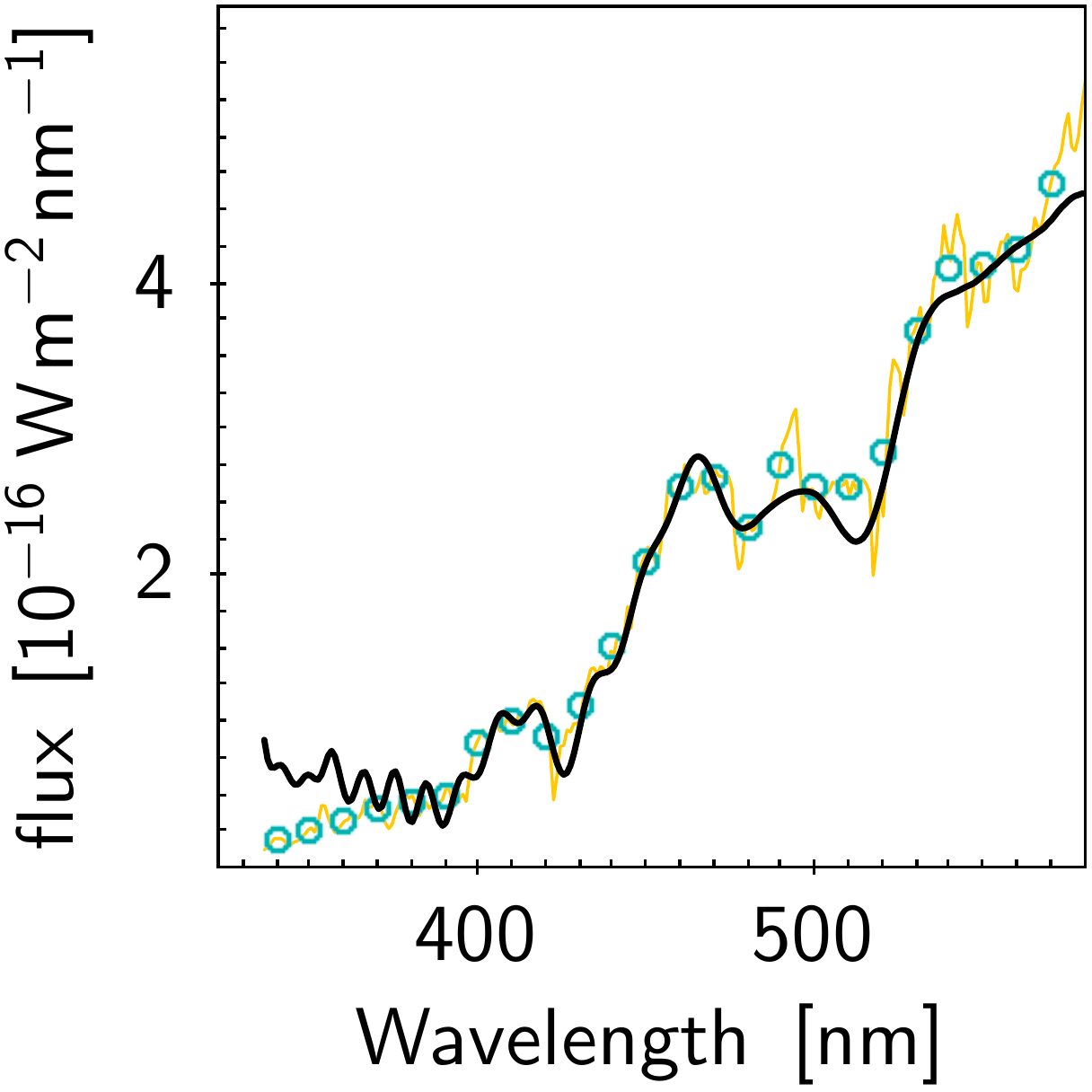}
    \includegraphics[width=(\columnwidth)/3]{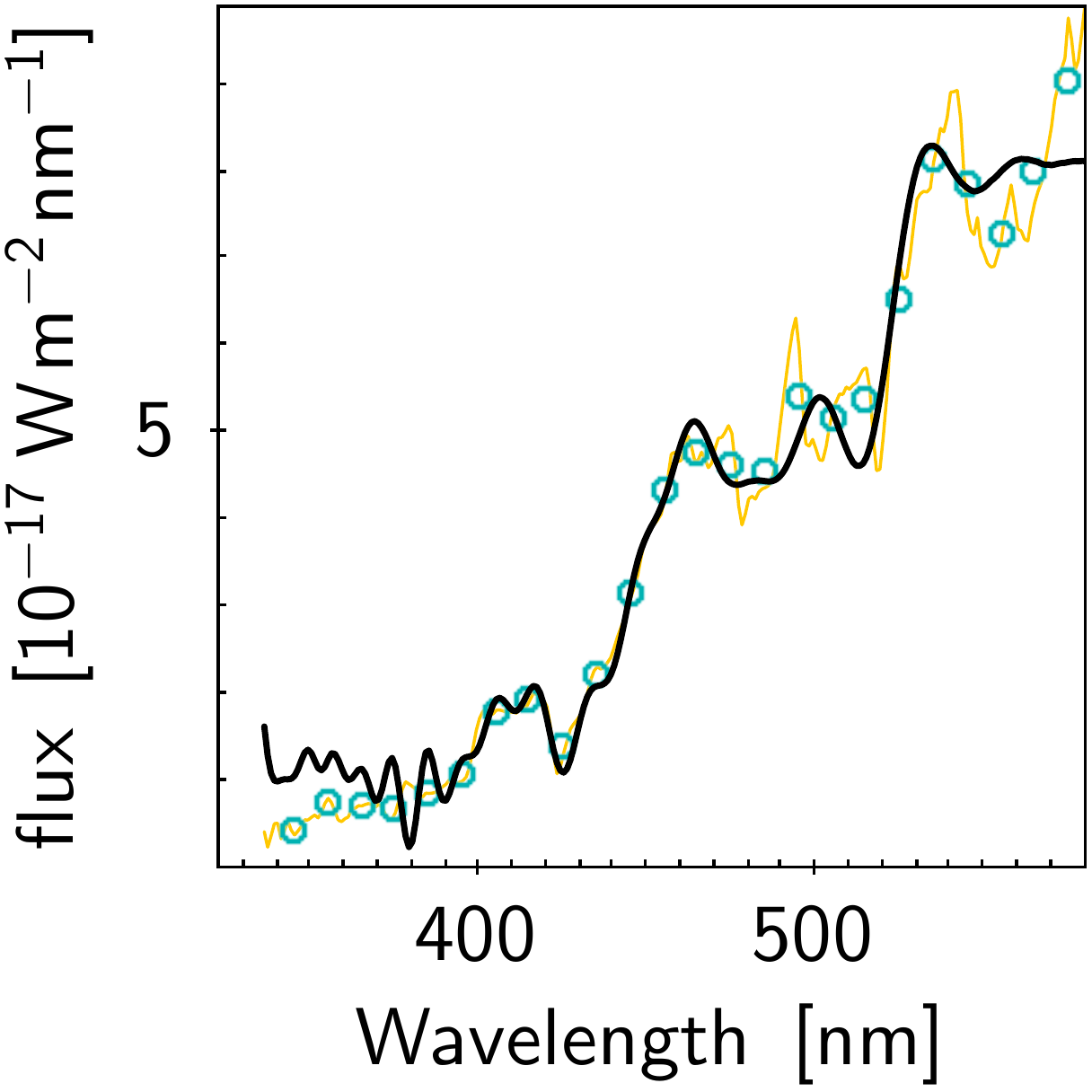}}
    \caption{Comparison between ECS and model SED for SPSS with colour index \bprp respectively equal to (from \emph{left} to \emph{right}, from \emph{top} to \emph{bottom}) 0.06, 0.27, 0.83, 1.03, 1.07, 1.35, 1.68, 2.26, and 2.65. ECS flux at wavelengths $\lambda<400$ nm is systematically lower than expected for bluer sources and higher than expected for the redder ones. The colour convention and symbols are the same as in \figref{ecsComp}.}
    \label{fig:ecs400nm}
\end{figure}
The systematic colour-dependent difference seen in ICS at wavelengths $\lambda < 400$ nm is still present and is illustrated in \figref{ecs400nm}: a sample of SPSS ECS with colour ranging from $\bprp = 0.06$ (\emph{top left} panel) to $\bprp = 2.65$ (\emph{bottom right} panel) is compared against the corresponding model SED. While the ECS is fainter than expected in the bluest source, it becomes systematically brighter than expected for redder sources. Moreover, the difference is higher at smaller wavelengths, following the same behaviour shown in \figref{samplBlueColourTerm} for ICS.

\subsection{Effect of the reconstruction error on ECS}\label{sec:invForwExperiment}

\begin{figure}[]
    \centerline{
    \includegraphics[width=(\columnwidth)-5mm]{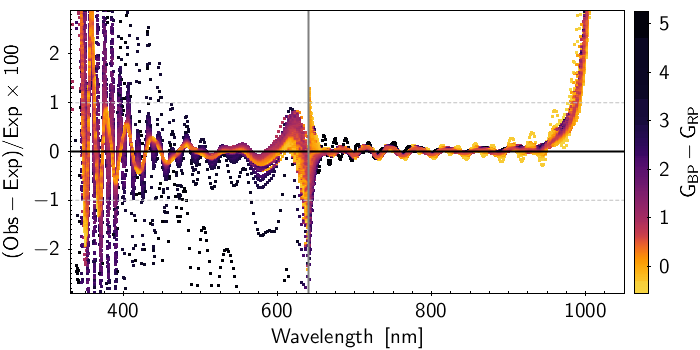}}
    \centerline{
    \includegraphics[width=(\columnwidth)-5mm]{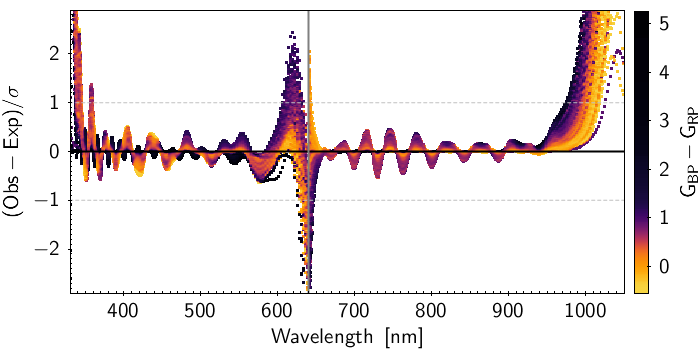}}
    \caption{Effect of the reconstruction error on ECS.\emph{Top}: Percentage residuals between mean BP/RP fluxes and corresponding forward modelled mean spectra computed from the ECS and instrument models as a function of wavelength; the plot uses data for the full set of SPSS, PVL, and NGSL samples. \emph{Bottom}: Residuals are normalised by the model error. In both cases, the colour map encodes the \bprp colour index, while horizontal dashed lines placed at $\pm1$ provide a reference. The vertical grey line at $\lambda=640$ sets the separation between BP and RP data.}
    \label{fig:invForw}
\end{figure}

In \secref{invbases} we evaluate the numerical reconstruction error on inverse bases by forward modelling their image through the instrument model and comparing the resulting functions with the original Hermite functions. As discussed in \secref{specresol}, well-behaved Gaussian noise on the instrument model can badly propagate to inverse bases, creating systematic errors in ECS. To look for such systematic effects, we performed the following test: using measured BP/RP mean spectra coefficients of SPSS, PVL, and NGSL sources, we computed the corresponding ECS, and then we forward modelled the BP/RP mean spectra corresponding to these ECS in sample space and compared the results with the original \gaia BP/RP sampled mean spectra. This comparison is shown in \figref{invForw} as percentage residuals (\emph{top} panel) and as normalised residuals (\emph{bottom} panel) as a function of wavelength. In the second case, the normalisation is done by dividing the residuals by the error on the model spectra\footnote{Errors on the measured BP/RP spectra are irrelevant here because we start from the noisy spectrum and compare back to it.}. In both cases, residuals are computed as observation minus model and the colour-coding indicates source colour. The percentage residuals show a strong systematic difference in RP for wavelengths $\lambda\gtrsim950$ nm, while in the remaining RP wavelength range, residuals are extremely well behaved, and are generally confined to the $\pm0.1\%$ region. In comparison, BP shows more important ripples confined to the range $\pm0.2\%$ for wavelengths in the interval $[480, 600]$ nm, progressively worsening outside this interval. It is interesting to notice the colour dependence of the amplitude of the  ripples from the \bprp colour index, with redder sources showing more extended ripples than bluer ones. The mean behaviour of the residuals at the shortest wavelength is more clear in the bottom panel, where the normalisation by the error has the effect of smoothing the ripples in the blue part of the BP spectrum, particularly for the reddest sources: residuals reveal a significant increasing systematic difference for wavelengths $\lambda\lesssim350$ nm. Residuals for RP at longer wavelengths show a trend with colour index; this is expected considering the larger flux errors at these wavelengths for blue sources than for redder ones. Finally, it is worth noting that 
these systematic effects have a much smaller amplitude than those seen in the ICS comparison in \figref{samplSpecCompByMag} and consequently their amplitude in wavelength space could possibly be overwhelmed by other systematic effects, such as those mentioned in relation to 
\figref{ecsPercRes} and \figref{ecsNormRes}.

\subsection{Synthetic photometry on ECS}\label{sec:synthphot}

The availability of calibrated low-resolution SEDs means various applications are possible, such as performing wide- and medium-band synthetic photometry in any photometric system whose passbands are fully enclosed in the $330-1050$ nm wavelength range covered by \gaia \xp spectra. The science verification paper by \cite{Montegriffo2022} is dedicated to this specific application. Here, we use this technique for the validation of our calibrations because it allows comparisons with completely independent reference data. In the following, we illustrate  two different test cases: a comparison with Landolt standard photometry in the Johnson-Kron-Cousins \citep[JKC hereafter,][]{Bessel05} system and a comparison with Hipparcos photometry in the $B_T$, $V_T$, and $H_p$ bands \citep{Hippaphot1997}. In both cases, we first show a comparison between synthetic photometry realised on external SED data and synthetic photometry on ECS data computed for the SPSS, PVL, and NGSL samples to verify the consistency between the two scales. We then test the ECS synthetic photometry against external photometric standards to allow for independent validation of the ECS flux scale. 
The acronym HRS is used to distinguish external high-resolution spectra from ECS data.

\subsubsection{Landolt photometry}{\label{sec:landolt}}

\begin{figure}
\centerline{
\includegraphics[width=\columnwidth/2]{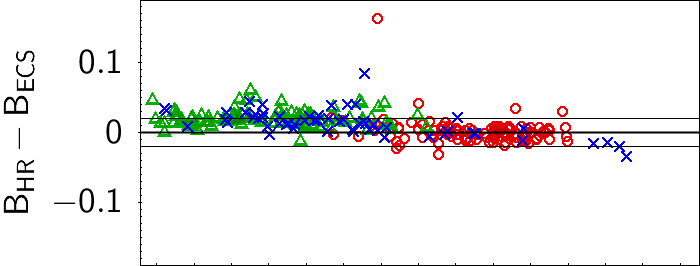}
\includegraphics[width=\columnwidth/2]{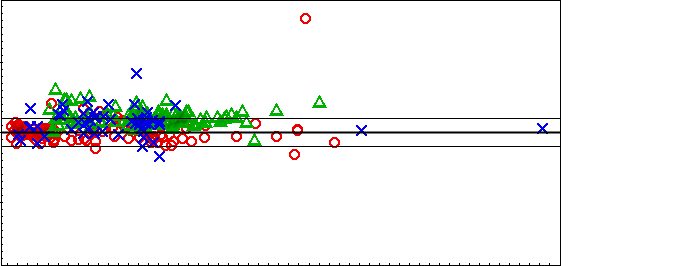}}
\centerline{
\includegraphics[width=\columnwidth/2]{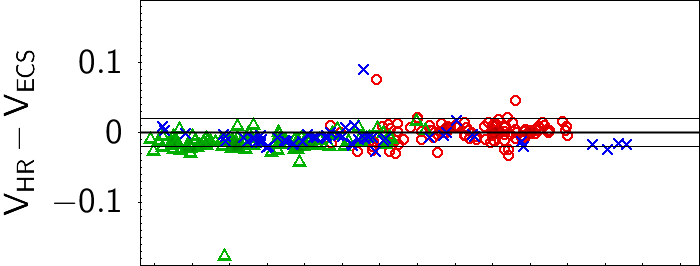}
\includegraphics[width=\columnwidth/2]{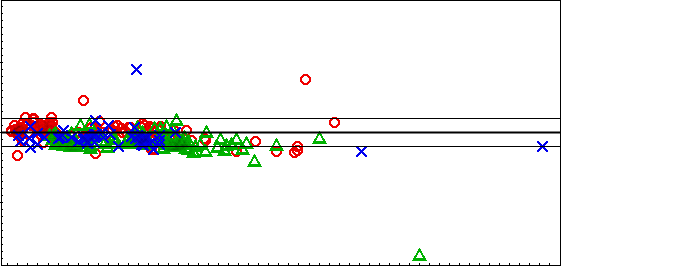}}
\centerline{
\includegraphics[width=\columnwidth/2]{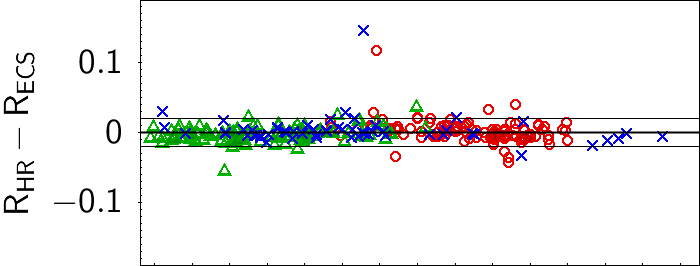}
\includegraphics[width=\columnwidth/2]{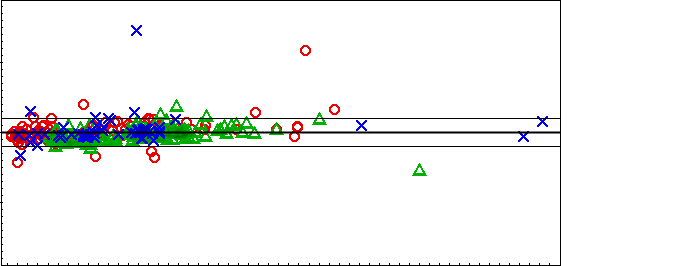}}
\centerline{
\includegraphics[width=\columnwidth/2]{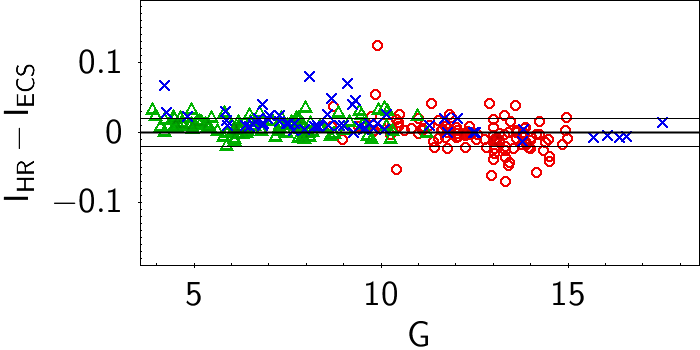}
\includegraphics[width=\columnwidth/2]{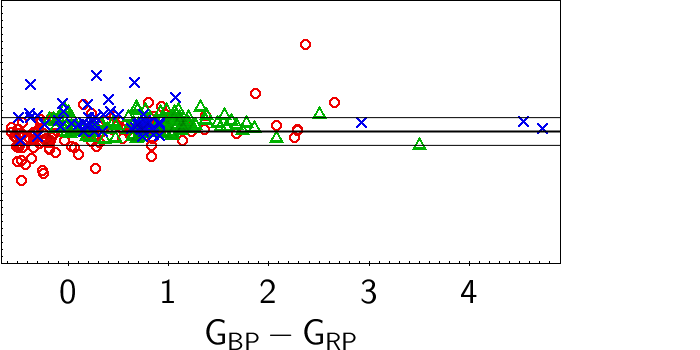}}
\caption{Comparison between synthetic photometry computed on external high-resolution spectra and that computed on ECS in the Johnson-Kron-Cousins photometric system. Data are from SPSS (open red circles), PVL (blue crosses), and NGSL (open green triangles) samples.
The comparison is shown as a function of G magnitude (\emph{left}) and colour index \bprp (\emph{right}). Two horizontal lines at $\pm0.02$ are shown for reference.}
\label{fig:land1}
\end{figure}

A first comparison is made by computing synthetic photometry in the JKC system  \citep{johnson53, johnson63, kron53, cousins73,cousins83,cousins84} for a set of nearly 32\,800 sources belonging to the Landolt collection of standard stars
\citep{Landolt92,landolt07a,landolt07b,landolt09,landolt13,clem13,clem16}; for a detailed description of this reference sample and the selection criteria applied to the original collection please refer to \citet{Pancino22arXiv} and references therein. The comparison is performed in the B, V, R, and I passbands only because U band does not fall entirely within the wavelength range covered by \gaia \xp spectra: passband response curves are taken from \citet{bessell}. 
Before comparing synthetic ECS photometry with Landolt data, it is useful to check the impact of systematic effects in ECS on the synthetic photometry itself. To do this, we compare the synthetic photometry computed on external reference SEDs for SPSS, PVL, and NGSL with the corresponding synthetic photometry computed on ECS (\figref{land1}). Residuals are shown as a function of G magnitude and colour index \bprp  on the left and right panels, respectively. As expected, the most relevant systematic differences are seen in the $G\lesssim11$ bright range (covered mostly by NGSL sources, plotted as green open triangles, and PVL data, plotted as blue crosses): synthetic ECS magnitudes are brighter than their HRS counterparts by nearly 0.02 mag in B, are fainter than expected by about 0.01 mag in V and R bands, and are again brighter by $\simeq 0.01$ mag in I band. 
Plots against colour index do not show any relevant colour trend, except for the blue part of the I band where SPSS data (red open circles) create a cluster of points with a mean residual around $\simeq -0.01$ mag (however, we note that SPSS data are more noisy in I band compared to other filters). 

\begin{figure}
\centerline{
\includegraphics[width=\columnwidth/2]{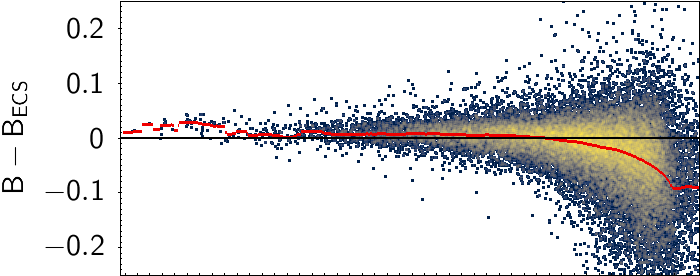}
\includegraphics[width=\columnwidth/2]{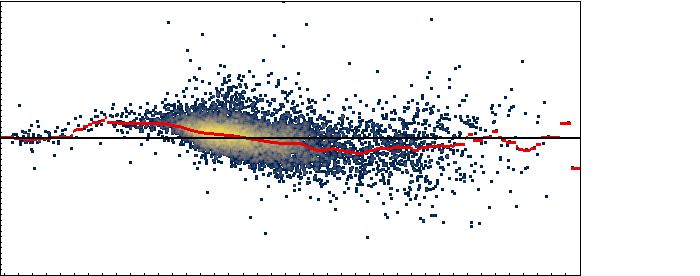}}
\centerline{
\includegraphics[width=\columnwidth/2]{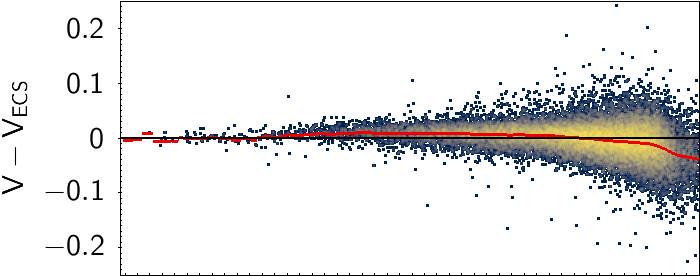}
\includegraphics[width=\columnwidth/2]{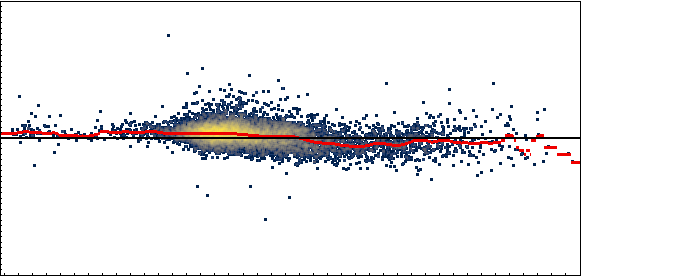}}
\centerline{
\includegraphics[width=\columnwidth/2]{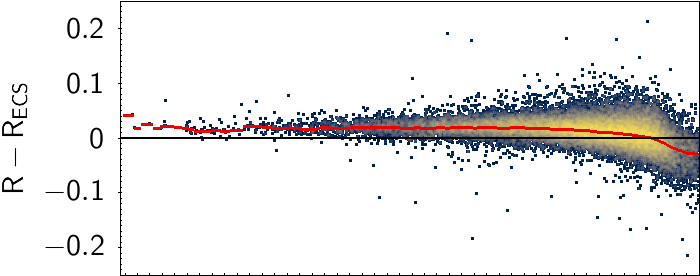}
\includegraphics[width=\columnwidth/2]{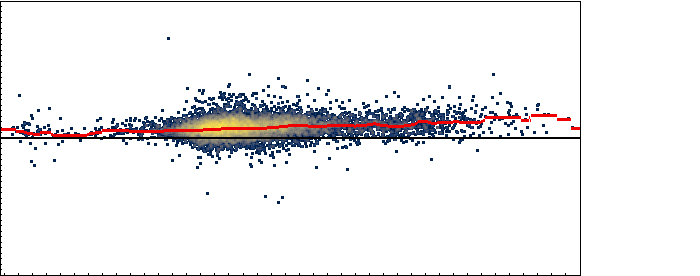}}
\centerline{
\includegraphics[width=\columnwidth/2]{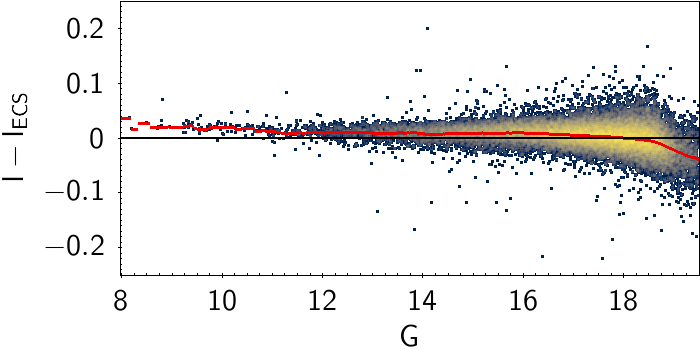}
\includegraphics[width=\columnwidth/2]{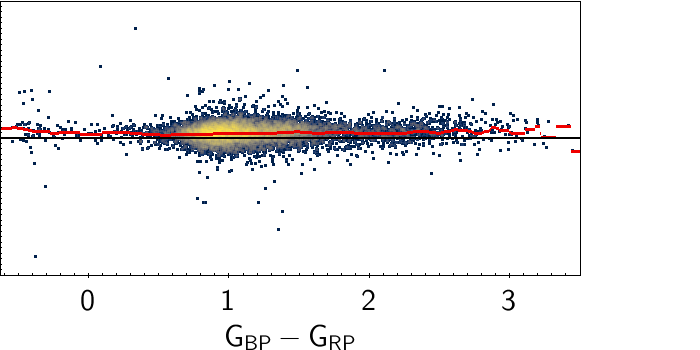}}
\caption{Comparison between Landolt standard BVRI magnitudes and synthetic BVRI photometry in the Johnson-Kron-Cousins system computed on externally calibrated \xp spectra for a sample of 32\,781 sources selected from the Landolt collection of standards. \emph{Left}: Residuals as a function of magnitude. The red lines represent the smoothed median distributions. \emph{Right}: Residuals  as a function of \bprp colour for sources with $G<17$.}
\label{fig:landolt}
\end{figure}

\afigref{landolt} shows residuals between the standard photometry and ECS synthetic photometry for the Landolt standards as a function of G magnitude (left panels) and \bprp colour index (right panels) for the B, V, R, and I bands of the JKC system.
Red curves represent the median ($P_{50}$) of the distributions. Residuals as a function of G magnitude reveal a magnitude-dependent term at the faint end (at $G\gtrsim16$): as discussed by \citet{Montegriffo2022}, this trend in residuals is interpreted as the result of a systematic overestimation of background in \gaia \xp spectra, which causes the actual \xp fluxes to be slightly underestimated. This effect was already revealed by \citet{Evans2018} by their comparison between \gdr 2 photometry and external data, where the \emph{hockey stick} term was first used to describe the observed feature, and by  \citet{Riello2021}, who compared \gband photometry and synthetic $G$ photometry computed on ECS (see their Fig. 23).
At magnitudes $G\lesssim11,$ residuals show offsets analogous to those seen in \figref{land1}. 
\begin{table}
\begin{center}
\caption{Offsets measured in the BVRI bands between Landolt standards and ECS synthetic photometry flux scale.}
\label{tab:offset}
\begin{tabular}{c|c|c|c|c}
\hline
 & B & V & R & I\\
\hline
$\Delta$ mag & 0.005 & 0.005 & 0.018 & 0.009\\
     stDev & 0.027 & 0.017 & 0.015 & 0.015\\
\hline
\end{tabular} 
\end{center}
\end{table}
However the overall accuracy estimated through the comparison with Landolt standards is excellent and ranges from $5$ mmag in B band to $18$ mmag in R band, while the standard deviations for sources with $G<17$ range from $27$ mmag for B to $15$ mmag for R band (values are reported in \tabref{offset}).
To minimise the noise due to the hockey stick effect and better trace the genuine colour terms, plots as a function of \bprp colour index refer only to sources with $G<17$.
These residuals reveal a significant colour term in the B filter causing differences of up to $0.06$ mag in the considered colour range. This colour term could be related to systematic effects impacting the blue part of ECS, considering that the B transmission curve extends to wavelengths $\lambda < 400$ nm. A smaller colour term is still visible in the V and R bands while residuals in I band are almost flat with differences within 0.01 mag.

\subsubsection{Hipparcos photometry}{\label{sec:hipparcos}}

The Hipparcos mission\footnote{\tt https://www.cosmos.esa.int/web/hipparcos/home}, in addition to state-of-the art astrometry for the time, provided space-based photometry in three wide bands for the $\simeq 10^5$ stars included in the catalogue \citep{Perryman1997,Hippaphot1997}. The Hipparcos $B_T$ and $V_T$ passbands are similar to JKC B and V, respectively, while the $H_p$ band ranges from $\simeq 340$~nm to $\simeq 900$~nm, peaking around $\simeq 450$~nm. Hipparcos photometry is generally recognised as a benchmark of excellent precision \citep[see e.g.][]{Bessel05}, and 95\% of the stars have uncertainties of $<0.01$~mag in $H_p$. The comparison with \gaia photometry is limited by the relatively bright magnitude limit of the Hipparcos catalogue, where the vast majority of stars have $G<11.0$. 
The original passband curves \citep{Hippaphot1997} were revised by \citet{bessell} 
who provide improved  transmissivity curves
that we adopt for the comparisons shown in this section.

\begin{figure}
\centerline{
\includegraphics[width=\columnwidth/2]{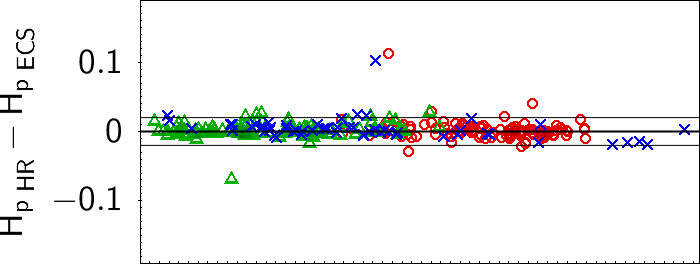}
\includegraphics[width=\columnwidth/2]{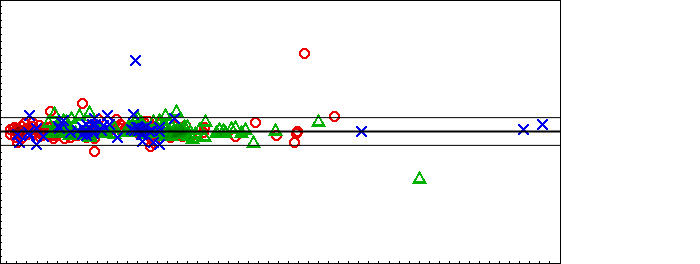}}
\centerline{
\includegraphics[width=\columnwidth/2]{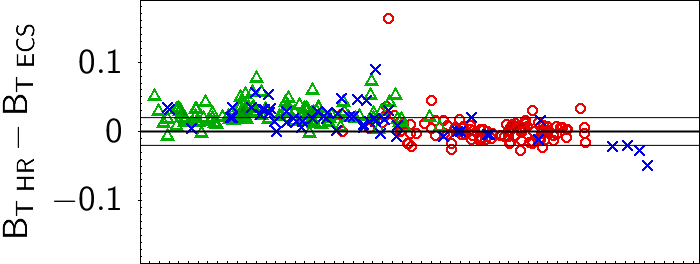}
\includegraphics[width=\columnwidth/2]{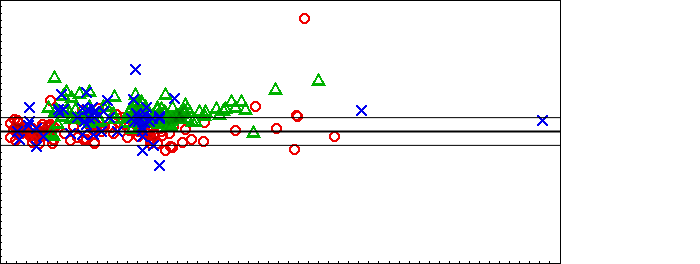}}
\centerline{
\includegraphics[width=\columnwidth/2]{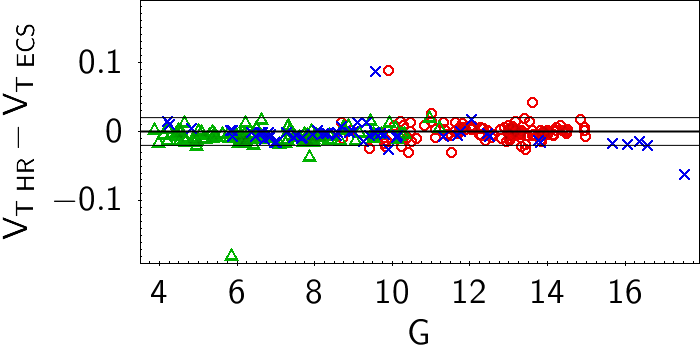}
\includegraphics[width=\columnwidth/2]{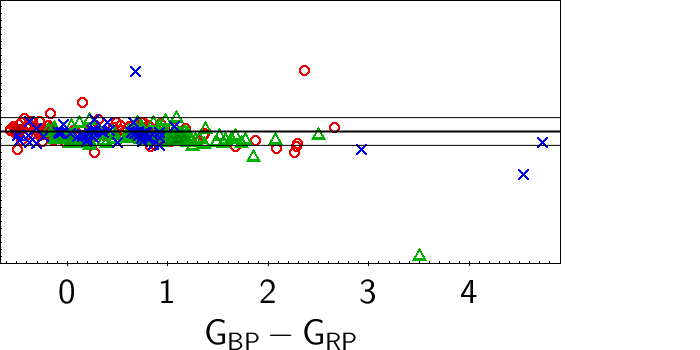}}
\caption{Comparison between synthetic photometry computed on external high-resolution spectra and on ECS in the Hipparcos $H_p$ and Tycho $B_T, V_T$ photometric system. SPSSs are represented as open red circles, PVLs as blue crosses, and NGSLs as open green triangles.
The difference between the two sets of magnitudes is shown as a function of G magnitude (\emph{left}) and colour index \bprp (\emph{right}). Two horizontal lines at $\pm0.02$ are shown for reference.}
\label{fig:hipp1}
\end{figure}

\afigref{hipp1} shows the comparison between HRS and ECS synthetic photometry computed on SPSS, PVL, and NGSL spectra. Symbols are the same as \figref{land1}.
As can be seen, thanks to its large width, the $H_p$ photometry shows excellent agreement between the two sets, with no magnitude or colour trends. On the other hand, the $B_T$ filter, covering the more problematic BP wavelength region at $\lambda<400$ nm, shows differences of up to 0.02 mag for sources brighter than $G\lesssim10.5$  ---the magnitude range covered by Hipparcos catalogue--- while the agreement is still excellent at fainter magnitudes. $V_T$ photometry shows better agreement, with synthetic ECS magnitudes being slightly fainter ($\sim0.01$ mag) than their HRS counterparts at the bright end. 

\begin{figure}
\centerline{
\includegraphics[width=\columnwidth/1]{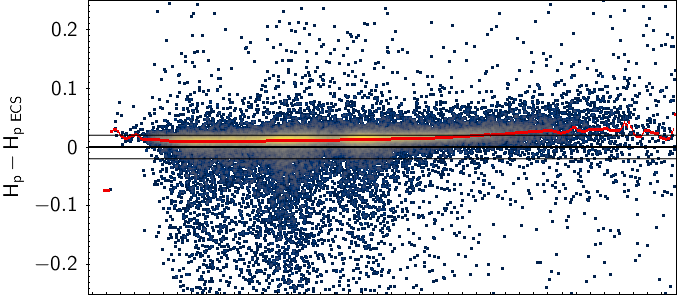}}
\centerline{
\includegraphics[width=\columnwidth/1]{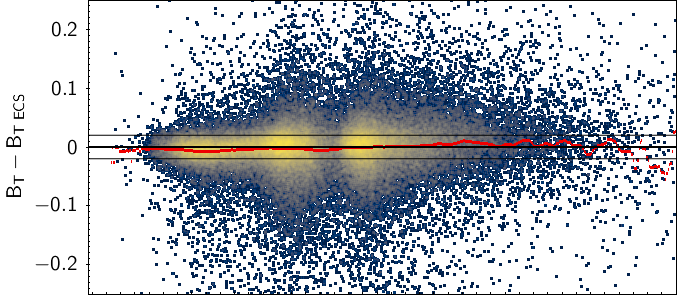}}
\centerline{
\includegraphics[width=\columnwidth/1]{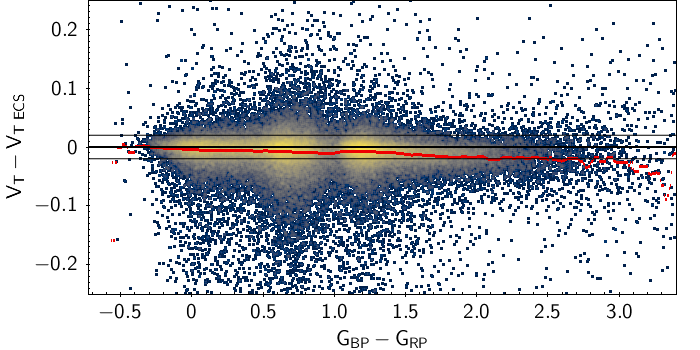}}
\caption{Residuals between Hipparcos/Tycho standard magnitudes and synthetic photometry from ECS computed assuming revised \citet{bessell} passbands. Red curves are the median distributions; horizontal lines at $\pm0.02$ are shown for reference.}
\label{fig:hippResid}
\end{figure}

The original Hipparcos catalogue, once cross-correlated with the \gdr3 catalogue, gives a list of 99525 matches that reduce to a list of 88662 sources when the selection criteria for published \xp spectra are applied \citep{Fouesneau_dr3,DeAngeli2022}. 
According to \citet{Hippaphot1997}, the magnitude scales for Hipparcos passbands  were chosen such that $H_p = V_T = V$ and $B_T = B$  at $B-V = 0$, and therefore in order to compare these magnitude scales to synthetic ECS magnitude, we need to adjust our zero points accordingly, with the following relations (see details in \appref{hipApp}):
\begin{eqnarray}
H_p = H_{p\ ECS}+0.022 \\
B_T = B_{T\ ECS}+0.054 \\
V_T = V_{T\ ECS}+0.011.
\end{eqnarray}
Residuals between standard Hipparcos/Tycho magnitudes and synthetic ECS magnitudes computed through Bessell passbands are shown in \figref{hippResid} as a function of \bprp colour index. The $H_p$ panel (\emph{top}) reveals a small colour term spanning an excursion of less than $2$~mmag along the whole colour range. At $\bprp = 0.0,$ the median distribution (red curve) shows an offset of roughly $10$~mmag. Due to the rather large width of the $H_p$ band, the collected flux is larger than that from the other two bands, resulting in a very narrow sequence of residuals with a standard deviation of $\sim12$~mmag. 
 $B_T$ band residuals are almost flat along the whole colour range, with a standard deviation of $\sim34$~mmag.
Good agreement is also found in $V_T$ band, with a 
small colour term spanning an excursion similar to that of $H_p$ band but opposite in sign;
the standard deviation of the  residuals is $\sim26$~mmag.
\begin{figure}[]
    \centerline{
    \includegraphics[width=(\columnwidth)/1]{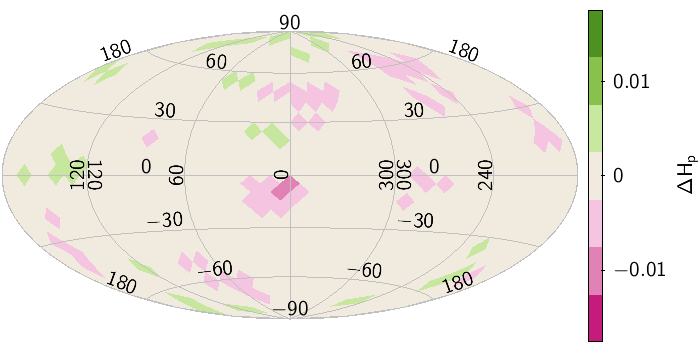}}
    \centerline{
    \includegraphics[width=(\columnwidth)/1]{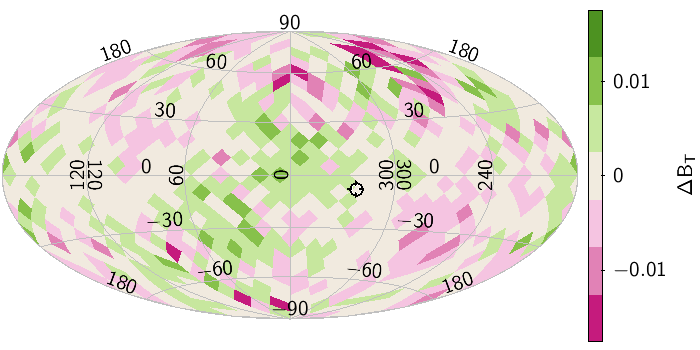}}
    \centerline{
    \includegraphics[width=(\columnwidth)/1]{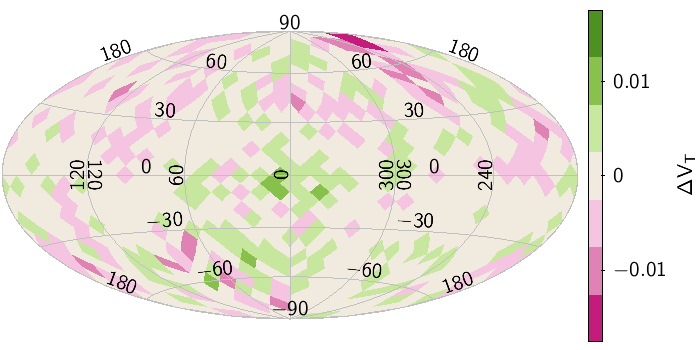}}
\caption{HEALPiX level 3 maps in Galactic coordinates (Hammer Aitoff projection) of the median residuals between standard Hipparcos/Tycho magnitudes and synthetic magnitudes from ECS, after correction for the small colour trends shown in Fig.~\ref{fig:hippResid}. To avoid spurious signals, we limited the comparison to stars with $-0.5<\bprp<3.0$, $|\Delta {\rm mag}|<0.2$, and $|C^{\star}|<0.05$.
The $\simeq 80000$ sources involved are not evenly distributed but cover the entire sky at this resolution: beige pixels correspond to absolute median differences $\lesssim 2.5$~mmag and not to empty pixels.}
    \label{fig:skymap}
\end{figure}
The comparison with Hipparcos photometry also provides a unique opportunity to test the spatial homogeneity of \xp spectrophotometry over the entire sky with space-based data. In order to show the residuals between the Hipparcos and \gaia synthetic magnitude scales, excluding colour systematic effects, we fitted each median distribution of \figref{hippResid} with a fourth degree polynomial function in \bprp colour and subtracted this term from $\Delta  {\rm mag}$: \afigref{skymap} shows the HEALPiX level 3 maps in Galactic coordinates (Hammer Aitoff projections) of residuals in the three bands, limiting the representation to sources with $|C^{\star}|<0.05$ \footnote{See \citet{Riello2021} for $C^{\star}$ definition}, 
$-0.5<\bprp<3.0$ and $|\Delta {\rm mag}|<0.2$ to avoid spurious signals due to outliers. The maps show that the synthetic photometry from \gaia \xp spectra matches the original Hipparcos photometry to within $\pm 5.0$~mmag in the vast majority of cases, in all three Hipparcos bands. Fluctuations of amplitude $\gtrsim 10.0$~mmag are quite rare. It is  interesting to compare these maps with the sky density of objects with BP/RP spectral data represented in the top left panel of Figure 28 of \citet{DeAngeli2022}: sky regions with higher residuals in the $B_T$ map correspond to regions with a lower number of BP/RP observations. 


\section{Known problems and caveats}\label{sec:caveats}

\subsection{Saturation}\label{sec:saturation}

Saturation can affect observations of bright sources. Even though an ad hoc gating scheme has been designed and configured to reduce the effective exposure time and avoid saturation, large uncertainties in the on-board magnitude estimates at the bright end imply that gate activation is not always optimal. Moreover, the saturation level is different between different CCDs and dispersion differences across the focal plane will also play a role. A bright source can therefore have a mixture of saturated and non-saturated BP/RP observed spectra. The size of the effect on combined mean spectra will depend on the fraction of saturated versus non-saturated epoch spectra and on the size of the effects on single observations.
\begin{figure}[]
    \centerline{
    \includegraphics[width=(\textwidth)/4]{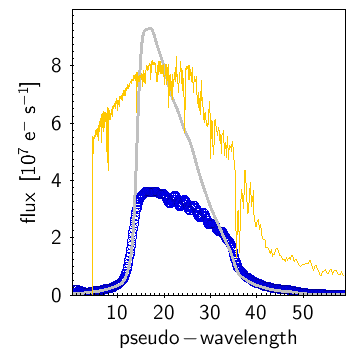}
    \includegraphics[width=(\textwidth)/4]{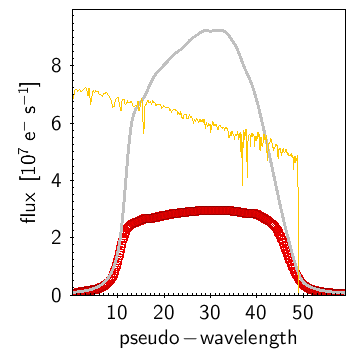}}
    \centerline{
    \includegraphics[width=(\textwidth)/4]{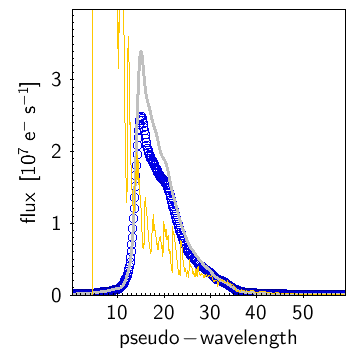}
    \includegraphics[width=(\textwidth)/4]{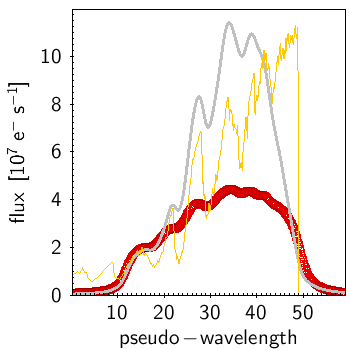}}
    \caption{Examples of heavily saturated spectra taken from the NGSL set: sources \texttt{Gaia\,DR3\,2067518817314952576} (\emph{top}) with $G=2.11$ and $\bprp = 0.99$ and  \texttt{Gaia\,DR3\,2106630885454013184} (\emph{bottom}) with $G=2.36$ and $\bprp = 2.42$. Colour coding and symbols are the same as in \figref{samplSpecComp1}.}
    \label{fig:satExamples}
\end{figure}
As a consequence, the shape of a saturated spectrum may not always be recognisable as a spectrum with a cropped, sharp, flat section at the top of it, corresponding to  some fixed flux threshold, but it can 
exhibit a very complicated behaviour,
as can be seen in \figref{satExamples} where a couple of examples are shown.

\begin{figure*}[]
    \centerline{
    \includegraphics[width=(\textwidth)/2]{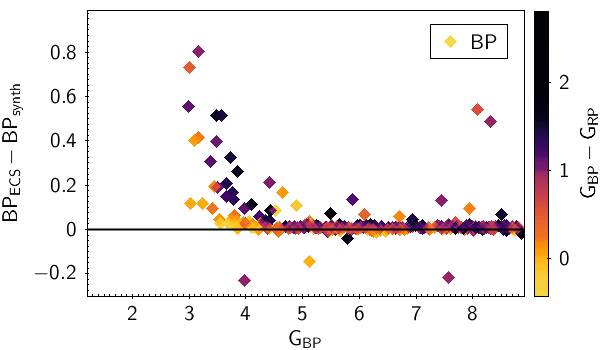}
    \includegraphics[width=(\textwidth)/2]{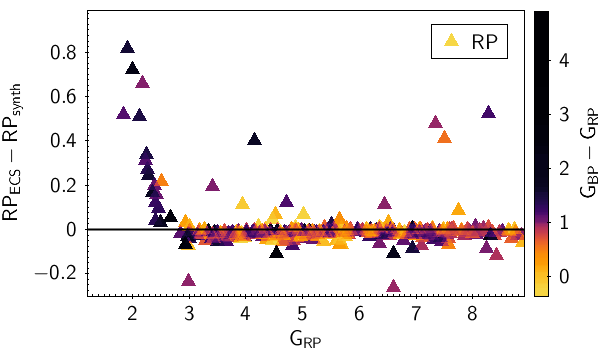}}
    \centerline{
    \includegraphics[width=(\textwidth)/2]{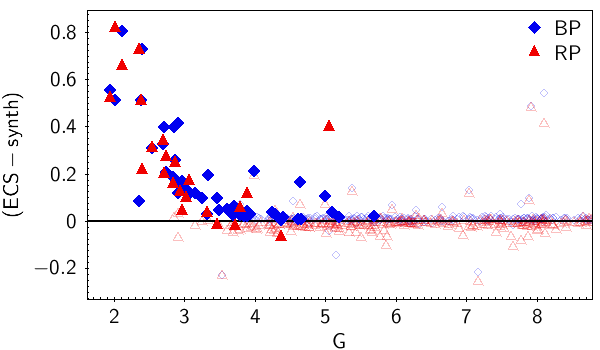}
    \includegraphics[width=(\textwidth)/2]{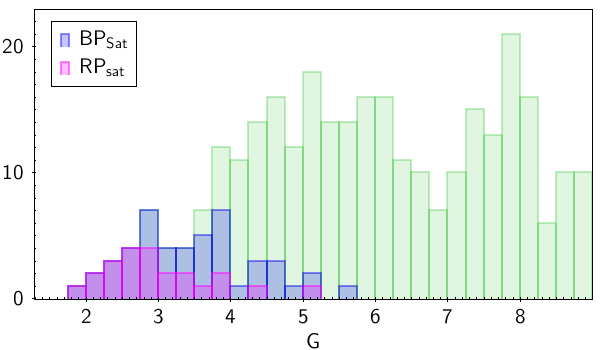}}
    \caption{Saturation in \xp spectra.
    \emph{Top}: Difference between synthetic magnitudes computed on NGSL SEDs and the corresponding synthetic magnitude computed on \gaia ECS, used as an indicator of saturation in \gaia spectra. Differences are computed for the BP band (\emph{left}) and RP (\emph{right}). \emph{Bottom}: Delta synthetic magnitudes measured on both BP and RP spectra against the G magnitude of the source (\emph{left}); filled symbols represent spectra with clear signs of saturation (selected manually) in their spectra shape. 
    Right panel: distributions of BP and RP saturated spectra against the distribution for unsaturated sources.}
    \label{fig:saturation}
\end{figure*}
The NGSL sample (containing very bright sources up to $G=1.97$) represents an ideal dataset for assessing the effect of saturation in \gaia spectra. We use synthetic photometry to measure the saturation level present in spectra by comparing the synthetic flux in \xp measured on the \gaia ECS with the synthetic flux obtained on NGSL SEDs. The difference between the corresponding synthetic magnitudes is represented in \figref{saturation} as a function of the source magnitude in the same filter: results for BP and RP are shown respectively in left and right top panels. 
Saturation in BP spectra starts around $\rm BP\simeq4.5$ depending on the colour, with redder sources being increasingly affected. This behaviour is due to the properties of the BP dispersion relation:
in the case of a red source as in the example shown in \figref{satExamples} (lower panel), most of the source flux is concentrated in a few pixels. These are the ones that will saturate in the case of a sufficiently bright source. At the same flux level, a bluer/hotter source has less chance of saturating because photons are distributed over a wider range of samples.
In the case of the RP, saturation is found at magnitudes $\rm RP\lesssim 3$ and affects mostly red sources. 
In the bottom left panel of \figref{saturation}, data for both BP and RP are plotted against the $G$ magnitude. Sources showing signs of saturation in their BP or RP spectra (differences in shape between observed ICS and model prediction) were selected manually and are visualised in the plot as filled symbols while well behaved sources are represented as open transparent ones. 
The bottom right panel of \figref{saturation} shows the distribution with G magnitude of all NGSL objects and those that show saturation in BP or RP: from the histogram, we can conclude that saturation is not significant at $G>5$ , and that most sources will have saturated BP spectra  at $G\lesssim4,$  and it is only  $G\lesssim3.5$ that saturation will also occur in most RP spectra.
A final word of caution should be added here to say that given the very small number of sources brighter than the thresholds indicated in this section and the fact that \gaia was not designed to cover such bright magnitudes, further investigation is needed to completely understand the origin of the discrepancies observed in \figref{saturation}. Other effects such as CCD non-linearities and larger uncertainties in the source astrometry could play a role.

\subsection{Wiggles in ECS errors}\label{sec:errorsWiggles}

\begin{figure}[]
    \centerline{
    \includegraphics[width=(\columnwidth)-5mm]{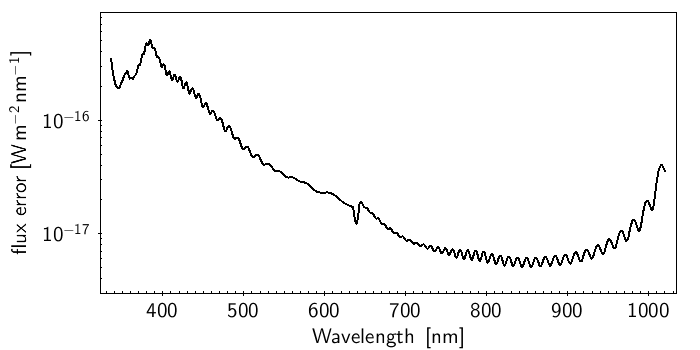}}
\caption{Flux error distribution for the externally calibrated spectrum of source \texttt{Gaia\,DR3\,1435896975388228224}. Errors are characterised by the presence of wiggles.}
    \label{fig:errorWiggles}
\end{figure}
The ECS flux error distribution against wavelength shows wiggles, as seen in \figref{errorWiggles} for source \texttt{Gaia\,DR3\,1435896975388228224}. This is visible for all sources. Considering that the flux error is computed as the squared root of the diagonal elements of the covariance matrix on the ECS (see \equref{ecsCovMatrix}) and that the flux error distribution of BP/RP ICS does not show such wiggles, we deduce that the effect is a numerical problem introduced by the inversion process and may be due to poor conditioning of the inverse basis design matrices.
\par~\par
~

\section{Summary and conclusions}

This paper describes the process leading to the external calibration of \gaia \xp low-resolution spectra. 
We derived an instrument model for the \xp spectrophotometers  
that allows us to forward-model observed \gaia mean spectra starting from empirical or theoretical SEDs. 
The model has been optimised using a large set of primary flux calibrators (SPSS) and secondary calibrators (PVL and emission lines sources). 

Through inverse modelling, we generated a set of inverse bases allowing the reconstruction of externally calibrated low-resolution SEDs for the 220 million sources with published BP/RP spectra within \gdr 3. These reconstructed SEDs are flux- and wavelength calibrated and have an increased resolution \wrt the original mean spectra 
due to mitigation of the LSF smearing effect. Errors associated to calibrated spectra appear to be underestimated, 
especially for sources brighter than magnitude $G\simeq12$. 

Comparison with external data reveals the presence of systematic effects in the wavelength region below $\lambda<400$ nm. These systematic effects are a function of the colour of the sources and, to a lower degree, their magnitude. Furthermore, smaller magnitude-dependent systematic effects impacting sources brighter than $G\simeq12$ are present in the wavelength range $560-600$ nm and in the red part of the spectra at $\lambda \simeq 950$ nm. Externally calibrated spectra are occasionally affected by the presence of numerical artefacts in the form of ripples 
produced by the basis inversion process. 

Calibrated spectra were also validated by comparing  synthetic photometry in the Johnson-Kron-Cousins system with Landolt standard sources 
and in the Hipparcos $H_p$ and Tycho-2 $B_T,V_T$ passbands with the Hipparcos catalogue. In both cases, we find excellent agreement overall, with typical accuracy ranging from $5$ to $20$ mmag and precision between $15$ and $\sim30$~mmag, as measured by the standard deviation of the residuals over wide colour and magnitude ranges. 
The comparison with Hipparcos photometry demonstrates the excellent degree of spatial homogeneity of the synthetic photometry with a relative accuracy of less than $2.7$ mmag over $90\%$ of the entire sky.
Additional validation by comparison with a large data set of reliable external photometry is provided in \citet{Montegriffo2022}.
Finally, caution must be exercised when using the spectra of very bright sources because ofv the presence of saturation, 
especially for sources brighter than $G\simeq4$.


\begin{acknowledgements}
We are very grateful to an anonymous Referee for a prompt and constructive report, that improved the quality of the manuscript. We would also like to thank C. Babusiaux for kindly reviewing an earlier version of this manuscript.

This work presents results from the European Space Agency (ESA) space mission \gaia. \gaia\ data are being processed by the \gaia\ Data Processing and Analysis Consortium (DPAC). Funding for the DPAC is provided by national institutions, in particular the institutions participating in the \gaia\ MultiLateral Agreement (MLA). The \gaia\ mission website is \url{https://www.cosmos.esa.int/gaia}. The \gaia\ archive website is \url{https://archives.esac.esa.int/gaia}. 

Acknowledgments are given in Appendix \ref{sec:app_ack}.
\end{acknowledgements}


\bibliographystyle{aa} 
\bibliography{refs,dpac} 


\begin{appendix} 

\section{Acknowledgements}
\label{sec:app_ack}
The \gaia\ mission and data processing have financially been supported by, in alphabetical order by country:

\begin{itemize}
\item the Algerian Centre de Recherche en Astronomie, Astrophysique et G\'{e}ophysique of Bouzareah Observatory;
\item the Austrian Fonds zur F\"{o}rderung der wissenschaftlichen Forschung (FWF) Hertha Firnberg Programme through grants T359, P20046, and P23737;
\item the BELgian federal Science Policy Office (BELSPO) through various PROgramme de D\'{e}veloppement d'Exp\'{e}riences scientifiques (PRODEX) grants and the Polish Academy of Sciences - Fonds Wetenschappelijk Onderzoek through grant VS.091.16N, and the Fonds de la Recherche Scientifique (FNRS), and the Research Council of Katholieke Universiteit (KU) Leuven through grant C16/18/005 (Pushing AsteRoseismology to the next level with TESS, GaiA, and the Sloan DIgital Sky SurvEy -- PARADISE);  
\item the Brazil-France exchange programmes Funda\c{c}\~{a}o de Amparo \`{a} Pesquisa do Estado de S\~{a}o Paulo (FAPESP) and Coordena\c{c}\~{a}o de Aperfeicoamento de Pessoal de N\'{\i}vel Superior (CAPES) - Comit\'{e} Fran\c{c}ais d'Evaluation de la Coop\'{e}ration Universitaire et Scientifique avec le Br\'{e}sil (COFECUB);
\item the Chilean Agencia Nacional de Investigaci\'{o}n y Desarrollo (ANID) through Fondo Nacional de Desarrollo Cient\'{\i}fico y Tecnol\'{o}gico (FONDECYT) Regular Project 1210992 (L.~Chemin);
\item the National Natural Science Foundation of China (NSFC) through grants 11573054, 11703065, and 12173069, the China Scholarship Council through grant 201806040200, and the Natural Science Foundation of Shanghai through grant 21ZR1474100;  
\item the Tenure Track Pilot Programme of the Croatian Science Foundation and the \'{E}cole Polytechnique F\'{e}d\'{e}rale de Lausanne and the project TTP-2018-07-1171 `Mining the Variable Sky', with the funds of the Croatian-Swiss Research Programme;
\item the Czech-Republic Ministry of Education, Youth, and Sports through grant LG 15010 and INTER-EXCELLENCE grant LTAUSA18093, and the Czech Space Office through ESA PECS contract 98058;
\item the Danish Ministry of Science;
\item the Estonian Ministry of Education and Research through grant IUT40-1;
\item the European Commission’s Sixth Framework Programme through the European Leadership in Space Astrometry (\href{https://www.cosmos.esa.int/web/gaia/elsa-rtn-programme}{ELSA}) Marie Curie Research Training Network (MRTN-CT-2006-033481), through Marie Curie project PIOF-GA-2009-255267 (Space AsteroSeismology \& RR Lyrae stars, SAS-RRL), and through a Marie Curie Transfer-of-Knowledge (ToK) fellowship (MTKD-CT-2004-014188); the European Commission's Seventh Framework Programme through grant FP7-606740 (FP7-SPACE-2013-1) for the \gaia\ European Network for Improved data User Services (\href{https://gaia.ub.edu/twiki/do/view/GENIUS/}{GENIUS}) and through grant 264895 for the \gaia\ Research for European Astronomy Training (\href{https://www.cosmos.esa.int/web/gaia/great-programme}{GREAT-ITN}) network;
\item the European Cooperation in Science and Technology (COST) through COST Action CA18104 `Revealing the Milky Way with \gaia (MW-Gaia)';
\item the European Research Council (ERC) through grants 320360, 647208, and 834148 and through the European Union’s Horizon 2020 research and innovation and excellent science programmes through Marie Sk{\l}odowska-Curie grant 745617 (Our Galaxy at full HD -- Gal-HD) and 895174 (The build-up and fate of self-gravitating systems in the Universe) as well as grants 687378 (Small Bodies: Near and Far), 682115 (Using the Magellanic Clouds to Understand the Interaction of Galaxies), 695099 (A sub-percent distance scale from binaries and Cepheids -- CepBin), 716155 (Structured ACCREtion Disks -- SACCRED), 951549 (Sub-percent calibration of the extragalactic distance scale in the era of big surveys -- UniverScale), and 101004214 (Innovative Scientific Data Exploration and Exploitation Applications for Space Sciences -- EXPLORE);
\item the European Science Foundation (ESF), in the framework of the \gaia\ Research for European Astronomy Training Research Network Programme (\href{https://www.cosmos.esa.int/web/gaia/great-programme}{GREAT-ESF});
\item the European Space Agency (ESA) in the framework of the \gaia\ project, through the Plan for European Cooperating States (PECS) programme through contracts C98090 and 4000106398/12/NL/KML for Hungary, through contract 4000115263/15/NL/IB for Germany, and through PROgramme de D\'{e}veloppement d'Exp\'{e}riences scientifiques (PRODEX) grant 4000127986 for Slovenia;  
\item the Academy of Finland through grants 299543, 307157, 325805, 328654, 336546, and 345115 and the Magnus Ehrnrooth Foundation;
\item the French Centre National d’\'{E}tudes Spatiales (CNES), the Agence Nationale de la Recherche (ANR) through grant ANR-10-IDEX-0001-02 for the `Investissements d'avenir' programme, through grant ANR-15-CE31-0007 for project `Modelling the Milky Way in the \gaia era’ (MOD4Gaia), through grant ANR-14-CE33-0014-01 for project `The Milky Way disc formation in the \gaia era’ (ARCHEOGAL), through grant ANR-15-CE31-0012-01 for project `Unlocking the potential of Cepheids as primary distance calibrators’ (UnlockCepheids), through grant ANR-19-CE31-0017 for project `Secular evolution of galaxies' (SEGAL), and through grant ANR-18-CE31-0006 for project `Galactic Dark Matter' (GaDaMa), the Centre National de la Recherche Scientifique (CNRS) and its SNO \gaia of the Institut des Sciences de l’Univers (INSU), its Programmes Nationaux: Cosmologie et Galaxies (PNCG), Gravitation R\'{e}f\'{e}rences Astronomie M\'{e}trologie (PNGRAM), Plan\'{e}tologie (PNP), Physique et Chimie du Milieu Interstellaire (PCMI), and Physique Stellaire (PNPS), the `Action F\'{e}d\'{e}ratrice \gaia' of the Observatoire de Paris, the R\'{e}gion de Franche-Comt\'{e}, the Institut National Polytechnique (INP) and the Institut National de Physique nucl\'{e}aire et de Physique des Particules (IN2P3) co-funded by CNES;
\item the German Aerospace Agency (Deutsches Zentrum f\"{u}r Luft- und Raumfahrt e.V., DLR) through grants 50QG0501, 50QG0601, 50QG0602, 50QG0701, 50QG0901, 50QG1001, 50QG1101, 50\-QG1401, 50QG1402, 50QG1403, 50QG1404, 50QG1904, 50QG2101, 50QG2102, and 50QG2202, and the Centre for Information Services and High Performance Computing (ZIH) at the Technische Universit\"{a}t Dresden for generous allocations of computer time;
\item the Hungarian Academy of Sciences through the Lend\"{u}let Programme grants LP2014-17 and LP2018-7 and the Hungarian National Research, Development, and Innovation Office (NKFIH) through grant KKP-137523 (`SeismoLab');
\item the Science Foundation Ireland (SFI) through a Royal Society - SFI University Research Fellowship (M.~Fraser);
\item the Israel Ministry of Science and Technology through grant 3-18143 and the Tel Aviv University Center for Artificial Intelligence and Data Science (TAD) through a grant;
\item the Agenzia Spaziale Italiana (ASI) through contracts I/037/08/0, I/058/10/0, 2014-025-R.0, 2014-025-R.1.2015, and 2018-24-HH.0 to the Italian Istituto Nazionale di Astrofisica (INAF), contract 2014-049-R.0/1/2 to INAF for the Space Science Data Centre (SSDC, formerly known as the ASI Science Data Center, ASDC), contracts I/008/10/0, 2013/030/I.0, 2013-030-I.0.1-2015, and 2016-17-I.0 to the Aerospace Logistics Technology Engineering Company (ALTEC S.p.A.), INAF, and the Italian Ministry of Education, University, and Research (Ministero dell'Istruzione, dell'Universit\`{a} e della Ricerca) through the Premiale project `MIning The Cosmos Big Data and Innovative Italian Technology for Frontier Astrophysics and Cosmology' (MITiC);
\item the Netherlands Organisation for Scientific Research (NWO) through grant NWO-M-614.061.414, through a VICI grant (A.~Helmi), and through a Spinoza prize (A.~Helmi), and the Netherlands Research School for Astronomy (NOVA);
\item the Polish National Science Centre through HARMONIA grant 2018/30/M/ST9/00311 and DAINA grant 2017/27/L/ST9/03221 and the Ministry of Science and Higher Education (MNiSW) through grant DIR/WK/2018/12;
\item the Portuguese Funda\c{c}\~{a}o para a Ci\^{e}ncia e a Tecnologia (FCT) through national funds, grants SFRH/\-BD/128840/2017 and PTDC/FIS-AST/30389/2017, and work contract DL 57/2016/CP1364/CT0006, the Fundo Europeu de Desenvolvimento Regional (FEDER) through grant POCI-01-0145-FEDER-030389 and its Programa Operacional Competitividade e Internacionaliza\c{c}\~{a}o (COMPETE2020) through grants UIDB/04434/2020 and UIDP/04434/2020, and the Strategic Programme UIDB/\-00099/2020 for the Centro de Astrof\'{\i}sica e Gravita\c{c}\~{a}o (CENTRA);  
\item the Slovenian Research Agency through grant P1-0188;
\item the Spanish Ministry of Economy (MINECO/FEDER, UE), the Spanish Ministry of Science and Innovation (MICIN), the Spanish Ministry of Education, Culture, and Sports, and the Spanish Government through grants BES-2016-078499, BES-2017-083126, BES-C-2017-0085, ESP2016-80079-C2-1-R, ESP2016-80079-C2-2-R, FPU16/03827, PDC2021-121059-C22, RTI2018-095076-B-C22, and TIN2015-65316-P (`Computaci\'{o}n de Altas Prestaciones VII'), the Juan de la Cierva Incorporaci\'{o}n Programme (FJCI-2015-2671 and IJC2019-04862-I for F.~Anders), the Severo Ochoa Centre of Excellence Programme (SEV2015-0493), and MICIN/AEI/10.13039/501100011033 (and the European Union through European Regional Development Fund `A way of making Europe') through grant RTI2018-095076-B-C21, the Institute of Cosmos Sciences University of Barcelona (ICCUB, Unidad de Excelencia `Mar\'{\i}a de Maeztu’) through grant CEX2019-000918-M, the University of Barcelona's official doctoral programme for the development of an R+D+i project through an Ajuts de Personal Investigador en Formaci\'{o} (APIF) grant, the Spanish Virtual Observatory through project AyA2017-84089, the Galician Regional Government, Xunta de Galicia, through grants ED431B-2021/36, ED481A-2019/155, and ED481A-2021/296, the Centro de Investigaci\'{o}n en Tecnolog\'{\i}as de la Informaci\'{o}n y las Comunicaciones (CITIC), funded by the Xunta de Galicia and the European Union (European Regional Development Fund -- Galicia 2014-2020 Programme), through grant ED431G-2019/01, the Red Espa\~{n}ola de Supercomputaci\'{o}n (RES) computer resources at MareNostrum, the Barcelona Supercomputing Centre - Centro Nacional de Supercomputaci\'{o}n (BSC-CNS) through activities AECT-2017-2-0002, AECT-2017-3-0006, AECT-2018-1-0017, AECT-2018-2-0013, AECT-2018-3-0011, AECT-2019-1-0010, AECT-2019-2-0014, AECT-2019-3-0003, AECT-2020-1-0004, and DATA-2020-1-0010, the Departament d'Innovaci\'{o}, Universitats i Empresa de la Generalitat de Catalunya through grant 2014-SGR-1051 for project `Models de Programaci\'{o} i Entorns d'Execuci\'{o} Parallels' (MPEXPAR), and Ramon y Cajal Fellowship RYC2018-025968-I funded by MICIN/AEI/10.13039/501100011033 and the European Science Foundation (`Investing in your future');
\item the Swedish National Space Agency (SNSA/Rymdstyrelsen);
\item the Swiss State Secretariat for Education, Research, and Innovation through the Swiss Activit\'{e}s Nationales Compl\'{e}mentaires and the Swiss National Science Foundation through an Eccellenza Professorial Fellowship (award PCEFP2\_194638 for R.~Anderson);
\item the United Kingdom Particle Physics and Astronomy Research Council (PPARC), the United Kingdom Science and Technology Facilities Council (STFC), and the United Kingdom Space Agency (UKSA) through the following grants to the University of Bristol, the University of Cambridge, the University of Edinburgh, the University of Leicester, the Mullard Space Sciences Laboratory of University College London, and the United Kingdom Rutherford Appleton Laboratory (RAL): PP/D006511/1, PP/D006546/1, PP/D006570/1, ST/I000852/1, ST/J005045/1, ST/K00056X/1, ST/\-K000209/1, ST/K000756/1, ST/L006561/1, ST/N000595/1, ST/N000641/1, ST/N000978/1, ST/\-N001117/1, ST/S000089/1, ST/S000976/1, ST/S000984/1, ST/S001123/1, ST/S001948/1, ST/\-S001980/1, ST/S002103/1, ST/V000969/1, ST/W002469/1, ST/W002493/1, ST/W002671/1, ST/W002809/1, and EP/V520342/1.
\end{itemize}

The \gaia\ project and data processing have made use of:
\begin{itemize}
\item the Set of Identifications, Measurements, and Bibliography for Astronomical Data \citep[SIMBAD,][]{2000AAS..143....9W}, the `Aladin sky atlas' \citep{2000A&AS..143...33B,2014ASPC..485..277B}, and the VizieR catalogue access tool \citep{2000A&AS..143...23O}, all operated at the Centre de Donn\'{e}es astronomiques de Strasbourg (\href{http://cds.u-strasbg.fr/}{CDS});
\item the National Aeronautics and Space Administration (NASA) Astrophysics Data System (\href{http://adsabs.harvard.edu/abstract_service.html}{ADS});

\item the SPace ENVironment Information System (SPENVIS), initiated by the Space Environment and Effects Section (TEC-EES) of ESA and developed by the Belgian Institute for Space Aeronomy (BIRA-IASB) under ESA contract through ESA’s General Support Technologies Programme (GSTP), administered by the BELgian federal Science Policy Office (BELSPO);
\item the Spanish Virtual Observatory (https://svo.cab.inta-csic.es) project funded by MCIN/AEI/10.13039/501100011033/ through grant PID2020-112949GB-I00 \citep{SVO};
\item the software products \href{http://www.starlink.ac.uk/topcat/}{TOPCAT}, \href{http://www.starlink.ac.uk/stil}{STIL}, and \href{http://www.starlink.ac.uk/stilts}{STILTS} \citep{2005ASPC..347...29T,2006ASPC..351..666T};
\item MATLAB \citep{MATLAB:2018};
\item Matplotlib \citep{Hunter:2007};
\item IPython \citep{PER-GRA:2007};  
\item Astropy, a community-developed core Python package for Astronomy \citep{2018AJ....156..123A};
\item  NumPy \citep{harris2020array};
\item pyphot
(http://github.com/mfouesneau/pyphot);
\item the \hip-2 catalogue \citep{2007A&A...474..653V}. The \hip and \tyc catalogues were constructed under the responsibility of large scientific teams collaborating with ESA. The Consortia Leaders were Lennart Lindegren (Lund, Sweden: NDAC) and Jean Kovalevsky (Grasse, France: FAST), together responsible for the \hip Catalogue; Erik H{\o}g (Copenhagen, Denmark: TDAC) responsible for the \tyc Catalogue; and Catherine Turon (Meudon, France: INCA) responsible for the \hip Input Catalogue (HIC);  
\item the \tyctwo catalogue \citep{2000A&A...355L..27H}, the construction of which was supported by the Velux Foundation of 1981 and the Danish Space Board;
\item The Tycho double star catalogue \citep[TDSC,][]{2002A&A...384..180F}, based on observations made with the ESA \hip astrometry satellite, as supported by the Danish Space Board and the United States Naval Observatory through their double-star programme;
\item the VizieR catalogue access tool (CDS, Strasbourg, France).
\end{itemize}

This publication made extensive use of the online authoring Overleaf platform (\url{https://www.overleaf.com/}).\\

\section{Data products}\label{sec:dataProducts}

The \xp data are made available via the \gaia archive\footnote{https://gea.esac.esa.int/archive/}.
\xp mean spectra are published for 219,197,643 \textbf{sources}.
This list includes mostly sources with \gband magnitude brighter than 17.65~mag and more than 15 CCD transits contributing to the generation of the mean spectra for both \xp. Other selection criteria were also applied; see \cite{DeAngeli2022} for more details.

All \xp spectra are provided in their continuous representation. See Sect. 4 and Appendices A and B in \cite{DeAngeli2022} for a description of the data format and for download instructions.
For a subset of sources with \xp spectral data including only sources brighter than $G=15$~mag (the exact list can be obtained by selecting entries with \texttt{gaia\_source.has\_xp\_sampled='t'}) \xp spectra are also provided in the sampled representation on a default grid as externally calibrated spectra in absolute flux and wavelength systems. These can be retrieved via DataLink from the archive interface or programmatically selecting the sampled type (\texttt{XP mean sampled spectra} on the web interface and \texttt{retrieval\_type=’XP\_SAMPLED’} when using \texttt{astroquery} from Python).

Externally calibrated sampled spectra for all sources, optionally on a user-defined wavelength grid, can be obtained via \gaiaxpy \citep{DeAngeli2022b}, a Python package developed to help users of the \xp spectral data.

Inverse bases computed for a default wavelength sampling (XpMerge/XpSampling tables) are distributed in a dedicated Cosmos page:
\url{https://www.cosmos.esa.int/web/gaia/web/gaia/dr3-xpmergexpsampling}
\gaiaxpy provides the possibility to sample ECS on a user-defined wavelength grid.

The IM sampled on a pre-defined wavelength/pseudo-wavelength grid is available on a dedicated Cosmos page:
\url{https://www.cosmos.esa.int/web/gaia/dr3-bprp-instrument-model}: on the same page users can retrieve the tabulated dispersion functions and the response for BP and RP instruments.
\gaiaxpy provides the possibility to forward model mean BP and RP sampled spectra via the IM on a user-defined pseudo-wavelength grid: optionally \gaiaxpy is capable to project these spectra into coefficients space allowing for the simulation of spectra in their continuous representation.

\section{LSF model implementation}\label{sec:lsfDetailed}

The current model for the PSF/LSF representation is based on the preliminary studies and developments described in several technical notes by \citet{LL:LL-068,LL:LL-084,LL:LL-089}. Here we sketch the basic ideas while a complete description of the model is available in \citet{LL:PMN-012}.
The optical PSF $P^O_\lambda$ describes the instantaneous two dimensional intensity distribution in the image of a point source on the CCD: it depends on the optical diffraction of the instrument and the optical aberrations and it is computed as the Fourier transform of the complex amplitude of the incident wavefront in the pupil plane $A$.
Let  $(x, y)$ be linear coordinates in the pupil plane in meters, and $(u, v)$ the corresponding angular coordinates in the image plane in radians, then for a given WFE map $w(x, y)$ we get 
\begin{equation}
A(x, y) = \left\{
\begin{array}{l@{\quad}l}
e^{i\frac{2\pi}{\lambda} w(x, y)} & {\rm for} (x, y) \in {\rm pupil}\\~\\
0                & {\rm otherwise,} 
\end{array}\right.
\end{equation}
and 
\begin{equation}
    P^O_\lambda(u, v) = \frac{1}{\lambda^2 D_x D_y} \left[ \int \int_{-\infty}^{\infty} A(x,y) \,e^{i \frac{2 \pi}{\lambda}(xu + yv)}  \, {\rm d}x \, {\rm d}y\right]^2
    \label{eq:opsf}
,\end{equation}
where $D_x$ and $D_y$ are the pupil dimensions along x and y: \gaia telescopes have two rectangular primary mirrors with dimensions $D_x=1.4510$ m and $D_y = 0.5016$ m.

\begin{figure}[t]
    \centering
    \includegraphics[width=(\textwidth-4mm)/2]{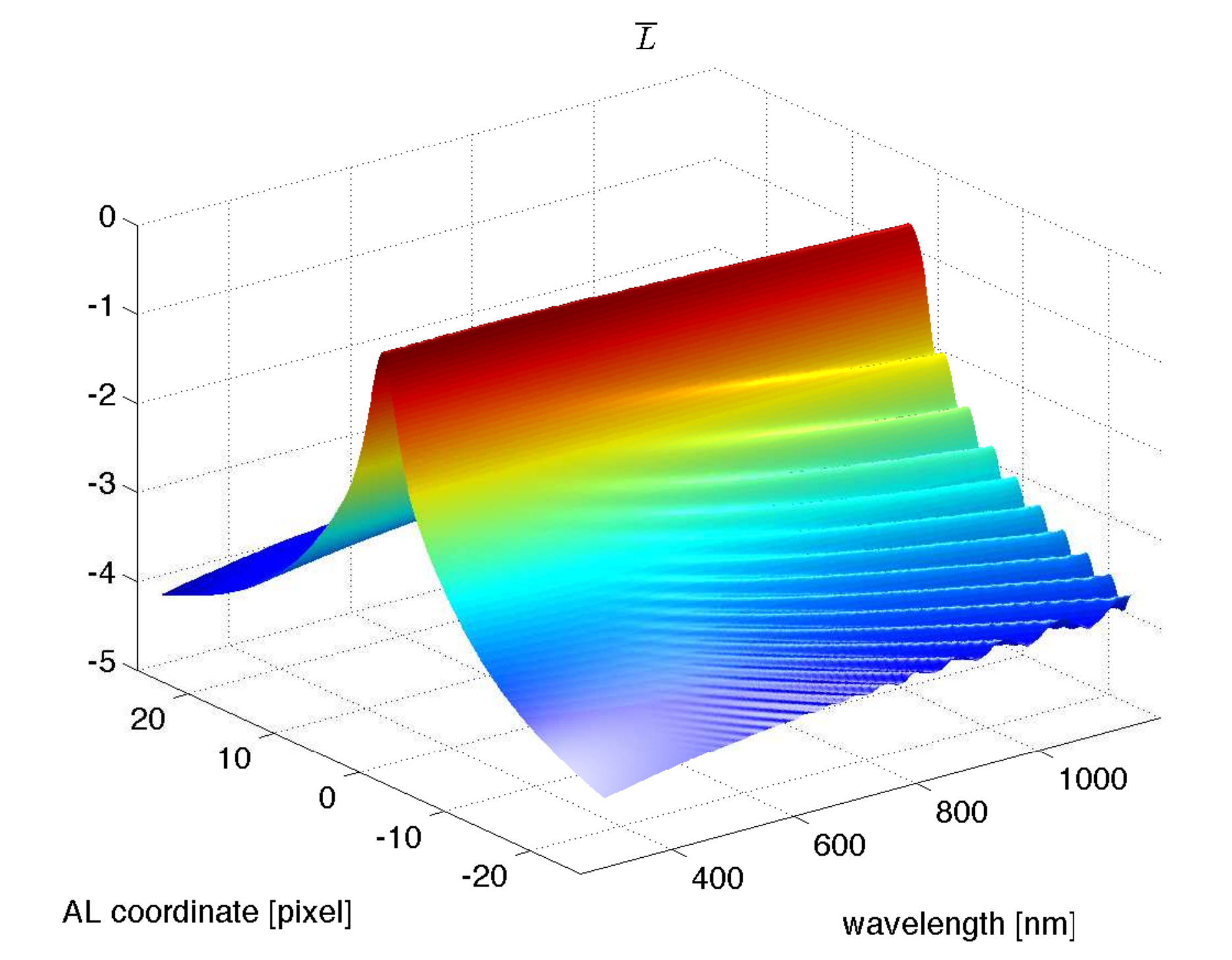}
    \caption{Mean LSF model $\overline{L}$ in logarithmic scale.}
    \label{fig:meanLsf}
\end{figure}

The WFE map depends on the FoV and field angles, but is independent from wavelength and should be relatively stable over time.
However, it is important to understand that the optical PSF is never observed directly: what is observed is the effective PSF, which takes into account the discretised nature of the data as well as the additional smearing introduced by the TDI technique and the effects due to charge diffusion.
The effective PSF $P_m$ can be computed as the inverse Fourier transform of the product between the modulation transfer function (MTF) $M$ and the optical transfer function $O_m$ 
(by definition $O_m$ is the Fourier Transform of the optical PSF, \equref{opsf}).
The total MTF must combine the effect of several spatial response functions, i.e.:
\begin{itemize}
    \item the AL pixel integration (a rectangle of width $p_u$ equal to the CCD AL pixel size expressed in radians);
    \item the AL four-phase TDI charge transfer (a rectangle of width $p_u/n_u$, where $n_p = 4$ is the number of phases per pixel);
    \item the AC pixel integration (a rectangle of width $p_v$ equal to the CCD AC pixel size expressed in radians);
    \item the charge diffusion (modelled as a bivariate normal distribution with diffusion width $\sigma_u = \sigma_v = 4 \mu m$);
\end{itemize}
\begin{equation}
    M(f_x, f_y) = \mathrm{sinc} (\pi f_x p_u) \,
                  \mathrm{sinc} \left(\pi f_x \frac{p_u}{n_p}\right) \,
                  \mathrm{sinc}(\pi f_y p_v) \,
                  e^{-2\pi^2\left(\sigma^2_u f^2_x + \sigma^2_v f^2_y\right)}
,\end{equation}
where $f_x$ and $f_y$ are the spatial frequencies AL and AC.
The effective PSF can then finally computed as:
\begin{equation}
    P_m(u, v) = \frac{1}{4\pi^2} \int \int_{-\infty}^{\infty} O_m(f_x,f_y)\,
    M(f_x, f_y)\,
    e^{-i2\pi (f_x u + f_y v)}  \, {\rm d}f_x \, {\rm d}f_y
    \label{eq:epsf}
.\end{equation}
By definition, the PSF is normalised to unity. The effective LSF $L_m(u)$ is obtained by summing \equref{epsf} over the AC direction.
Among all the effects cited here, the WFE maps modelling optical aberrations are the most uncertain
and as explained in \secref{lsf} numerical maps measured by Airbus are unusable. 
Instead we model optical WFE maps as a linear combination of normalised Legendre polynomials up to the $6^{th}$ order: these maps are randomly generated and scaled to have a root mean square (RMS) uniformly distributed between 40 and 60 nm.
Using this setup we generated up to 5000 random numerical WFE maps to compute the corresponding numerical LSFs: in practice each LSF is sampled on a discrete wavelength grid of $c$
points unevenly distributed between ~288 nm and ~1150 nm with a sampling scheme chosen to guarantee the same relative accuracy of the PSF/LSF computation over the entire wavelength range. Moreover these monochromatic LSFs are sampled over a regular grid of $r$
points centred on $u=0$ and with a spatial resolution high enough to sample the images at least to the Nyquist frequency. Each numerical LSF is  included twice by reversing the AL axis in order to preserve the symmetry of the problem: LSFs are arranged into a stack of matrices $L_i \in \mathbb{R}^{r\times c}$ with $i = 1,\cdots, 10000$.
From this stack 
we calculate the mean LSF $\overline{L}$:
\begin{equation}
\label{eq:lsfMean}
\overline{L} = \frac{1}{n}\sum_{i=1}^n L_i
,\end{equation}
and then we compute a stack of residuals
\begin{equation}
\label{lLsfResidual}
\widetilde{L_i} = L_i - \overline{L}
,\end{equation}
that is reduced to a set of two-dimensional basis functions by means of GPCA.
In practice, an optimal $(\ell_1, \ell_2)$-dimensional axis system, with $\ell_1 < r$, $\ell_2 < c$,  defined by two matrices $U\! \in\mathbb{R}^{r\times \ell_1}$ and $W\! \in\mathbb{R}^{c\times \ell_2}$ with orthonormal columns, 
is derived such that the projections of the data points $\widetilde{L_i}$ onto this axis system have the maximum variance over all the possible $(\ell_1, \ell_2)$-dimensional axis systems, where the variance is defined as:
\begin{equation}
\label{eq:lsfLVariance}
var(U, W) = \frac{1}{n-1}\sum_{i=1}^n \|U^T \cdot \widetilde{L_i}  \cdot W  \|_F
.\end{equation}
The symbol $\|.\|_F$ denotes the Frobenius norm of a matrix.
For given $U$ and $W$ matrices, the projection of $\widetilde{L_i} $ can be computed as
\begin{equation}
\label{eq:projCoeffs}
D_i = U^T \cdot \widetilde{L_i} \cdot W~~~~~~\textrm{with}~~D_i\in \mathbb{R}^{\ell_1 \times \ell_2}
.\end{equation}
From $D_i$ we can reconstruct $\widetilde{L_i} $ by setting
\begin{equation}
\label{eq:lTilde}
\widetilde{L_i}  \approx U \cdot D_i \cdot W^T
,\end{equation}
then 
\begin{equation}
\label{eq:reconstruction}
L_i \approx U \cdot D_i \cdot W^T + \overline{L}
,\end{equation}
or, in extended notation:
\begin{equation}
\label{eq:extendedNotation}
L(u_i, \lambda_j) = \overline{L}_{u_i, \lambda_j} + \sum_{m=1}^{\ell_1}  \sum_{n=1}^{\ell_2} d_{m,n} \cdot U_{i, m} \cdot W_{j, n}
.\end{equation}
The numerical mean LSF $\overline{L}$ is represented in \figref{meanLsf} in logarithmic scale; the second term of \equref{extendedNotation} is the component of the model that is fitted during the optimisation process and represents the deviation of the current model from the mean LSF.

The reconstruction error for $L_i$ is then
\begin{equation}
    \label{eq:lsfError}
     E_i = \| \widetilde{L_i} - U \cdot D_i \cdot W^T \|_F = \|\widetilde{L_i} - U \cdot U^T \cdot\widetilde{L_i} \cdot W \cdot W^T \|_F
.\end{equation}
The RMS error is defined as:
\begin{equation}
\label{eq:lsfLRMSE}
RMSE = \sqrt{\frac{1}{n} \sum_{i=1}^n {E_i}^2}
,\end{equation}
and it measures the average reconstruction error.
\begin{figure}[t]
    \centering
    \includegraphics[width=(\textwidth-4mm)/2]{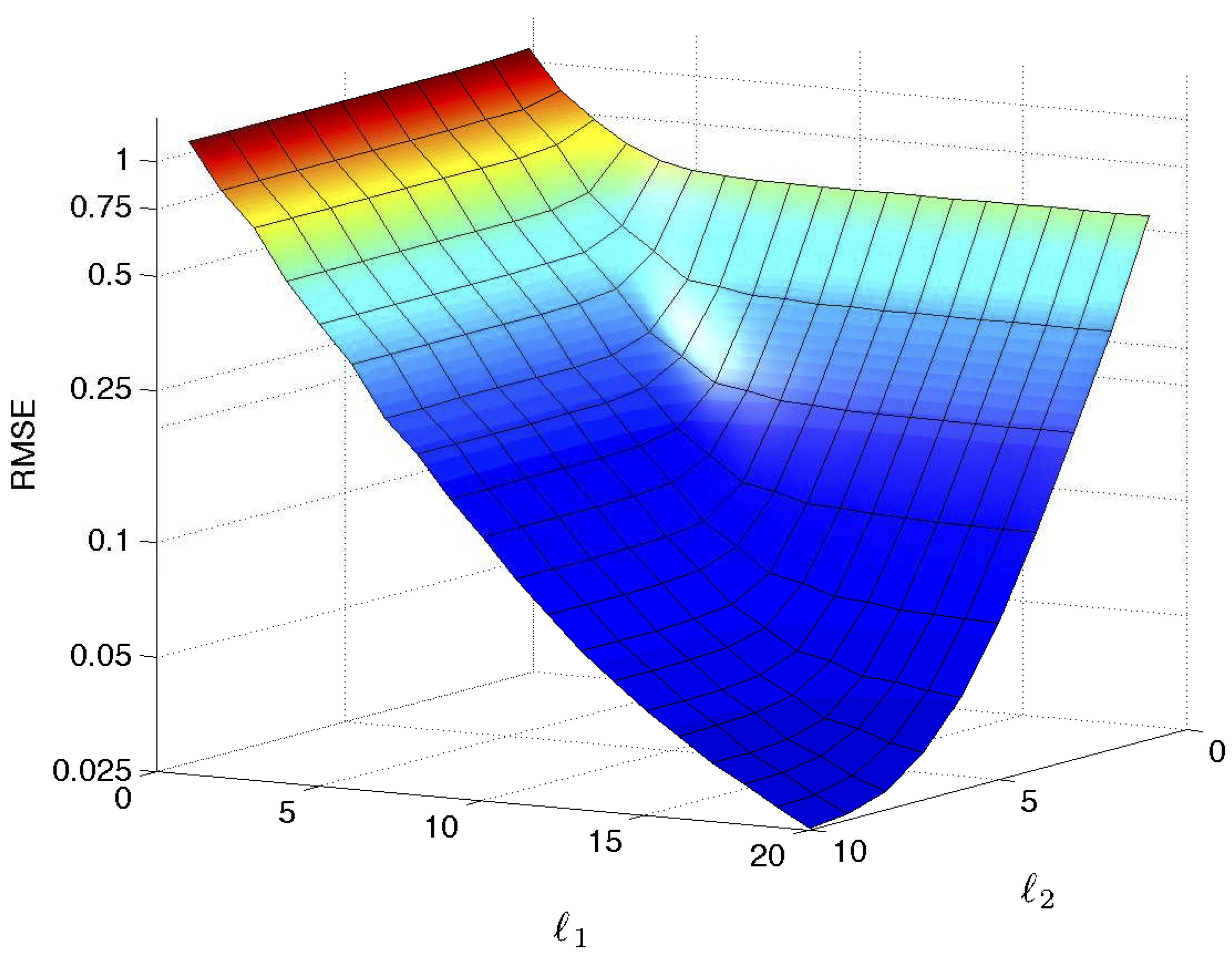}
    \caption{Empirical RMS error for the set of 10000 simulated LSFs measured as function of order $(\ell_1, \ell_2)$.}
    \label{fig:rms}
\end{figure}
Figure \ref{fig:rms} shows the RMS error that has been measured on the complete set of 10000 LSF as a function of $(\ell_1, \ell_2)$: 
the shape of the surface suggests that the LSF wavelength modelling should require about half dimensions with respect to the AL dependency modelling.
The optimal subspace dimensions for GPCA has been set to $(\ell_1, \ell_2) = (20, 10)$. 
Columns of matrices $U$ and $W$ can be interpreted as basis functions to model the dependencies along the spatial coordinate $u$ and the wavelength coordinate $\lambda$ respectively.
The numerical LSF given by \equref{extendedNotation} can be easily 2D-interpolated to continuous variables $(u, \lambda)$ by 1D-interpolation of U and W bases separately. To ensure that the interpolation for the U bases satisfies the 'shift invariant sum' condition ---i.e. preserves the underlying function normalisation independently from the sub-pixel position of the sampling grid--- these bases were then fitted with an S-spline model \citep{LL:LL-084}. The interpolation for $W$ bases is less problematic and is achieved by a cubic spline. The numerical bases are extrapolated in the AL direction by a smooth transition to a Lorentzian profile in the wings, while in the wavelength domain extrapolation is not needed, being the wavelength range of \xp photometers fully covered by the $W$ bases. Finally, an efficient interpolation scheme based on GPCA decomposition has been implemented also for the mean $\overline{L}$, thus enabling to exploit a continuous LSF model as:
\begin{equation}
L(u, \lambda) = \overline{L}(u, \lambda) + \sum_{m=1}^{\ell_1}  \sum_{n=1}^{\ell_2} d_{m,n} \cdot U_{m}(u) \cdot W_{n}(\lambda)
.\end{equation}

\section{Projection of sampled spectra into coefficient space}\label{sec:projCoeffs}

To exploit the projection of spectra from sample space to  coefficients space it is convenient to adopt matrix formalism.
Let 
\begin{equation}
   \textbf{b}^T = {b_1, b_2, \dots, b_{N}}
\end{equation}
represent the vector of $(1\times N)$ coefficients; then a spectrum $\textbf{s}$ sampled on a given grid $\textbf{u}$ of $M$ points will be given by
\begin{equation}
    \textbf{s} = \mathrm{D} \cdot \textbf{b}
,\end{equation}
where $\mathrm{D}\! \in\mathbb{R}^{M\times N}$ is the design matrix of the Hermite functions sampled on the $\textbf{u}$ grid. In principle the design matrix D could be rectangular, and hence not invertible; however we can always compute a pseudo-inverse matrix $\mathrm{D}^\dagger\! \in\mathbb{R}^{N\times U}$ such that 
\begin{equation}
   \textbf{b} = \mathrm{D}^\dagger \cdot \textbf{s}
,\end{equation}
by first multiplying both sides of \equref{sampledSpec}  by $\mathrm{D}^T$ from the left and then by multiplying the results 
by the inverse of matrix $\mathrm{D}^T\cdot \mathrm{D}$ thus obtaining:
\begin{equation}
   \label{eq:pseudoInverse}
\mathrm{D}^\dagger = \left(\mathrm{D}^T\cdot \mathrm{D}\right)^{-1}\cdot \mathrm{D}^T
.\end{equation}
The pseudo-inverse can be computed in a numerically stable way through a singular value decomposition of matrix $\mathrm{D}$ itself:
\begin{equation}
    \mathrm{D} = U\cdot S\cdot V^T
,\end{equation}
where
\begin{equation}
    U^T\cdot U=I
,\end{equation}
and
\begin{equation}
    V^T\cdot V=I
.\end{equation}
The pseudo-inverse is then given by
\begin{equation}
    \mathrm{D}^\dagger = V\cdot S^{-1}\cdot U^T
.\end{equation}

\section{Hipparcos/Tycho synthetic photometry zero points}\label{sec:hipApp}

\begin{figure}
\centerline{
\includegraphics[width=\columnwidth/1]{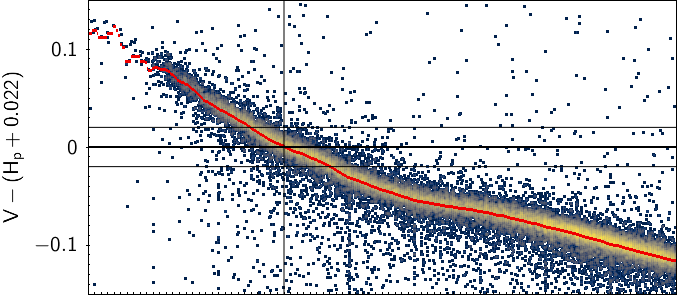}}
\centerline{
\includegraphics[width=\columnwidth/1]{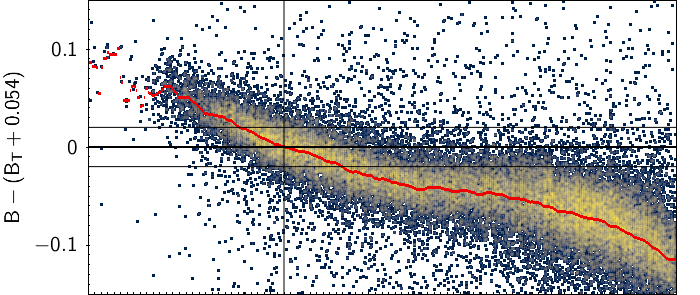}}
\centerline{
\includegraphics[width=\columnwidth/1]{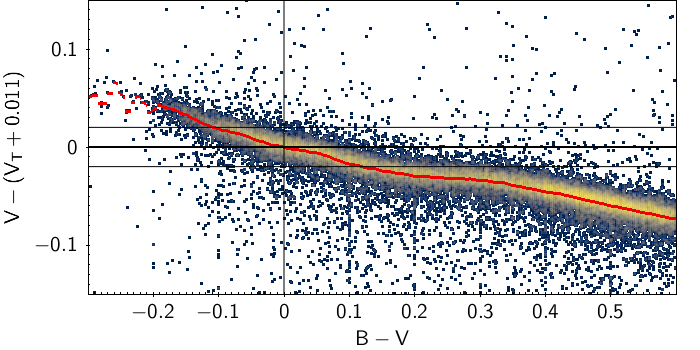}}
\caption{Calibration of the zero point for the Hipparcos/Tycho passbands.}
\label{fig:hipZp}
\end{figure}

Magnitude scales for Hipparcos passbands were chosen such that $H_p = V_T = V$ and $B_T = B$  at $B-V = 0$ \citep{Hippaphot1997}. In order to reproduce these scales we need to adjust the zero points of the synthetic ECS photometry accordingly. In order to do so 
we have evaluated the offsets $\delta_i$ to let the median of the distributions $V-(H_p+\delta_1)$, $B-(B_T+\delta_2)$ and $V-(V_T+\delta_3)$ to be equal to zero at $B-V=0$. Residuals with the zero point correction applied are shown on \figref{hipZp} while  
the assumed zero point values can be retrieved in \secref{hipparcos}.
 
\newpage
\section{\gaia-related acronyms}\label{sec:acronyms}

\begin{table}[hp]
    \caption{\gaia-related acronyms used in the paper. Each acronym is also defined at its first occurrence in the paper.}
    \label{tab:acronyms}
    \centering
    \begin{tabular}{l|l|l}\hline\hline
Acronym & Description & See \\\hline

AC & ACross scan direction & \secref{instrumentModel} \\
AL & ALong scan direction & \secref{instrumentModel} \\
BP & Blue Photometer & \secref{introduction} \\
CCD(s) & Charge Coupled Device(s) & \secref{instrumentModel} \\
CTI & Charge Transfer Inefficiency & \secref{instrumentModel} \\
DEA & Differential Evolution Algorithm & \secref{processing} \\
DR & Data Release  & \secref{introduction} \\
DS &  Defence and Space  & \secref{dispersion} \\
ECS & Externally Calibrated Spectrum & \secref{introduction} \\
ELS & Emission Line Sources  & \secref{calibrators} \\
ESA & European Space Mission & \secref{introduction} \\
FoV(s) &  Field(s) of View & \secref{dispersion} \\
FWHM & Full Width at Half Maximum & \secref{specresol} \\
G, G$_{BP}$, G$_{BP}$ & Integrated \gaia magnitudes &\secref{calibrators} \\
\multirow{2}{*}{GPCA} & Generalised Principal  & \multirow{2}{*}{\secref{lsf}} \\
 & Component Analysis &  \\
HRS & High Resolution Spectrum & \secref{synthphot} \\
\multirow{2}{*}{ICS} & Internally Calibrated BP/RP   & \multirow{2}{*}{\secref{specresol}} \\
 & mean Spectra &\\
IM & Instrument model & \secref{introduction} \\
JKC & Johnson-Kron-Cousins  & \secref{synthphot} \\
LSF &  Line Spread Function  & \secref{introduction} \\
MTF & Modulation Transfer Function & \secref{lsfDetailed} \\
NGSL &  Next Generation Spectral Library  & \secref{calibrators} \\
PSF & Point Spread Function & \secref{instrumentModel} \\
PVL & Passband Validation Library   & \secref{calibrators} \\
QSO & Quasi-stellar object &  \secref{calibrators} \\
RMS & Root Mean Square & \secref{lsfDetailed} \\
SDSS & Sloan Digital Sky Survey &\secref{calibrators} \\
SED & Spectral Energy Distribution   & \secref{introduction} \\
SPD & Spectral Photon Distribution   & \secref{overview} \\
\multirow{2}{*}{SPSS} & Spectro-Photometric Standard   & \multirow{2}{*}{\secref{calibrators}} \\
& Stars & \\
RP & Red Photometer & \secref{introduction} \\
TDI & Time Delayed Integration & \secref{instrumentModel} \\
WC(s) & Window Class or strategy & \secref{instrumentModel} \\
WFE & WaveFront Error & \secref{lsf} \\

\end{tabular}
\end{table}

\end{appendix}

\end{document}